\documentclass[preprint, 5p]{elsarticle}
\title{Efficient flexible boundary conditions for long dislocations}
\author[]{M.~Hodapp}
\address{\vspace{0.5em}Skolkovo Institute of Science and Technology (Skoltech), Center for Energy Science and Technology, Moscow (RU)}
\usepackage{amssymb}
\usepackage{mathrsfs}
\usepackage{amsthm}
\usepackage[nonumberlist,nopostdot,sort=def,nowarn,acronym]{glossaries} 
\usepackage{glossary-mcols} 
\usepackage{color,upgreek}
\usepackage[ruled,lined]{algorithm2e}
\usepackage{setspace}
\usepackage{pifont}
\usepackage{array}
\usepackage{multirow}
\usepackage{multicol}
\usepackage{arydshln}
\usepackage{titlesec}
\usepackage{appendix}
\usepackage{fancybox,framed} 
\usepackage{manfnt} 
\usepackage{scrextend}
\usepackage{float}
\usepackage{enumitem}
\usepackage{mathtools} 
\usepackage{textcomp} 
\usepackage[bottom]{footmisc}
\usepackage{subdepth}
\usepackage{empheq}
\usepackage{hyperref}
\usepackage[dvipsnames]{xcolor}
\newcommand{\sty}[1]{\boldsymbol{#1}}
\newcommand{\styy}[1]{\mathbb{#1}}
\newcommand{\ubar}[1]{\mkern 0.5mu\underline{\mkern-0.5mu#1\mkern-0.5mu}\mkern 0.5mu}
\newcommand{\uubar}[1]{\ubar{\ubar{#1}}}
\let\mepsilon\epsilon
\let\epsilon\varepsilon

\let\theta\vartheta
\let\mrho\rho
\let\rho\varrho
\let\mphi\phi
\let\phi\varphi
\let\Gamma\varGamma
\let\Delta\varDelta
\let\Theta\varTheta
\let\Lambda\varLambda
\let\Xi\varXi
\let\Pi\varPi
\let\Sigma\varSigma
\let\Upsilon\varUpsilon
\let\Phi\varPhi
\let\Psi\varPsi

\let\Omega\varOmega

\newcommand{\grad}[3]{\nabla_{#1}^{#2} #3} 
\newcommand{\var}[3]{\delta_{#1}^{#2} #3} 

\newcommand{\nrm}[1]{\| #1 \|}
\newcommand{\abs}[1]{\vert #1 \vert}

\newcommand{\FT}[1]{\mathcal{F}\left\{#1\right\}} 
\newcommand{\FTinv}[1]{\mathcal{F}^{-1}\left\{#1\right\}} 

\newcommand{\bmb}{\sty{b}}
\newcommand{\bmc}{\sty{c}}

\newcommand{\bme}{\sty{e}}
\newcommand{\bmf}{\sty{f}}

\newcommand{\bmk}{\sty{k}}

\newcommand{\bmm}{\sty{m}}
\newcommand{\bmn}{\sty{n}}

\newcommand{\bmp}{\sty{p}}

\newcommand{\bmr}{\sty{r}}

\newcommand{\bmt}{\sty{t}}
\newcommand{\bmu}{\sty{u}}
\newcommand{\bmv}{\sty{v}}

\newcommand{\bmx}{\sty{x}}

\newcommand{\bmA}{\sty{A}}
\newcommand{\bmB}{\sty{B}}
\newcommand{\bmC}{\sty{C}}
\newcommand{\bmD}{\sty{D}}

\newcommand{\bmF}{\sty{F}}
\newcommand{\bmG}{\sty{G}}

\newcommand{\bmK}{\sty{K}}
\newcommand{\bmL}{\sty{L}}
\newcommand{\bmM}{\sty{M}}

\newcommand{\bmP}{\sty{P}}
\newcommand{\bmQ}{\sty{Q}}

\newcommand{\bmalpha}{\sty{\alpha}}

\newcommand{\bmmeps}{\sty{\mepsilon}}

\newcommand{\bmZero}{\sty{0}}

\newcommand{\bbN}{\styy{N}}

\newcommand{\bbP}{\styy{P}}

\newcommand{\bbR}{\styy{R}}

\newcommand{\bbZ}{\styy{Z}}

\newcommand{\clB}{\mathcal{B}}

\newcommand{\clE}{\mathcal{E}}
\newcommand{\clF}{\mathcal{F}}
\newcommand{\clG}{\mathcal{G}}

\newcommand{\clI}{\mathcal{I}}

\newcommand{\clL}{\mathcal{L}}

\newcommand{\clO}{\mathcal{O}}
\newcommand{\clP}{\mathcal{P}}

\newcommand{\clR}{\mathcal{R}}

\newcommand{\sT}{\mathsf{T}} 

\newcommand{\scH}{\mathscr{H}}

\newcommand{\scS}{\mathscr{S}}

\newcommand{\rma}{\mathrm{a}}

\newcommand{\rmc}{\mathrm{c}}
\newcommand{\rmd}{\mathrm{d}}
\newcommand{\rme}{\mathrm{e}}

\newcommand{\rmh}{\mathrm{h}}
\newcommand{\rmi}{\mathrm{i}}

\newcommand{\rmp}{\mathrm{p}}

\newcommand{\rmx}{\mathrm{x}}

\newcommand{\uf}{\ubar{f}}

\newcommand{\uu}{\ubar{u}}
\newcommand{\uv}{\ubar{v}}
\newcommand{\uw}{\ubar{w}}

\newcommand{\uuG}{\uubar{G}}

\newcommand{\uuR}{\uubar{R}}

\usepackage{accents}
\robustify{\accentset} 

\newglossary{symbols}{sym}{sbl}{List of Symbols}
\newglossary[glignoredl]{ignored}{glignored}{glignoredin}{Ignored Glossary} 

\newcommand{\mynewglossary}[4]{%
 \newglossaryentry{#2}{type=#1, name={#3}, description={#4}, sort={#2}}%
 \expandafter\newcommand\expandafter{\csname #2\endcsname}{\gls{#2}\xspace}%
}

\newcommand{\mynewglossaryB}[7]{%
 \newglossaryentry{#2}{type=#1, name={#5, #6}, user1={#5}, user2={#6}, description={#7}, sort={#2}}
 \expandafter\newcommand\expandafter{\csname #3\endcsname}{\glsuseri{#2}\xspace}%
 \expandafter\newcommand\expandafter{\csname #4\endcsname}{\glsuserii{#2}\xspace}%
}

\newcommand{\mynewglossaryC}[9]{%
 \newglossaryentry{#2}{type=#1, name={#6, #7, #8}, user1={#6}, user2={#7}, user3={#8}, description={#9}, sort={#2}}
 \expandafter\newcommand\expandafter{\csname #3\endcsname}{\glsuseri{#2}\xspace}%
 \expandafter\newcommand\expandafter{\csname #4\endcsname}{\glsuserii{#2}\xspace}%
 \expandafter\newcommand\expandafter{\csname #5\endcsname}{\glsuseriii{#2}\xspace}%
}

\newcommand{\indDomC}  {\mathrm{c}}                                     
\newcommand{\indDomA}  {\mathrm{a}}                                     
\newcommand{\indDomP}  {\mathrm{p}}                                     
\newcommand{\indHarm}  {\mathrm{h}}                                     
\newcommand{\indAHarm} {\mathrm{ah}}                                    
\newcommand{\indDomI}  {\mathrm{i}}                                     
\newcommand{\indDomIm} {\textnormal{\ensuremath{\mathrm{i\texttt{-}}}}} 
\newcommand{\indDomIpl}{\textnormal{\ensuremath{\mathrm{i\texttt{+}}}}} 
\newcommand{\indCGF}   {\mathrm{cgf}}                                   
\newcommand{\indLGF}   {\mathrm{lgf}}                                   

\mynewglossary{symbols}{real}     {\ensuremath{\bbR}}   {Set of real numbers}
\mynewglossary{symbols}{nat}      {\ensuremath{\bbN}}   {Set of natural numbers}
\mynewglossary{symbols}{integ}    {\ensuremath{\bbZ}}   {Set of integers}
\mynewglossary{ignored}{euclTwo}  {\ensuremath{\real^2}}{Two-dimensional Euclidean space}
\mynewglossary{ignored}{euclThree}{\ensuremath{\real^3}}{Three-dimensional Euclidean space}
\mynewglossary{ignored}{euclD}    {\ensuremath{\real^d}}{$d$-dimensional Euclidean space}

\mynewglossary{symbols}{bnrmVec}{\ensuremath{\bmn}}   {Unit normal vector}
\mynewglossary{symbols}{bperm}  {\ensuremath{\bmmeps}}{Third-order permutation tensor (Levi-Civita tensor)}
\mynewglossary{symbols}{dirac}  {\ensuremath{\delta}} {Dirac delta function}

\mynewglossary{symbols}{lat}       {\ensuremath{\Lambda}}              {Bravais lattice}
\mynewglossary{ignored}{latA}      {\ensuremath{\lat^\indDomA}}        {Atomic lattice}
\mynewglossary{ignored}{latC}      {\ensuremath{\lat^\indDomC}}        {Virtual lattice associated with a continuum domain}
\mynewglossary{ignored}{latP}      {\ensuremath{\lat^\indDomP}}        {Virtual lattice associated with a pad domain}
\mynewglossary{symbols}{domRef}    {\ensuremath{\Omega_0}}             {Reference placement of a continuous body}
\mynewglossary{ignored}{domRefA}   {\ensuremath{\domRef^\indDomA}}     {Reference continuum domain associated with an atomistic region}
\mynewglossary{ignored}{domRefC}   {\ensuremath{\domRef^\indDomC}}     {Reference continuum domain}
\mynewglossary{ignored}{domRefP}   {\ensuremath{\domRef^\indDomP}}     {Reference pad domain}
\mynewglossary{symbols}{dom}       {\ensuremath{\Omega}}               {Current placement of a Continuous body}
\mynewglossary{ignored}{domA}      {\ensuremath{\dom^\indDomA}}        {Continuum domain associated with an atomistic region}
\mynewglossary{ignored}{domC}      {\ensuremath{\dom^\indDomC}}        {Continuum domain}
\mynewglossary{ignored}{domP}      {\ensuremath{\dom^\indDomP}}        {Pad domain}
\mynewglossary{ignored}{domRefCore}{\ensuremath{\domRef^\mathrm{core}}}{Dislocation core domain (reference configuration)}
\mynewglossary{symbols}{domCore}   {\ensuremath{\dom^\mathrm{core}}}   {Dislocation core region}
\mynewglossary{symbols}{intf}      {\ensuremath{\Gamma_\rmi}}          {Artificial sharp interface between the atomistic and the continuum domain}

\mynewglossary{ignored}{mpRef}  {\ensuremath{X}}               {Scalar material point in the reference configuration}
\mynewglossary{ignored}{mpRefB} {\ensuremath{Y}}               {Scalar material point in the reference configuration (2)}
\mynewglossary{symbols}{bmpRef} {\ensuremath{\sty{\mpRef}}}    {Material point in the reference configuration}
\mynewglossary{ignored}{bmpRefB}{\ensuremath{\sty{\mpRefB}}}   {Material point in the reference configuration (2)}
\mynewglossary{ignored}{mpCur}  {\ensuremath{x}}               {Scalar material point in the current configuration}
\mynewglossary{ignored}{mpCurB} {\ensuremath{y}}               {Scalar material point in the current configuration (2)}
\mynewglossary{symbols}{bmpCur} {\ensuremath{\sty{\mpCur}}}    {Material point in the current configuration}
\mynewglossary{ignored}{bmpCurB}{\ensuremath{\sty{\mpCurB}}}   {Material point in the current configuration (2)}
\mynewglossary{symbols}{motion} {\ensuremath{\chi}}            {Motion of a material point \bmpRef}
\mynewglossary{ignored}{displ}  {\ensuremath{u}}               {Scalar displacement field}
\mynewglossary{ignored}{bdispl} {\ensuremath{\sty{\displ}}}    {Displacement field}
\mynewglossary{ignored}{bdisplA}{\ensuremath{\bdispl^\indDomA}}{Atomistic displacement associated with \latA}
\mynewglossary{ignored}{bdisplC}{\ensuremath{\bdispl^\indDomC}}{Atomistic displacement associated with \domC}
\mynewglossary{ignored}{bdisplP}{\ensuremath{\bdispl^\indDomP}}{Atomistic displacement associated with \latP}
\mynewglossary{ignored}{vel}    {\ensuremath{v}}               {Scalar velocity}
\mynewglossary{ignored}{bvel}   {\ensuremath{\sty{\vel}}}      {Velocity vector}

\mynewglossaryC{symbols}{atos} {ato} {atoB} {atoC} {\ensuremath{\xi}}      {\ensuremath{\eta}}      {\ensuremath{\zeta}}      {Atomic indices}
\mynewglossaryC{symbols}{batos}{bato}{batoB}{batoC}{\ensuremath{\sty{\xi}}}{\ensuremath{\sty{\eta}}}{\ensuremath{\sty{\zeta}}}{Positions of atoms \ato, \atoB, \atoC}

\mynewglossary{ignored}{defGrad}   {\ensuremath{F}}                   {Scalar deformation gradient}
\mynewglossary{symbols}{bdefGrad}  {\ensuremath{\sty{\defGrad}}}      {Deformation gradient}
\mynewglossary{symbols}{bdefGradE} {\ensuremath{\bdefGrad^\rme}}      {Elastic contribution of \bdefGrad}
\mynewglossary{symbols}{bdefGradP} {\ensuremath{\bdefGrad^\rmp}}      {Plastic contribution of \bdefGrad}
\mynewglossary{ignored}{strnCG}    {\ensuremath{C}}                   {Scalar right Cauchy-Green strain tensor}
\mynewglossary{symbols}{bstrnCG}   {\ensuremath{\sty{\strnCG}}}       {Right Cauchy-Green strain tensor}
\mynewglossary{ignored}{strnLag}   {\ensuremath{E}}                   {Scalar right Lagrange strain tensor}
\mynewglossary{ignored}{bstrnLag}  {\ensuremath{\sty{\strnLag}}}      {Lagrange strain tensor}
\mynewglossary{ignored}{strnSml}   {\ensuremath{\epsilon}}            {Scalar small strain tensor}
\mynewglossary{symbols}{bstrnSml}  {\ensuremath{\sty{\strnSml}}}      {Small strain tensor}
\mynewglossary{symbols}{bstrnSmlEl}{\ensuremath{\bstrnSml^\rme}}      {Elatic part of the small strain tensor}
\mynewglossary{symbols}{bstrnSmlPl}{\ensuremath{\bstrnSml^\rmp}}      {Plastic part of the small strain tensor}
\mynewglossary{symbols}{bstrnSmlNs}{\ensuremath{\bstrnSml^\mathrm{ns, p}}}{Nonsingular plastic strain}
\mynewglossary{ignored}{bdist}     {\ensuremath{\sty{\beta}}}         {Distortion tensor}
\mynewglossary{symbols}{bdistEl}   {\ensuremath{\sty{\beta}^\rme}}    {Elastic distortion}
\mynewglossary{symbols}{bdistPl}   {\ensuremath{\sty{\beta}^\rmp}}    {Plastic distortion}
\mynewglossary{symbols}{bdistNs}   {\ensuremath{\sty{\beta}^\mathrm{p, ns}}}{Nonsingular plastic distortion}

\mynewglossary{symbols}{Efree}   {\ensuremath{\psi}}               {Helmholtz free energy density}
\mynewglossary{symbols}{Etot}    {\ensuremath{\Pi}}                {Total energy of a mechanical system}
\mynewglossary{ignored}{EtotA}   {\ensuremath{\Etot^\indDomA}}     {Total atomistic energy}
\mynewglossary{ignored}{EtotC}   {\ensuremath{\Etot^\indDomC}}     {Total continuum energy}
\mynewglossary{symbols}{EtotPot} {\ensuremath{\Etot^\mathrm{pot}}} {Potential energy of a mechanical system}
\mynewglossary{symbols}{EtotKin} {\ensuremath{\Etot^\mathrm{kin}}} {Kinetic energy of a mechanical system}
\mynewglossary{symbols}{EtotInt} {\ensuremath{\Etot^\mathrm{int}}} {Internal energy of a mechanical system}
\mynewglossary{symbols}{EtotExt} {\ensuremath{\Etot^\mathrm{ext}}} {External energy of a mechanical system}
\mynewglossary{symbols}{EtotEl}  {\ensuremath{\Etot^\rme}}         {Elastic part of \Etot}
\mynewglossary{symbols}{EtotCore}{\ensuremath{\Etot^\mathrm{core}}}{Dislocation core energy}
\mynewglossary{symbols}{Eato}    {\ensuremath{\clE_\xi}}           {Site energy of an atom \ato}
\mynewglossary{symbols}{Edd}     {\ensuremath{W}}                  {Energy per unit length of the dislocation line}
\mynewglossary{symbols}{Eel}     {\ensuremath{\Edd^\mathrm{e}}}    {Elastic energy per unit length of the dislocation line}
\mynewglossary{symbols}{Ecore}   {\ensuremath{\Edd^\mathrm{core}}} {Core energy per unit length of the dislocation line}

\mynewglossary{ignored}{stressSml}  {\ensuremath{\sigma}}          {Scalar Cauchy stress}
\mynewglossary{symbols}{bstressSml} {\ensuremath{\sty{\stressSml}}}{Cauchy stress tensor}
\mynewglossary{symbols}{stressShear}{\ensuremath{\tau}}            {Scalar shear stress}
\mynewglossary{symbols}{bstressPK}  {\ensuremath{\bmP}}            {First Piola-Kirchhoff stress tensor}
\mynewglossary{symbols}{btract}     {\ensuremath{\bmt}}            {Traction vector}
\mynewglossary{symbols}{bstressEsh} {\ensuremath{\bmB}}            {Eshelby stress tensor}

\mynewglossary{ignored}{force} {\ensuremath{f}}                    {Scalar force}
\mynewglossary{symbols}{bforce}{\ensuremath{\sty{\force}}}         {Force vector}
\mynewglossary{ignored}{bfA}   {\ensuremath{\bforce^\indDomA}}     {Atomic force vector}
\mynewglossary{ignored}{bfC}   {\ensuremath{\bforce^\indDomC}}     {Continuum force vector}
\mynewglossary{symbols}{bfbody}{\ensuremath{\bforce^\mathrm{body}}}{Body force}
\mynewglossary{symbols}{bfPK}  {\ensuremath{\bforce^\mathrm{pk}}}  {Peach-Koehler force}
\mynewglossary{symbols}{bfCore}{\ensuremath{\bforce^\mathrm{core}}}{Force on \DDcur due to core energy contribution \EtotCore}

\mynewglossary{ignored}{stiffMat} {\ensuremath{C}}               {Scalar material stiffness}
\mynewglossary{symbols}{bstiffMat}{\ensuremath{\styy{\stiffMat}}}{Material stiffness tensor}
\mynewglossary{symbols}{shearMod} {\ensuremath{\mu}}             {Shear modulus}
\mynewglossary{symbols}{poisRatio}{\ensuremath{\nu}}             {Poisson ratio}
\mynewglossary{ignored}{green}    {\ensuremath{G}}               {Scalar entry of the Green tensor}
\mynewglossary{symbols}{bgreen}   {\ensuremath{\sty{\green}}}    {Second-order Green tensor}

\mynewglossary{symbols}{slipPlane}     {\ensuremath{\scS}}                        {Slip plane}
\mynewglossary{ignored}{DDref}         {\ensuremath{\Gamma}}                      {Discrete dislocation line with respect to a reference configuration}
\mynewglossary{ignored}{DDrefA}        {\ensuremath{\DDref^\indDomA}}             {Discrete dislocation line in the atomistic domain with respect to a reference configuration}
\mynewglossary{ignored}{DDrefC}        {\ensuremath{\DDref^\indDomC}}             {Discrete dislocation line in the continuum domain with respect to a reference configuration}
\mynewglossary{symbols}{DDrefHyb}      {\ensuremath{\DDref^\mathrm{hyb}}}         {Hybrid discrete dislocation line with respect to a reference configuration}
\mynewglossary{symbols}{DDcur}         {\ensuremath{\gamma}}                      {Discrete dislocation line}
\mynewglossary{ignored}{DDa}           {\ensuremath{\DDcur^\indDomA}}             {Discrete dislocation line in the atomistic domain}
\mynewglossary{ignored}{DDc}           {\ensuremath{\DDcur^\indDomC}}             {Discrete dislocation line in the continuum domain}
\mynewglossary{symbols}{DDhyb}         {\ensuremath{\DDcur^\mathrm{hyb}}}         {Hybrid discrete dislocation line}
\mynewglossary{symbols}{DDvir}         {\ensuremath{\DDcur^\mathrm{v}}}           {Virtual subset of a discrete dislocation line outside a bounded domain}
\mynewglossary{ignored}{mpDDref}       {\ensuremath{S}}                           {Scalar material point on \DDref}
\mynewglossary{ignored}{mpDDcur}       {\ensuremath{s}}                           {Scalar material point on \DDcur}
\mynewglossary{symbols}{bmpDDref}      {\ensuremath{\sty{\mpDDref}}}              {Material point on \DDref}
\mynewglossary{symbols}{bmpDDcur}      {\ensuremath{\sty{\mpDDcur}}}              {Material point on \DDcur}
\mynewglossary{symbols}{btanVecDD}     {\ensuremath{\bmt}}                        {Unit tangent vector of the dislocation line}
\mynewglossary{symbols}{bnrmVecDD}     {\ensuremath{\bmm}}                        {Unit normal vector of the dislocation line}
\mynewglossary{symbols}{burgers}       {\ensuremath{b}}                           {Magnitude of the Burgers vector}
\mynewglossary{symbols}{bburgers}      {\ensuremath{\sty{\burgers}}}              {Burgers vector}
\mynewglossary{symbols}{bdislDens}     {\ensuremath{\bmalpha}}                    {Dislocation density tensor}
\mynewglossary{symbols}{cutOffCore}    {\ensuremath{r_\mathrm{cut}^\mathrm{core}}}{Dislocation core cut-off radius}
\mynewglossary{symbols}{charAngle}     {\ensuremath{\theta}}                      {Character angle of the dislocation line}
\mynewglossary{ignored}{EcoreParam}    {\ensuremath{E^\mathrm{core}}}             {Scalar core energy parameter}
\mynewglossary{ignored}{spreadIso}     {\ensuremath{w}}                           {Isotropic spreading function of the Burgers vector}
\mynewglossary{ignored}{spreadIsoW}    {\ensuremath{a}}                           {Isotropic core spreading width}
\mynewglossary{symbols}{bmobility}     {\ensuremath{\bmM}}                        {Dislocation mobility tensor}
\mynewglossary{symbols}{bdrag}         {\ensuremath{\bmD}}                        {Dislocation drag tensor}
\mynewglossary{ignored}{displDD}       {\ensuremath{\tilde{\displ}}}              {Scalar displacement of an isolated dislocation in an unbounded domain}
\mynewglossary{ignored}{displCorr}     {\ensuremath{\hat{\displ}}}                {Scalar corrective displacement}
\mynewglossary{ignored}{bdisplDD}      {\ensuremath{\tilde{\bdispl}}}             {Displacement field of due to a dislocation}
\mynewglossary{ignored}{bdisplCorr}    {\ensuremath{\hat{\bdispl}}}               {Corrective displacement field}
\mynewglossary{ignored}{bstressSmlDD}  {\ensuremath{\tilde{\bstressSml}}}         {Stress field of an isolated dislocation in an unbounded domain}
\mynewglossary{ignored}{bstressSmlCorr}{\ensuremath{\hat{\bstressSml}}}           {Corrective stress field}

\mynewglossary{ignored}{cutOff}  {\ensuremath{r_\mathrm{cut}}}{Cut-off radius}
\mynewglossary{ignored}{intRg}   {\ensuremath{\clR}}          {Interaction range}
\mynewglossary{ignored}{intRgAto}{\ensuremath{\clR_{\ato}}}   {Interaction range of atom \bato}
\mynewglossary{ignored}{latConst}{\ensuremath{a_0}}           {Lattice constant}

\mynewglossary{symbols}{intvar}{\ensuremath{\hat{\alpha}}}{Vector of internal state variables}

\mynewglossary{symbols}{Prob}      {\ensuremath{\clP}}                       {Shorthand notation of a problem definition}
\mynewglossary{ignored}{Pa}        {\ensuremath{\Prob^\indDomA}}             {Atomistic problem}
\mynewglossary{ignored}{Pc}        {\ensuremath{\Prob^\indDomC}}             {Continuum problem}
\mynewglossary{symbols}{Pp}        {\ensuremath{\Prob^\rmp}}                 {Physical problem}
\mynewglossary{symbols}{PcP}       {\ensuremath{\Prob^{\indDomC/\rmp}}}      {Physical subproblem}
\mynewglossary{symbols}{PcDD}      {\ensuremath{\Prob^{\indDomC/\rmd\rmd}}}  {Material subproblem}
\mynewglossary{symbols}{Pcadd}     {\ensuremath{\Prob^\mathrm{cadd}}}        {CADD problem}
\mynewglossary{symbols}{PcaddApprx}{\ensuremath{\tilde{\Prob}^\mathrm{cadd}}}{Approximate CADD problem}

\mynewglossary{symbols}{bdisplDDt}   {\ensuremath{\tilde{\bdispl}^\mathrm{t}}}         {Core template for an infinite straight dislocation}
\mynewglossary{symbols}{bdisplDDcorr}{\ensuremath{\Delta\tilde{\bdispl}^\mathrm{corr}}}{Core template correction}
\mynewglossary{symbols}{transNode}   {\ensuremath{\bmpDDcur^\mathrm{trans}}}           {Transmission node}
\mynewglossary{symbols}{domDetn}     {\ensuremath{\dom^\mathrm{detn}}}                 {Detection domain}

\mynewglossary{ignored}{latI}  {\ensuremath{\lat^\indDomI}}  {Interface region}
\mynewglossary{ignored}{latIm} {\ensuremath{\lat^\indDomIm}} {Interface (-) region}
\mynewglossary{ignored}{latIpl}{\ensuremath{\lat^\indDomIpl}}{Interface (+) region}

\mynewglossary{ignored}{bdisplI}  {\ensuremath{\bdispl^\indDomI}}  {Atomistic displacement associated with \latI}
\mynewglossary{ignored}{bdisplIm} {\ensuremath{\bdispl^\indDomIm}} {Atomistic displacement associated with \latIm}
\mynewglossary{ignored}{bdisplIpl}{\ensuremath{\bdispl^\indDomIpl}}{Atomistic displacement associated with \latIpl}

\mynewglossary{ignored}{diffOp} {\ensuremath{\clL}}                         {Differential operator}
\mynewglossary{ignored}{Lh}    {\ensuremath{\diffOp_\indHarm}}    {Linearized operator}
\mynewglossary{ignored}{Lah}   {\ensuremath{\diffOp_\indAHarm}}   {Anharmonic operator}
\mynewglossary{ignored}{La}   {\ensuremath{\diffOp^\indDomA}}               {Nonlinear atomistic operator}
\mynewglossary{ignored}{LaH}    {\ensuremath{\diffOp_\indHarm^\indDomA}}    {Linearized atomistic operator}
\mynewglossary{ignored}{LaAH}   {\ensuremath{\diffOp_\indAHarm^\indDomA}}   {Anharmonic atomistic operator}
\mynewglossary{ignored}{Lc}     {\ensuremath{\diffOp_\indHarm^\indDomC}}             {Continuum operator}
\mynewglossary{ignored}{Laa}    {\ensuremath{\diffOp_\indHarm^{\indDomA/\indDomA}}}  {Coupling operator}
\mynewglossary{ignored}{Lcc}    {\ensuremath{\diffOp_\indHarm^{\indDomC/\indDomC}}}  {Coupling operator}
\mynewglossary{ignored}{Lac}    {\ensuremath{\diffOp_\indHarm^{\indDomA/\indDomC}}}  {Coupling operator}
\mynewglossary{ignored}{Lca}    {\ensuremath{\diffOp_\indHarm^{\indDomC/\indDomA}}}  {Coupling operator}
\mynewglossary{ignored}{Lic}    {\ensuremath{\diffOp_\indHarm^{\indDomI/\indDomC}}}  {Coupling operator}
\mynewglossary{ignored}{Lii}    {\ensuremath{\diffOp_\indHarm^{\indDomI/\indDomI}}}  {Coupling operator}
\mynewglossary{ignored}{Lci}    {\ensuremath{\diffOp_\indHarm^{\indDomC/\indDomI}}}  {Coupling operator}
\mynewglossary{ignored}{Liipl}  {\ensuremath{\diffOp_\indHarm^{\indDomI/\indDomIpl}}}{Coupling operator}
\mynewglossary{ignored}{Lipla}  {\ensuremath{\diffOp_\indHarm^{\indDomIpl/\indDomA}}}{Coupling operator}
\mynewglossary{ignored}{Lipli}  {\ensuremath{\diffOp_\indHarm^{\indDomIpl/\indDomI}}}{Coupling operator}
\mynewglossary{ignored}{greenOp}{\ensuremath{\clG}}                         {Green operator}
\mynewglossary{ignored}{Ga}     {\ensuremath{\greenOp^\indDomA}}            {Atomistic Green operator}
\mynewglossary{ignored}{Gc}     {\ensuremath{\greenOp^\indDomC}}            {Continuum Green operator}
\mynewglossary{ignored}{Gaa}    {\ensuremath{\greenOp^{\indDomA/\indDomA}}} {Coupling operator}
\mynewglossary{ignored}{Gcc}    {\ensuremath{\greenOp^{\indDomC/\indDomC}}} {Coupling operator}
\mynewglossary{ignored}{Gac}    {\ensuremath{\greenOp^{\indDomA/\indDomC}}} {Coupling operator}
\mynewglossary{ignored}{Gca}    {\ensuremath{\greenOp^{\indDomC/\indDomA}}} {Coupling operator}
\mynewglossary{ignored}{Gci}    {\ensuremath{\greenOp^{\indDomC/\indDomI}}} {Coupling operator}
\mynewglossary{ignored}{Gcipl}    {\ensuremath{\greenOp^{\indDomC/\indDomIpl}}} {Coupling operator}
\mynewglossary{ignored}{Gipli}    {\ensuremath{\greenOp^{\indDomIpl/\indDomI}}} {Coupling operator}
\mynewglossary{ignored}{Giipl}    {\ensuremath{\greenOp^{\indDomI/\indDomIpl}}} {Coupling operator}
\mynewglossary{ignored}{Giplipl}    {\ensuremath{\greenOp^{\indDomIpl/\indDomIpl}}} {Coupling operator}
\mynewglossary{ignored}{Gii}    {\ensuremath{\greenOp^{\indDomI/\indDomI}}} {Coupling operator}
\mynewglossary{ignored}{Gpi}    {\ensuremath{\greenOp^{\indDomP/\indDomI}}} {Coupling operator}
\mynewglossary{ignored}{Gic}    {\ensuremath{\greenOp^{\indDomI/\indDomC}}} {Coupling operator}

\mynewglossary{ignored}{bK}       {\ensuremath{\sty{K}}}        {Nodal stiffness tensor}
\mynewglossary{ignored}{greenCGF} {\ensuremath{\green^\indCGF}} {Scalar continuum Green function}
\mynewglossary{ignored}{bgreenCGF}{\ensuremath{\bgreen^\indCGF}}{Continuum Green tensor}
\mynewglossary{ignored}{greenLGF} {\ensuremath{\green^\indLGF}} {Scalar lattice Green function}
\mynewglossary{ignored}{bgreenLGF}{\ensuremath{\bgreen^\indLGF}}{Lattice Green tensor}

\mynewglossary{ignored}{forceOp}{\ensuremath{\clF}}{Force operator}
\mynewglossary{ignored}{hdiffOp}{\ensuremath{\hat{L}}}{Matrix representation of \diffOp}
\mynewglossary{ignored}{hgreenOp}{\ensuremath{\hat{G}}}{Matrix representation of \greenOp}
\mynewglossary{ignored}{hforceOp}{\ensuremath{\hat{F}}}{Matrix representation of \forceOp}

\newglossary{newsymbols}{newsym}{newsbl}{List of Symbols}
\glsmoveentry{integ}{newsymbols}
\glsmoveentry{real}{newsymbols}
\glsmoveentry{euclD}{newsymbols}
\glsmoveentry{lat}{newsymbols}
\glsmoveentry{latConst}{newsymbols}
\glsmoveentry{atos}{newsymbols}
\glsmoveentry{batos}{newsymbols}
\glsmoveentry{bdispl}{newsymbols}
\glsmoveentry{bdisplDD}{newsymbols}
\glsmoveentry{bforce}{newsymbols}
\glsmoveentry{intRgAto}{newsymbols}
\glsmoveentry{Eato}{newsymbols}
\glsmoveentry{Etot}{newsymbols}
\glsmoveentry{bstiffMat}{newsymbols}
\glsmoveentry{bK}{newsymbols}
\glsmoveentry{bgreenCGF}{newsymbols}
\glsmoveentry{bgreenLGF}{newsymbols}
\glsmoveentry{burgers}{newsymbols}
\glsmoveentry{cutOff}{newsymbols}
\glsmoveentry{diffOp}{newsymbols}
\glsmoveentry{Lh}{newsymbols}
\glsmoveentry{Lah}{newsymbols}
\glsmoveentry{greenOp}{newsymbols}
\glsmoveentry{forceOp}{newsymbols}
\glsmoveentry{hdiffOp}{newsymbols}
\glsmoveentry{hgreenOp}{newsymbols}
\glsmoveentry{hforceOp}{newsymbols}
\glsmoveentry{Prob}{newsymbols}

\makeglossaries

\glsdisablehyper 
\renewcommand{\glossarysection}[2][]{} 
\glssetwidest[0]{\hspace{3.5em}}

\newtheorem{rem}{Remark}

\newproof{prf}{Proof}

\newtheorem*{rem2*}{Important remark}

\titleformat*{\section}{\bfseries}

\newcommand*{\textcal}[1]{%
 \textit{\large\fontfamily{qzc}\selectfont#1}%
}

\newlength{\commentWidth}

\makeatletter
\def\blfootnote{\gdef\@thefnmark{}\@footnotetext}
\makeatother

\newcommand{\mrm}[1]{\mathrm{#1}}
\newcommand{\inv}[1]{\ensuremath{{#1}^{-1}}}

\newcommand{\latInf}{\ensuremath{\lat_\infty}}

\renewcommand{\a}{\indDomA}
\newcommand{\p}{\indDomP}
\renewcommand{\i}{\indDomI}
\newcommand{\ipl}{\indDomIpl}

\newcommand{\ipli}{\ensuremath{{\ipl|\i}}}
\newcommand{\pipl}{\ensuremath{{\p|\ipl}}}
\newcommand{\ii}{\ensuremath{{\i|\i}}}
\newcommand{\iipl}{\ensuremath{{\i|\ipl}}}
\newcommand{\iplpl}{\ensuremath{{\ipl\textnormal{\texttt{+}}}}}
\newcommand{\ipliplpl}{\ensuremath{{\ipl|\iplpl}}}

\newcommand{\lgf}{\mrm{lgf}}
\newcommand{\cgf}{\mrm{cgf}}

\renewcommand{\c}{\indDomC}

\newcommand{\ip}{\indDomIpl}

\renewcommand{\aa}{\ensuremath{{\a|\a}}}
\newcommand{\cc}{\ensuremath{{\c|\c}}}
\newcommand{\ac}{\ensuremath{{\a|\c}}}
\newcommand{\ca}{\ensuremath{{\c|\a}}}

\newcommand{\ap}{\ensuremath{{\a|\p}}}

\renewcommand{\L}{\diffOp\xspace}

\newcommand{\Lcpl}{\ensuremath{\diffOp_\mathrm{cpl}}\xspace} 
\newcommand{\G}{\greenOp\xspace}

\newcommand{\Id}{\ensuremath{\clI}\xspace}
\newcommand{\nullOp}{\ensuremath{\textcal{0}\,}\xspace}

\begin{document}

\begin{frontmatter}
 \blfootnote{\textit{E-mail address:} \href{mailto:m.hodapp@skoltech.ru}{m.hodapp@skoltech.ru}}
 \begin{abstract}
  We present a novel efficient implementation of the flexible boundary condition (FBC) method, initially proposed by Sinclair et al., for large single-periodic problems.
  Efficiency is primarily achieved by constructing a hierarchical matrix ($\scH$-matrix) representation of the periodic Green matrix, reducing the complexity for updating the boundary conditions of the atomistic problem from quadratic to almost linear in the number of pad atoms.
  In addition, our implementation is supported by various other tools from numerical analysis, such as a residual-based transformation of the boundary conditions to accelerate the convergence.
  We assess the method for a comprehensive set of examples, relevant for predicting mechanical properties, such as yield strength or ductility, including dislocation bow-out, dislocation-precipitate interaction, and dislocation cross-slip.
  The main result of our analysis is that the FBC method is robust, easy-to-use, and up to two orders of magnitude more efficient than the current state-of-the-art method for this class of problems, the periodic array of dislocations (PAD) method, in terms of the required number of per-atom force computations when both methods give similar accuracy.%
  This opens new prospects for large-scale atomistic simulations---without having to worry about spurious image effects that plague classical boundary conditions.
 \end{abstract}
\end{frontmatter}


\section{Introduction}

The advancements in hard- and software technology during the past decades have shifted the field of materials science towards a computer-assisted discipline making use of, in particular, atomistic simulations.
Atomistic simulations can be used to study the nucleation, motion, and interaction of crystalline defects, e.g., vacancies, dislocations, grain boundaries, voids, or cracks.
In general, the goal of such studies is then to relate the behavior of those defects to macroscopic mechanical properties, e.g., yield strength, ductility, etc.

One major class of defects are line defects: the \emph{dislocations}.
It is well-understood that dislocations are the main carrier of plasticity in metals and their behavior is therefore intrinsically tied to any of the underlying strengthening and hardening mechanisms for this class of materials \citep{argon_strengthening_2007}.
A representative behavior of long dislocations on the atomic-scale can be simulated with the periodic array of dislocations (PAD) method \citep{daw_embedded-atom_1993,osetsky_atomic-level_2003}, where the periodic length in the dislocation line direction defines the intrinsic material length scale via the spacing of, e.g., obstacles (precipitates, voids, etc.).
In addition to periodic boundary conditions in the dislocation line direction, the PAD method uses periodic boundary conditions in the dislocation glide direction and free surfaces in the direction normal to the glide plane.
However, this particular choice of boundary conditions can introduce large image stresses, with spurious effects on the dislocation motion, as demonstrated by \citet{szajewski_analysis_2015}.
In particular, \citet{szajewski_analysis_2015} have shown that, for a dislocation bowing around periodic obstacles, \emph{all} side lengths of the simulation cell must be increased \emph{equally} when varying the periodic length---but keeping the maximum bow-out constant---in order to maintain comparable accuracy in the final position of the dislocation.
This implies that the PAD method scales cubically with the number of atoms which is very inefficient.
To reduce this computational burden, conventional atomistic/continuum (A/C) coupling methods (e.g., \citep{kohlhoff_new_1989,tadmor_quasicontinuum_1996,knap_analysis_2001,curtin_atomistic/continuum_2003,xiao_bridging_2004,shimokawa_matching_2004,xiong_coarse-grained_2011,kochmann_meshless_2014,ortner_energy-based_2014,fang_blended_2020}) can be used to restrict atomistic resolution to some small part around the dislocation core, but scaling the side lengths of the computational domain with the periodic length is still required.

A natural approach that avoids the scaling issue of PAD boundary conditions is to use A/C coupling methods with \emph{semi-infinite} continuum domains using boundary element methods (BEMs) \citep{li_efficient_2009,li_atomistic-based_2012,dedner_coupling_2017,hodapp_flexible_2018,hodapp_lattice_2019}.
To solve the coupled problem, \citet{li_efficient_2009,li_atomistic-based_2012} further proposed an alternating Schwarz method which iterates between the atomistic problem and the BEM.
A potentially more efficient method was developed by Hodapp et al. \citep{hodapp_flexible_2018,hodapp_lattice_2019} who proposed a monolithic Newton-GMRes solver with Hessian stabilization.
However, the latter method is very difficult to parallelize \emph{and} to integrate into existing molecular dynamics codes.
The latter is a major concern since developers of A/C coupling methods are rarely users of their own codes which is likely the reason why many interesting approaches have been left unnoticed.

For coupling multiple codes, a much more convenient choice are domain decomposition methods, notably, in the field of A/C coupling, the \emph{flexible boundary condition (FBC) method}, originally developed by Sinclair and coworkers in the 1970s \citep{sinclair_improved_1971,sinclair_influence_1975,sinclair_flexible_1978},
and newer related variants thereof \citep{gallego_harmonic/anharmonic_1993,yavari_theory_2006} (another related method, independently developed specifically for contact problems, is ``Green function molecular dynamics''; see, e.g., \citep{campana_practical_2006,monti_greens_2021}).
However, an analysis of the FBC method has been developed only recently by \citet{ehrlacher_analysis_2016} and \citet{hodapp_analysis_2021} who demonstrated its excellent convergence properties.
In particular, \citet{hodapp_analysis_2021} showed that the FBC method can essentially be considered as an iteration between a local anharmonic problem, the atomistic problem, and a global harmonic (continuum) problem which yields improved convergence rates over the classical alternating Schwarz method.
As such, it shares analogies with recently developed global-local formulations for continuum fracture (e.g., \citep{aldakheel_multilevel_2021}).
Thereby, the atomistic problem interacts with the harmonic problem through the displacements in the pad domain, the boundary of the atomistic problem.
Vice versa, the harmonic problem interacts with the atomistic problem through incompatibility forces which arise at the artificial interface due to the mismatch between both models.
In every iteration the solution of the harmonic problem $\bdispl(\bato)$ at a pad atom $\bato$ can then be directly obtained by summing up the solutions due to the incompatibility forces $\bmf(\batoB)$ at all interface atoms $\batoB$, i.e.,
\begin{equation}\label{eq:u=Gf}
 \bdispl(\bato) = \sum \bmG(\bato - \batoB)\bmf(\batoB),
\end{equation}
making use of the lattice Green function $\bmG(\bato - \batoB)$ of the harmonic problem.
Evidently, the FBC method converges when the incompatibility forces become sufficiently small.
The FBC method therefore has similarities with BEMs since the incompatibility forces are the only actual additional degrees of freedom.%
\footnote{additional to the anharmonic/atomistic degrees of freedom}

Hitherto, the FBC method has yet primarily been be applied to problems requiring only a rather small number of atoms of $\clO(10^3)$--$\clO(10^4)$, such as straight or kinked dislocations (e.g., in \citep{sinclair_flexible_1978,rao_greens_1998}); additionally, it has seen pronounced attention in context of quantum-mechanical/molecular mechanics coupling (e.g., \citep{woodward_flexible_2002,tan_dislocation_2019,andreoni_ab_2020}).
This is primarily due to the fact that the Green matrix obtained from \eqref{eq:u=Gf} is dense and, thus, storing and multiplying it by a vector both scale quadratically with the number of pad atoms.
This is problematic for two reasons: i) storing the Green matrix rapidly requires several hundreds of Gigabytes, even if the size of the atomistic domain is still relatively small (of $\clO(10^5)$--$\clO(10^6)$ atoms) for today's standards (cf. \citep{hodapp_lattice_2019}), and ii) computing the matrix-vector multiplication \eqref{eq:u=Gf} can become more costly than solving the atomistic problem.
Combined with the fact that domain decomposition solvers require several back-and-forth iterations between the subproblems and, so, more force computations than monolithic solvers makes the FBC method potentially so inefficient that any advantage over the PAD method disappears---even though the PAD method requires much more atoms.
This is likely the reason why, to date, materials scientists considered the FBC method as impractical for large-scale problems (see, e.g., \citep{bacon_chapter_2009}).

On the other hand, we have demonstrated in \citep{hodapp_flexible_2018,hodapp_lattice_2019} that Green matrices can be efficiently approximated using the framework of hierarchical matrices \citep[$\scH$-matrices,][]{tyrtyshnikov_mosaic-skeleton_1996,hackbusch_sparse_1999}.
$\scH$-matrices approximate admissible off-diagonal matrix blocks using low-rank representations reducing the storage and arithmetic complexity from quadratic to linear-logarithmic without sacrificing the accuracy of the coupled problem, thanks to asymptotic smoothness of the lattice Green function.
However, these previous works were devoted to the analysis of infinite problems, mostly by means of idealized two-dimensional examples that are of limited interest to practitioners.
Therefore, we now develop a new efficient implementation of the formulation of the FBC method from \citep{hodapp_analysis_2021} for single-periodic problems which does not suffer from any of the aforementioned drawbacks.

The first major novelty compared to our previous works \citep{hodapp_flexible_2018,hodapp_lattice_2019,hodapp_analysis_2021} are three additional building blocks for the implementation.
First, we develop a highly accurate and efficient algorithm for computing the periodic lattice Green function from a summation of the fundamental lattice Green functions over only a few periodic images based on series acceleration.
Second, using the periodic lattice Green functions, we construct an $\scH$-matrix representation of the \emph{periodic} Green matrix and demonstrate the linear-logarithmic scaling in terms of the number of pad atoms.
Third, to reduce the number of domain decomposition iterations, we integrate a new residual-based relaxation method \citep{ramiere_iterative_2015} for the incompatibility forces which further speeds up the simulations by factor of $\sim$\,2--2.5 at basically no additional computational cost.

As a second major novelty, we assess the performance of the FBC method using a comprehensive and diverse set of examples for dislocation motion, relevant for predicting mechanical properties, e.g., yield strength or ductility, including dislocation bow-out, dislocation-precipitate interaction, and dislocation cross-slip.
On the basis of these examples, we show that the FBC method is not only much more efficient than the PAD method, when both methods give similar accuracy, but also more general as it naturally allows to incorporate elastic far-field contributions.
To the author's best knowledge, the examples therefore demonstrate for the first time that FBCs can now indeed be considered as a practical, efficient, and easy-to-use method for simulating long dislocations at realistic length scales of tens of nanometers and beyond.

\subsection*{Notation}

We denote zeroth-order tensors/scalars by normal letters, e.g., $a,\Etot$, first-order tensors/vectors by lowercase bold letters, e.g., $\bdispl,\bmv$, and second-order tensors by uppercase bold letters, e.g., $\bmL,\bmK$.
The Euclidean inner product between two tensorial quantities is denoted by $\langle \bullet, \bullet \rangle$ and its induced norm is $\| \bullet \|$.
All tensorial quantities are defined with respect to the usual orthonormal basis system $\{ \bme_i \in \real^3 \,\vert\, \langle \bme_i, \bme_j \rangle = \delta_{ij} \}_{i=1,...,3}$.
Non-tensorial vectors/matrices are denoted by underlined/double-underlined normal letters $\uu,\uuG$, etc.

Since we primarily work with discrete problems, we adopt the notation from \citep{hodapp_flexible_2018,hodapp_lattice_2019,hodapp_analysis_2021} specifically developed for this purpose.
Following \citep{hodapp_flexible_2018,hodapp_lattice_2019,hodapp_analysis_2021}, let $\bdispl,\bforce$ be functions defined on a discrete domain (lattice) $\lat$ with $\vert \lat \vert$ elements (atoms).
For operators acting on $\bdispl,\bforce$ we use the calligraphic symbols $\L,\G$.
In general, we have $\bforce$ given and are looking for $\bdispl$ such that $\forall\,\bato \in \lat \; \L[\bdispl](\bato) = \bforce(\bato)$, or in short form $\L[\bdispl] = \bforce \; \text{in} \; \lat$.
Additional notation is introduced on-demand throughout the manuscript.


\section{Flexible boundary condition method}

\subsection{Reference atomistic problem}

\begin{figure}[hbt]
 \centering
 \includegraphics[width=0.7\textwidth]{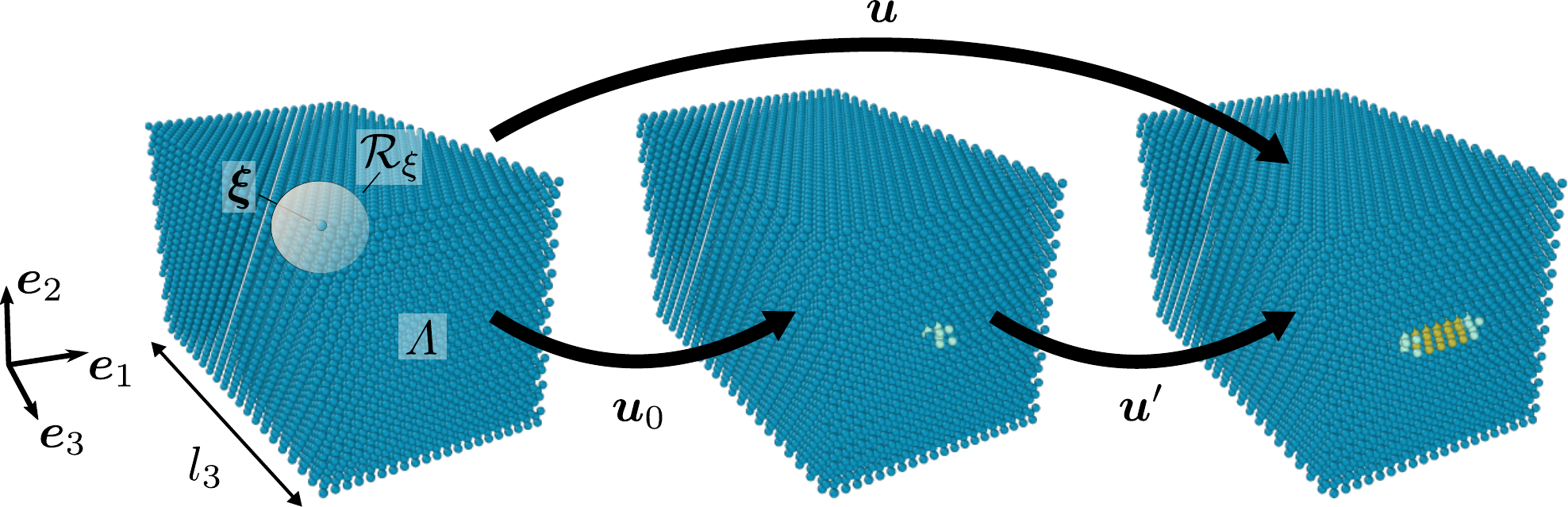}
 \caption{Schematic illustration of the atomistic reference problem}
 \label{fig:ato_problem}
\end{figure}

We let $\latInf$ be a Bravais lattice
\begin{equation}\label{eq:flexbc.bravais}
 \latInf := \left\{\, \sum_{i=1}^3 a_i\bmb_i + \bmc \,\bigg\vert\, a_i \in \integ \,\right\},
\end{equation}
where $\{ \bmb_i \in \real^3 \}_{i=1,...,3}$ is the set of basis vectors defining the lattice type, e.g., body-centered cubic (bcc) or face-centered cubic (fcc), and $\bmc$ is some constant.
The computational problem we consider in this work is periodic in one direction (here: $\rmx_3$-direction) and, therefore, our computational domain shown in Figure \ref{fig:ato_problem} is the subset
\begin{equation}
 \lat := \{\, \bato\in\latInf \,\vert\, 0 \le \ato_3 < l_3 \,\} \subset \latInf,
\end{equation}
where the periodic length $l_3$ is chosen to meet the periodicity of the lattice.

Every atom $\bato$ in $\lat$ is associated with a site energy $\Eato$ which depends on the periodic displacement
\begin{equation}
 \bdispl(\bato) \in \{\, \bmv : \latInf \rightarrow \real^3 \,\vert\, \bmv(\ato_1,\ato_2,\ato_3) = \bmv(\ato_1,\ato_2,\ato_3 \pm l_3)  \,\} 
\end{equation}
of atom $\bato$ relative to the displacements $\bdispl(\batoB)$ of all other atoms $\batoB$ within its local neighborhood $\intRgAto$ which usually extends to 1--2 lattice constants.
We abbreviate this collection of differential displacements as $\{ \bdispl(\batoB) - \bdispl(\bato) \}_{\batoB \in \intRgAto \setminus \bato} \equiv \{ \bdispl(\batoB) - \bdispl(\bato) \}$ and, hence, we write $\Eato = \Eato(\{ \bdispl(\batoB) - \bdispl(\bato) \})$.
We note that, besides the assumption of locality, there are no other restrictions on the structure of $\Eato$.
The total energy of the system then reads
\begin{equation}
 \Etot(\bdispl) = \Etot_\mrm{int}(\bdispl) + \Etot_\mrm{ext}(\bdispl),
\end{equation}
neglecting the cohesive energy, with the internal and external contributions, $\Etot_\mrm{int}$ and $\Etot_\mrm{ext}$, to the total energy given by
\begin{align}
    \Etot_\mrm{int}(\bdispl) =   \sum_{\bato \in \lat} \Eato(\{ \bdispl(\batoB) - \bdispl(\bato) \}),
 && \Etot_\mrm{ext}(\bdispl) = - \sum_{\bato \in \lat} \bforce_\mrm{ext}(\bato) \cdot \bdispl(\bato),
\end{align}
where $\bforce_\mrm{ext}$ is an external force.

With the above-stated definitions, we now define the computational problem.
We will focus on quasi-static problems in the following.
In general, we initialize the problem with an initial guess $\bdispl_0$ which carries infinite energy, e.g., the elastic solution of a dislocation, as shown in Figure \ref{fig:ato_problem}.
We then seek for (local) minimizers $\bdispl'$ of the (finite) energy difference (cf. \citep{luskin_atomistic--continuum_2013}
\begin{equation}\label{eq:flexbc.ato_problem}
 \bdispl' := \arg{\left\{ \, \underset{\bmv}{\min} \, \Etot(\bdispl_0 + \bmv) - \Etot(\bdispl_0) \, \right\}}
\end{equation}
such that the full solution is given by $\bdispl = \bdispl_0 + \bdispl'$.
We remark that problem \eqref{eq:flexbc.ato_problem} is not computable (unless being linear) and, so, additional approximations are required, typically truncating the infinite computational domain $\lat$ to one of finite dimension.

In what follows we will mostly work with the equivalent formulation of problem \eqref{eq:flexbc.ato_problem} using the force balance
\begin{equation}\label{eq:flexbc.euler-lagrange}\boxed{
 \begin{aligned}
  \; \L[\bdispl] &= \bforce_\mrm{ext} &\qquad& \text{in} \; \lat,
 \end{aligned}
 }
\end{equation}
with $\L[\bdispl](\bato) = \var{}{}{\Etot_\mrm{int}}(\bato)$, where $\var{}{}{\Etot_\mrm{int}}(\bato)$ is the derivative of $\Etot_\mrm{int}(\bdispl)$ with respect to $\bdispl$ at atom $\bato$.
The additional requirement for \eqref{eq:flexbc.euler-lagrange} being equivalent to \eqref{eq:flexbc.ato_problem} is the stability of the solution $\bdispl$, that is,
$\forall\,\bmv \; \langle \var{}{2}{\Etot_\mathrm{int}}(\bdispl)[\bmv], \bmv \rangle > 0$, which we assert to hold throughout this work.

\subsection{Coupled problem}
\label{sec:flexbc.coupled_problem}

In our atomistic problem, introduced in the previous section, we expect that significant nonlinearities arise only in the vicinity of a defect, such as a dislocation.
To avoid evaluating the expensive fully atomistic model in a very large computational domain motivates an atomistic/continuum (A/C) coupling which approximates the fully atomistic problem by restricting atomic resolution to some small part of the computational domain around the defect, and using a cheaper continuum model in the remainder domain.

\subsubsection{Continuum model}
\label{sec:flexbc.coupled_problem.con_model}

In the present work we use a local continuum model.
This choice is justified when the far-field is sufficiently smooth, which is the case for crystalline defects.
More precisely, it was shown in \citep{hodapp_analysis_2021} that, given the reference solution $\bdispl_\mrm{ref}$ to \eqref{eq:flexbc.ato_problem} and the solution of the coupled problem $\bdispl$, the error can be bounded by $\| \bdispl_\mrm{ref} - \bdispl \| \lesssim f(\grad{}{2}{\bdispl_\mrm{ref}}, ...)$, where $f$ is a function which depends on gradients of degree higher or equal than two.
This result is independent of the amount of ``nonlocality'' of the interatomic potential;
however, we note that the degree  of nonlocality in the continuum model can, on the other hand, influence the convergence rate of the coupled solver (cf. Remark \ref{rem:conv_rate}).

The continuum model we employ is based on a linearization around the defect-free reference lattice $\latInf$.
Therefore, in order to derive the continuum model, we assume that every atom ``sees'' a perfect environment and the site energy $\Eato$ of an atom $\bato$ is thus independent of $\bato$.
Expanding $\Eato$ around $\bdispl(\bato) = \bmZero$ to second order then gives the nonlocal harmonic site energy
\begin{equation}\label{eq:flexbc.Eato_hnl}
   \clE_{\mrm{hnl},\ato}(\{ \bdispl(\batoB) - \bdispl(\bato) \})
 = \frac{1}{2} \sum_{\batoB \in \intRg_{\ato}} \bmK_\mrm{hnl}(\bato - \batoB) \cdot \big( \bdispl(\bato) \otimes \bdispl(\batoB) \big),
\end{equation}
where
$\bmK_\mrm{hnl}(\bato - \batoB) = \sum_{\batoC \in \intRgAto}
\frac{\partial^2 \Eato(\{ \bdispl(\batoC) - \bdispl(\bato) \})}
     {\partial\bdispl(\bato)\partial\bdispl(\batoB)}$
is the interatomic force constant tensor.
In addition, we consider a linearization of $\bdispl(\bato)$ around $\bato$ such that
\begin{equation}\label{eq:flexbc.CBH}
 \bdispl(\batoB) \approx \bdispl(\bato) + \grad{}{}{\bdispl}(\bato)(\batoB - \bato).
\end{equation}
The latter approximation is commonly denoted as the Cauchy-Born hypothesis.
Using \eqref{eq:flexbc.Eato_hnl} and \eqref{eq:flexbc.CBH} leads to the local harmonic site energy
\begin{equation}\label{eq:flexbc.Eato_h}
   \clE_{\mrm{h},\ato}(\{ \bdispl(\batoB) - \bdispl(\bato) \})
 = \frac{1}{2} \, \bstiffMat \cdot (\grad{}{}{\bdispl(\bato)} \otimes \grad{}{}{\bdispl(\bato)})
 =
 \frac{1}{2} \sum_{\batoB \in \intRgAto^\mrm{h}} \bmK_\mrm{h}(\bato - \batoB) \cdot \big( \bdispl(\bato) \otimes \bdispl(\batoB) \big),
\end{equation}
where $\bstiffMat$ is the fourth-order material stiffness tensor and $\bmK_\mrm{h}(\bato - \batoB)$ is the local version of $\bmK_\mrm{hnl}(\bato - \batoB)$ which is nonzero only for $\batoB$ in the interaction range $\intRgAto^\mrm{h}$ typically comprising nearest neighbors.
Note that in \eqref{eq:flexbc.Eato_h}, the central term transforms into the term on the right hand side when using the definition of the gradient $\grad{}{}{\bdispl}$ \eqref{eq:appdx.def_u+ugrad} (for further details see Appendix A in \citep{hodapp_lattice_2019}).

With $\bmK_\mrm{h}(\bato - \batoB)$, we now define the linearization of $\L$
\begin{equation}
 \forall\,\bato\in\latInf \qquad \Lh[\bdispl](\bato) =
 \sum_{\batoB \in \intRgAto^\mrm{h}} \bmK_\mrm{h}(\bato - \batoB) \bdispl(\batoB)
\end{equation}
of which we will make use of in the following.

\subsubsection{Problem formulation}
\label{sec:flexbc.coupled_problem.formulation}

\begin{figure}[hbt]
 \centering
 \includegraphics[width=0.54\textwidth]{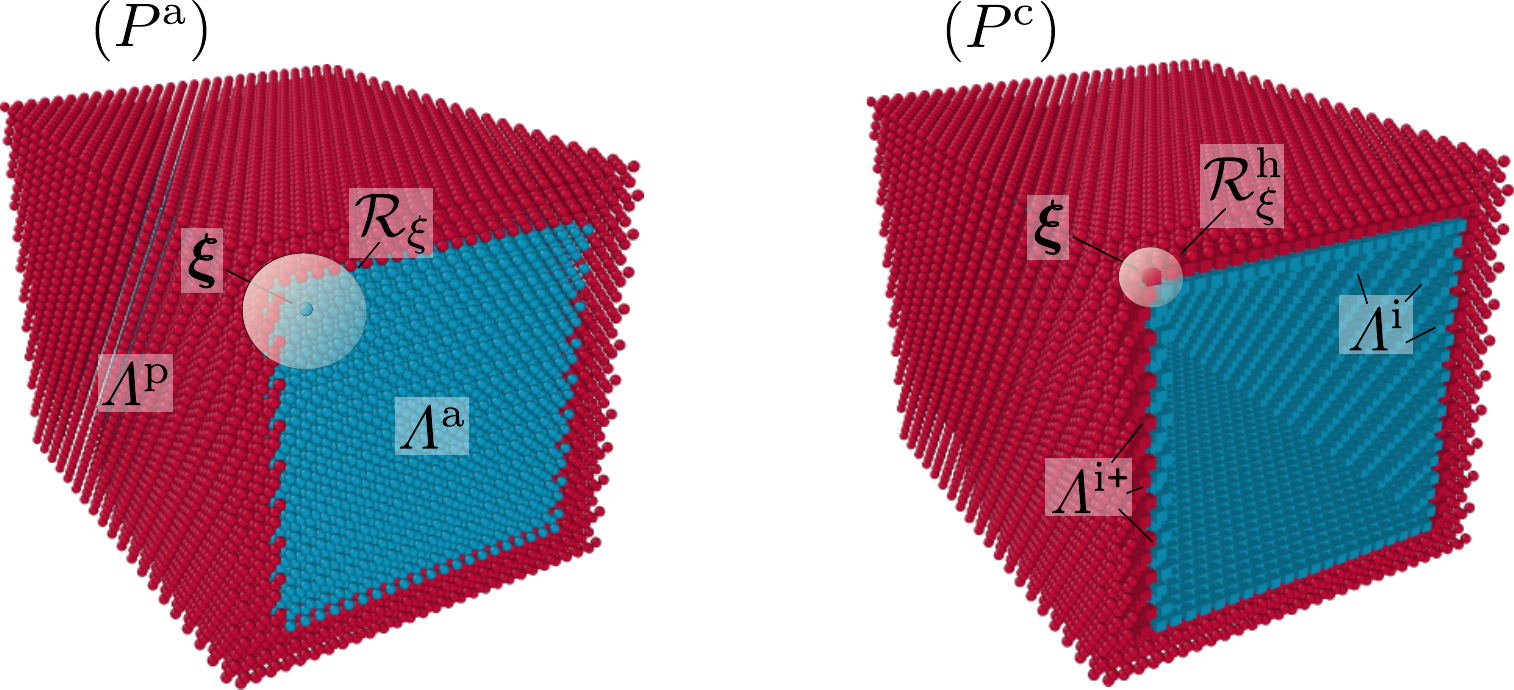}
 \caption{Schematic illustration of the coupled problem decomposed into an atomistic problem $(P^\rma)$ and a continuum problem $(P^\rmc)$}
 \label{fig:dom_decomp}
\end{figure}

The flexible boundary condition (FBC) method uses a force-based A/C coupling scheme (see, e.g., \citep{kohlhoff_new_1989}) which is described in the following.
To that end, we decompose the computational domain $\lat$ into an atomistic domain $\latA \subset \lat$ and a continuum domain $\latC := \lat \setminus \latA $ as shown in Figure \ref{fig:dom_decomp}.

In the vicinity of the artificial interface, atoms in $\latA$ can interact with continuum nodes from the pad domain $\latP \subset \latC$.
Thereby, the size of $\latP$ must depend on the interaction range $\intRgAto$ of the atomistic model to avoid spurious free surface effects.
On the other hand, the continuum model is local and, therefore, the continuum nodes can only interact with atoms on the interface layer $\lat^\i$.
This further implies that displacements of interface atoms do not exert forces in the entire pad domain but only in the subset $\lat^\ipl \subset \lat^\p$, schematically depicted by the cubical-shaped atoms in Figure \ref{fig:dom_decomp}.

Note that Figure \ref{fig:dom_decomp} is only a schematic illustration and we do not impose any restrictions on the size and shape of the atomistic domain.
Nevertheless, we emphasize that our focus here is mainly on problems where the periodic length $l_3$ is of the same order as the diameter of $\latA$ (see Section \ref{sec:results}).
In fact, this is the realm of applications where we expect our implementation of the FBC method to be most useful and more efficient than other existing methods.

The reader familiar with the original works of Sinclair et al. recognizes that the domains defined above correspond to the regions 1, 2 and 3, as defined in \citep{sinclair_flexible_1978}.
We nevertheless recall that
\begin{alignat*}{3}
 \latA & \qquad \Leftrightarrow \qquad & \text{region 1,} \\
 \lat^\ipl & \qquad \Leftrightarrow \qquad & \text{region 2,} \\
 \lat \setminus \lat^\ipl & \qquad \Leftrightarrow \qquad & \text{region 3,}
\end{alignat*}
with the only difference that in \citep{sinclair_flexible_1978} the continuum model is nonlocal.
If we would use a nonlocal continuum model as well, which is not essential for reasons described in the previous section, the domain $\lat^\ipl$ would be equivalent to $\lat^\p$.

Having the notation well-defined, we now write the coupled problem seeking for solutions $\bdispl := \{ \bdispl^\a, \bdispl^\c \}$ as follows
\begin{empheq}[box=\fbox]{align}\label{eq:flexbc.cpld_problem}
 (P^\a) \; \left\{
 \begin{aligned}
  \; \L[\{\bdispl^\a,\bdispl^\p\}] &= \bforce_\mrm{ext} &\qquad& \text{in} \; \lat^\a, \\
  \; \bdispl                       &= \bdispl^\c        &      & \text{in} \; \lat^\p,
 \end{aligned}
 \right.
 &&
 (P^\c) \; \left\{
 \begin{aligned}
  \; \Lh[\{\bdispl^\c,\bdispl^\i\}] &= \bforce_\mrm{ext}  &\qquad& \text{in} \; \lat^\c, \\
  \; \bdispl                        &= \bdispl^{\indDomA} &      & \text{on} \; \lat^\i.
 \end{aligned}
 \right.
\end{empheq}

\subsection{Domain decomposition algorithm}
\label{sec:flexbc.dd_algo}

\subsubsection{Sinclair's iteration equation}

The FBC method belongs to the class of domain decomposition solvers but stands out from other representatives of this class, e.g., the popular alternating Schwarz method, by offering significantly improved convergence rates \citep{hodapp_lattice_2019}.
The origin of the improved convergence behavior stems from its particular domain decomposition into a local anharmonic domain, i.e., the atomistic domain, and a \emph{global} harmonic domain.
In this case, it has been shown in \citep{hodapp_analysis_2021} that the convergence rate of the FBC method depends on the mismatch between the nonlocal atomistic and the local continuum model.
That is, if the atomistic model is not ``too nonlinear and nonlocal'', the FBC method converges fast.

To derive Sinclair's iteration equation we follow the procedure described in \citep{hodapp_lattice_2019,hodapp_analysis_2021}.
To that end, we split the operator associated with the coupled problem \eqref{eq:flexbc.cpld_problem}, that we denote as $\Lcpl$, and the solution $\bdispl$ into anharmonic and harmonic parts as follows
\begin{align}\label{eq:flexbc.Lcpl_split}
 \Lcpl = \Lah + \Lh, && \bdispl = \bdispl_\indAHarm + \bdispl_\indHarm.
\end{align}
For the anharmonic parts we require that
\begin{align}\label{eq:flexbc.Lah_uah}
 \Lah =
 \left\{\,
 \begin{aligned}
  \L - \Lh & \qquad \text{in} \; \latA, \\
  \nullOp  & \qquad \text{else},
 \end{aligned}
 \right.
 &&
 \bdispl_\indAHarm = \bmZero \qquad \text{in} \; \lat \setminus \latA.
\end{align}
Otherwise the splitting \eqref{eq:flexbc.Lcpl_split} is arbitrary and primarily used here as a means to construct the iteration equation---in practice we never need to compute neither $\bdispl_\indAHarm$ nor $\bdispl_\indHarm$ in $\latA$ as we shall see in the following.

Before proceeding, we need to introduce some additional notation to facilitate the decomposition of operators that we need for our algorithm.
To that end, if $\lat$ is decomposed into subdomains $\lat^\a,\lat^\c$, we write
$
\L[\bdispl] \equiv
\begin{psmallmatrix} \L^{\a|}[\bdispl] \\ \L^{\c|}[\bdispl] \end{psmallmatrix}
=
\begin{psmallmatrix} \bforce^\a \\ \bforce^\c \end{psmallmatrix}
$,
where the superscripted indicators ``$\a|$'',``$\c|$'' imply that $\L$ is acting on functions defined on $\lat$ and produces functions defined on $\lat^\a,\lat^\c$.
For a linear operator $\Lh$ (and also $\greenOp$ below) we may further split $\bdispl$ into contributions $\bdispl^\a,\bdispl^\c$ defined on $\lat^\a,\lat^\c$ and write $\Lh[\bdispl] = \Lh^{|\a}[\bdispl] + \Lh^{|\c}[\bdispl]$, where the superscripted indicators ``$|\a$'',``$|\c$'' imply that $\Lh$ is acting on $\bdispl^\a,\bdispl^\c$ and produces functions defined on $\lat$.
Using both splittings then yields the matrix notation
$
\Lh[\bdispl] \equiv
\begin{psmallmatrix} \Lh^{\a|}[\bdispl] \\ \Lh^{\c|}[\bdispl] \end{psmallmatrix}
=
\begin{psmallmatrix} \Lh^{\a|\a}[\bdispl] + \Lh^{\a|\c}[\bdispl] \\ \Lh^{\c|\a}[\bdispl] + \Lh^{\c|\c}[\bdispl] \end{psmallmatrix}
=
\begin{psmallmatrix} \Lh^{\a|\a} & \Lh^{\a|\c} \\ \Lh^{\c|\a} & \Lh^{\c|\c} \end{psmallmatrix}
\begin{bsmallmatrix} \bdispl^\a \\ \bdispl^\c \end{bsmallmatrix}
=
\begin{psmallmatrix} \bforce^\a \\ \bforce^\c \end{psmallmatrix}
$.

Using the above defined notation and the anharmonic/harmonic operator split \eqref{eq:flexbc.Lah_uah}, we first rewrite the coupled problem \eqref{eq:flexbc.Lcpl_split} in matrix notation and then regroup some of the terms as follows
\begin{equation}
 \begin{alignedat}{3}
  \Lcpl[\bdispl]
  &=
  \Lah[\bdispl] + \Lh[\bdispl_\indAHarm] + \Lh[\bdispl_\indHarm]
  &&=
  \bforce_\mathrm{ext} \\
  &=
  \begin{pmatrix} \L^{\a|}[\bdispl] - \Lh^{\a|}[\bdispl] \\ \bmZero \end{pmatrix}
  +
  \begin{pmatrix} \Lh^\aa & \nullOp^{\a|\c} \\ \Lh^\ca & \nullOp^{\c|\c} \end{pmatrix}
  \begin{bmatrix} \bdispl_\indAHarm^\indDomA \\ \bmZero \end{bmatrix}
  +
  \begin{pmatrix} \Lh^\aa & \Lh^{\a|\c} \\ \Lh^\ca & \Lh^{\c|\c} \end{pmatrix}
  \begin{bmatrix} \bdispl_\indHarm^\indDomA \\ \bdispl^\c \end{bmatrix}
  &&=
  \begin{pmatrix} \bforce_\mathrm{ext}^\indDomA \\ \bforce_\mrm{ext}^\c \end{pmatrix} \\
  &=
  \begin{pmatrix} \L^{\a|}[\bdispl] \\ \bmZero \end{pmatrix}
  +
  \begin{pmatrix} \Lh^\aa & \Lh^{\a|\c} \\ \Lh^\ca & \Lh^{\c|\c} \end{pmatrix}
  \begin{bmatrix} \bdispl_\indHarm^\indDomA \\ \bdispl^\c \end{bmatrix}
  +
  \begin{pmatrix} \Lh^\aa & \nullOp^{\a|\c} \\ \Lh^\ca & \nullOp^{\c|\c} \end{pmatrix}
  \begin{bmatrix} \bdispl_\indAHarm^\indDomA \\ \bmZero \end{bmatrix}
  -
  \begin{pmatrix} \Lh^{\a|}[\bdispl] \\ \bmZero \end{pmatrix}
  &&=
  \begin{pmatrix} \bforce_\mathrm{ext}^\indDomA \\ \bforce_\mrm{ext}^\c \end{pmatrix}.
 \end{alignedat}
\end{equation}
By moving $\bforce_\mrm{ext}$ to the left hand side, we obtain
\begin{equation}
 \underbrace{
  \begin{pmatrix} \L^{a|}[\bdispl] \\ \bmZero \end{pmatrix}
  - \begin{pmatrix} \bforce_\mathrm{ext}^\indDomA \\ \bmZero \end{pmatrix}
 }_{\textbf{\normalsize{(AH)}}}
 +
 \underbrace{
  \begin{pmatrix} \Lh^\aa & \Lh^{\a|\c} \\ \Lh^\ca & \Lh^{\c|\c} \end{pmatrix}
  \begin{bmatrix} \bdispl_\indHarm^\indDomA \\ \bdispl^\c \end{bmatrix}
  - \begin{pmatrix}
     \Lh^\aa[\bdispl_\indHarm] + \Lh^{\a|\c}[\bdispl] \\ 
     -\Lh^\ca[\bdispl_\indAHarm] + \bforce_\mathrm{ext}^\indDomC
    \end{pmatrix}
 }_{\textbf{\normalsize{(H)}}}
 = \begin{pmatrix} \bmZero \\ \bmZero \end{pmatrix}.
\end{equation}
The parts denoted by \textbf{(AH)} and \textbf{(H)} are the \emph{local anharmonic} and the \emph{global harmonic} problem, respectively.

The FBC method solves those two parts separately.
That is, starting from some initial guess $\bdispl_0$, for example one which solves the global harmonic problem (with $\Lh^\aa[\bdispl_\indHarm] + \Lh^{\a|\c}[\bdispl] = \bforce_\mathrm{ext}^\indDomA$ and $-\Lh^{\c|\a}[\bdispl_\indAHarm] = \bmZero$), we first solve \textbf{(AH)} for $\bdispl^\a$.
Then, we use the new anharmonic solution $\bdispl^\a_\indAHarm = \bdispl^\a - \bdispl^\a_{\indHarm,0}$ to update the source term of \textbf{(H)}.
The new pad displacements that develop after solving \textbf{(H)} will subsequently be used to update the boundary condition of \textbf{(AH)}, etc.
Writing this procedure formally, we arrive at Sinclair's iteration equation
\begin{empheq}[box=\fbox]{align}
 \textbf{(AH)}_{k+1} \;
 \left\{
 \begin{aligned}
  \; \L[\{ \bdispl^\a_{k+1}, \bdispl^\p_k \}] &= \bforce_\mrm{ext} &\;& \text{in} \; \lat^\a, \\
  \; \bdispl_k                                &= \bdispl^\c_k      &  & \text{in} \; \lat^\p,
 \end{aligned}
 \right.
 &&
 \textbf{(H)}_{k+1} \;
 \left\{
 \begin{aligned}
  \; \Lh[\bdispl_{\mrm{h},k+1}] &= \L^{|\a}[\bdispl_{\indHarm,0}] + \Lh^{|\c}[\bdispl_0]      &\;& \text{in} \; \latA, \\
  \; \Lh[\bdispl_{\mrm{h},k+1}] &= -\Lh^{|\a}[\bdispl_{\indAHarm,k+1}] + \bforce_\mrm{ext} &\;& \text{in} \; \lat^\c,
 \end{aligned}
 \right.
\end{empheq}
where $k \in \nat_0$ is the iteration index.

We point out that the initial guess does not necessarily need be the solution to the global harmonic problem due to the fact that the right hand side of \textbf{(H)} in $\latA$ can in principle be chosen arbitrarily (cf. \citep{hodapp_analysis_2021}).
If it is not available, or too cumbersome to compute, other classical choices are the continuum elasticity solution, or simply $\bdispl_0 = \bmZero$.
In any of these cases, we may, for practical reasons, universally choose $\bdispl_{\indHarm,0} = \bmZero$, thanks to the nonunique splitting---the right hand side of $\textbf{(H)}_{k+1}$ in $\latA$ is then always zero.

\begin{rem}[Convergence rate of the FBC method]\label{rem:conv_rate}
 The asymptotic convergence rate of the FBC method is neither impacted by the precise choice of the initial guess nor the choice of the right hand side of $\textnormal{\textbf{(H)}}_{k+1}$, but only depends on the choice of $\Lh$ \citep{hodapp_analysis_2021}.
 In fact, if $\Lh$ and $\L$ coincide, the FBC method is guaranteed to converge in two steps (cf. Corollary 1 in \citep{hodapp_analysis_2021}).
\end{rem}

\subsubsection{Practical algorithm}

Since the harmonic problem $\textbf{(H)}_{k+1}$ is infinite, we can conveniently obtain the solution $\bdispl_{\indHarm,k+1}$ using the lattice Green function.
Therefore, let $\G$ be the lattice Green operator corresponding to the \emph{periodic harmonic problem} such that
\begin{equation}
  \forall\,\bato\in\lat \; \wedge \; \forall\,\bmv \qquad
  (\greenOp\Lh)[\bmv](\bato)
  = \sum_{\batoB \in \lat} \left( \sum_{\batoC \in \latInf \,\vert\, \batoC-\batoB \in \intRg_{\atoC}^\rmh} \bgreen(\bato - \batoC) \bmK_\rmh(\batoC - \batoB) \right) \bmv(\batoB) \\
  = \clI[\bmv](\bato),
\end{equation}
where $\clI$ is the identity operator and $\bgreen$ the lattice Green function.
We postpone the details concerning the computation of $\bgreen$ to Section \ref{sec:impl.plgf} and proceed under the assumption that $\G$ is given.

Then, using the Green operator $\G$, we compute the solution $\bdispl_{\indHarm,k+1}$ as
\begin{equation}
   \begin{pmatrix} \bdispl^\a_{\mrm{h},k+1} \\ \bdispl^\c_{k+1} \end{pmatrix}
 = \inv{\begin{pmatrix} \Lh^\aa & \Lh^\ac \\ \Lh^\ca & \Lh^\cc \end{pmatrix}}
   \begin{bmatrix} \L^{\a|\a}[\bdispl_{\indHarm,0}] + \Lh^{\a|\c}[\bdispl_0] \\ - \Lh^\ca[\bdispl_{\mrm{ah},k+1}] + \bforce_\mrm{ext}^\c \end{bmatrix}
 = \begin{pmatrix} \G^\aa & \G^\ac \\ \G^\ca & \G^\cc \end{pmatrix}
   \begin{bmatrix} \L^{\a|\a}[\bdispl_{\indHarm,0}] + \Lh^{\a|\c}[\bdispl_0] \\ - \Lh^\ca[\bdispl_{\mrm{ah},k+1}] + \bforce_\mrm{ext}^\c \end{bmatrix}.
\end{equation}
Further, by subtracting $\bdispl_{\mrm{h},k}$ from $\bdispl_{\mrm{h},k+1}$, we obtain the more convenient representation
\begin{equation}\label{eq:flexbc.displ_update}
   \begin{pmatrix} \bdispl^\a_{\mrm{h},k+1} \\ \bdispl^\c_{k+1} \end{pmatrix}
 = \begin{pmatrix} \bdispl^\a_{\mrm{h},k} \\ \bdispl^\c_{k} \end{pmatrix}
 - \begin{pmatrix} \G^\aa & \G^\ac \\ \G^\ca & \G^\cc \end{pmatrix}
   \begin{bmatrix} \bmZero \\ \Lh^\ca[\bdispl_{\mrm{ah},k+1} - \bdispl_{\mrm{ah},k}] \end{bmatrix}.
\end{equation}
since the constant source term is now contained in $\bdispl_{\mrm{h},0}$.
In addition, to avoid computing $\bdispl_{\mrm{ah}}$, we rewrite ${\Lh^\ca[\bdispl_{\mrm{ah},k+1} - \bdispl_{\mrm{ah},k}]}$ as follows
\begin{equation}\label{eq:flexbc.finc}
 \begin{aligned}
  \Lh^\ca[\bdispl_{\mrm{ah},k+1} - \bdispl_{\mrm{ah},k}]
  &= \Lh^\ca[\bdispl_{\mrm{ah},k+1}] - \Lh^\ca[\bdispl_{\mrm{ah},k}] \\
  &= \Lh^\ca[\bdispl_{\mrm{ah},k+1}] + \Lh^\ca[\bdispl_{\mrm{h},k}] + \Lh^\cc[\bdispl_{k}] \\
  &= \Lh^\ca[\bdispl_{\mrm{ah},k+1} + \bdispl_{\mrm{h},k}] + \Lh^\cc[\bdispl_{k}] \\
  &= \Lh^\ca[\bdispl_{k+1}] + \Lh^\cc[\bdispl_{k}] \\
  &=: \bforce_\mrm{in},
 \end{aligned}
\end{equation}
where only the full solution appears (since $\bdispl_{k+1} = \bdispl_{\mrm{ah},k+1} + \bdispl_{\mrm{h},k}$ in $\latA$).
Here, we used the nomenclature from Sinclair et al. and defined \eqref{eq:flexbc.finc} as the \emph{``incompatibility force''} (or inhomogeneous force) $\bforce_\mrm{in}$ which arises after the atoms in $\latA$ have been relaxed.

Recognizing that our continuum model is local, we never need to evaluate $\bforce_\mrm{in}$ in the entire continuum domain since only the continuum nodes in $\lat^\ipl$ interact with the real atoms (cf. Section \ref{sec:flexbc.coupled_problem.formulation}).
For the same reason we do not require $\bdispl$ to be computed in the entire continuum domain---it suffices that we know it in the domain $\lat^\iplpl$ containing the $\lat^\ipl$-atoms and its nearest neighbors.
Hence, instead of \eqref{eq:flexbc.displ_update}, we only need to compute $\bforce_{\mrm{in},k+1}^\ipl = \Lh^\ipli[\bdispl_{k+1}] + \Lh^\ipliplpl[\bdispl_{k}]$ and subsequently update the pad displacements as follows
\begin{equation}\label{eq:flexbc.up=Gpipl*fipl}
 \bdispl^\p_{k+1} = \bdispl^\p_k - \greenOp^\pipl[\bforce_{\mrm{in},k+1}],
\end{equation}
where $\greenOp^\pipl$ is the Green operator which maps the incompatibility forces in $\lat^\ipl$ to displacements in $\lat^\p$.

Our implementation of the FBC method is summarized in Algorithm \ref{algo:fbc} with another two additions.
First, to improve the convergence rate, we rescale $\bforce^\ipl_{\mrm{in},k+1}$ after solving \textbf{(AH)} using residual-based relaxation, as described in the following section. 
Second, we have found that, since the problem is not fixed in space, the domains may start drifting rigidly from one iteration to another due to some numerical inaccuracies, possibly leading to a divergence of the solution.
To that end, we define the additional boundary condition on $(P^\a)$ in \eqref{eq:flexbc.cpld_problem}
\begin{equation}\label{eq:mean(u^p)=0}
 \frac{1}{\vert\lat^\p\vert} \sum_{\bato \in \lat^\p} \displ_i(\bato) = 0,
\end{equation}
where $\displ_i$ is the $i$-th component of $\bdispl$, i.e., the mean pad displacements are set to zero.
This is implemented by setting the mean of $\displ^\p_i$ after the $k$+1-th iteration, i.e., of $\displ^\p_{i,k+1}$, to zero and resolved the issue in all our numerical experiments.
Algorithm \ref{algo:fbc} terminates when some suitable convergence criterion is achieved, e.g., $\| \var{}{}{\Etot^\a}(\bdispl_k) \|$ being less than some prescribed tolerance.

\begin{algorithm}[hbt]
 \SetAlgoSkip{bigskip}
 \LinesNumbered
 \SetKwInput{Input}{Input}
 \SetKwInput{Output}{Output}
 \SetKwBlock{Repeat}{repeat}{end}
 \setstretch{1.2}
 \setlength{\commentWidth}{0.5\textwidth}
 \newcommand{\atcp}[1]{\tcp*[r]{\makebox[\commentWidth]{#1\hfill}}}
 \caption{Flexible boundary condition method with relaxation}
 \label{algo:fbc}
 \Input{initial guess $\bdispl_0$}
 $k \,\leftarrow\, 0$, optional: $\alpha_2 \,\leftarrow\, 1$;\\
 \While{\textbf{convergence not achieved}}{
  $\bdispl^\a_{k+1} \,\leftarrow\, \arg \, \left\{\, \underset{\bmv^\a}{\min} \, \Etot^\a(\{\bmv^\a,\bdispl^\p_k\}) \,\right\}$
   \atcp{solve \textnormal{\textbf{(AH)}}$_{k+1}$}
  $\bforce^\ip_{\mrm{in},k+1} \,\leftarrow\, \Lh^\ipli[\bdispl_{k+1}] + \Lh^\ipliplpl[\bdispl_{k}]$
   \atcp{compute incompatibility force}
  \If{\textbf{relaxation} \textnormal{\textbf{and}} $k \ge 2$}{
   $\alpha_{k+1} \,\leftarrow\, \texttt{Relax}(\bforce^\ip_{\mrm{in},k+1}, \bforce^\ip_{\mrm{in},k}, \alpha_k)$
    \atcp{residual-based relaxation}
   $\bforce^\ip_{\mrm{in},k+1} \,\leftarrow\, \alpha_{k+1} \bforce^\ip_{\mrm{in},k+1}$;
  }
  $\bdispl^\p_{k+1} \,\leftarrow\, \bdispl^\p_k - \greenOp^\pipl[\bforce_{\mrm{in},k+1}]$
   \atcp{solve \textnormal{\textbf{(H)}}$_{k+1}$}
  $\displ^\p_{i,k+1} \,\leftarrow\, \displ^\p_{i,k+1} - (1 / \vert\lat^\p\vert) \sum_{\bato \in \lat^\p} \displ_{i,k+1}(\bato)$;\\
  $k \,\leftarrow\, k+1$;\\
 }
 \Output{global solution $\bdispl_k$}
\end{algorithm}

\subsubsection{Residual-based relaxation}

One option to accelerate the convergence of domain decomposition methods is relaxation which aims at rescaling the transmission conditions between the subproblems.
For the FBC method, \citet{hodapp_analysis_2021} proposed a relaxation of the incompatibility force which resets $\bforce^\ip_{\mrm{in},k+1}$ to $\alpha_{k+1} \bforce^\ip_{\mrm{in},k+1}$, where $\alpha_{k+1}$ is the relaxation parameter, each time after solving the atomistic problem.
In \citep{hodapp_analysis_2021}, the optimal $\alpha_{k+1}$ was considered to be the one minimizing the difference between the solutions of two subsequent iterations as follows
\begin{equation}
 \alpha_{k+1} := \arg{\left\{ \, \underset{\alpha}{\min} \, \| \bdispl^\p_{\mrm{trial},k+2}(\alpha) - \bdispl^\p_{k+1}(\alpha) \| \, \right\}},
\end{equation}
where $\bdispl^\p_{k+1}(\alpha) = \bdispl^\p_k - \alpha\,\greenOp^\pipl[\bforce_{\mrm{in},k+1}]$ and $\bdispl^\p_{\mrm{trial},k+2}(\alpha)$ being a trial solution in the \mbox{$k$+$2$-th} iteration.
In \citep{hodapp_analysis_2021}, this trial solution has been computed by first solving a \emph{linearized version} of \textbf{(AH)}$_{k+2}$ with boundary condition $\bdispl^\p_{k+1}(\alpha)$, and then using its solution $\bdispl^\a_{\mrm{trial},k+2}$ to obtain $\bdispl^\p_{\mrm{trial},k+2}$.
This approach is therefore rather complicated to implement (efficiently) and most useful in situations when solving a linearized problem in $\lat^\a$ is significantly cheaper than solving the nonlinear problem (such as in the case of quantum mechanics/molecular mechanics coupling schemes, e.g., \citep{woodward_flexible_2002}).

Here, we propose to use a residual-based method \citep{ramiere_iterative_2015} which only requires to compute a couple of inner products between the incompatibility force from the current and previous iteration and is thus far easier to implement.
To that end, we reconsider the problem of solving $\bforce^\ip_{\mrm{in}} = \bmZero$ as a sequence $(\bforce^\ipl_k)_{k \ge 1}$, where $\sum_{l=1}^k \bforce^\ip_{\mrm{in},l}$ is the \emph{``total incompatibility force''}, which converges to $\bforce^\ipl$.
Using relaxation, the iteration equation for this sequence is 
\begin{equation}\label{eq:flexbc.fp_iter}
 \bforce^\ipl_{k+1} = \bforce^\ipl_k + \alpha_{k+1} \bforce^\ip_{\mrm{in},k+1},
\end{equation}
in which, obviously, $\bforce^\ip_{\mrm{in},k+1}$ plays the role of the residual.
Now, to obtain a good $\alpha_{k+1}$ which accelerates the sequence, we pose the assumptions that i) $(\bforce^\ipl_k)_{k \ge 1}$ converges linearly to $\bforce^\ipl$, and ii) $\bforce^\ipl_{k-1}$, $\bforce^\ipl_k$ and $\bforce^\ipl_{k+1}$ are sufficiently close to $\bforce^\ipl$ in the direction of some vector $\bmv$ that we define momentarily.
Then,
\begin{equation}\label{eq:steffensen_apprx}
 \frac
 {\langle \widehat{\bforce}{}^\ipl_{k+1} - \bforce^\ipl, \bmv \rangle}
 {\langle \widehat{\bforce}{}^\ipl_k - \bforce^\ipl, \bmv \rangle}
 \approx
 \frac
 {\langle \bforce^\ipl_k- \bforce^\ipl, \bmv \rangle}
 {\langle \bforce^\ipl_{k-1} - \bforce^\ipl, \bmv \rangle},
\end{equation}
where $\widehat{\bforce}{}^\ipl_k$ and $\widehat{\bforce}{}^\ipl_{k+1}$ are the solutions obtained without relaxation in the $k$-th and $k$+1-th iteration, respectively.
By solving \eqref{eq:steffensen_apprx} for $\langle \bforce^\ipl, \bmv \rangle$ we get
\begin{equation}
 \langle \bforce^\ipl, \bmv \rangle
 \approx
 \frac
 {\langle \widehat{\bforce}{}^\ipl_{k+1}, \bmv \rangle
  \langle \bforce^\ipl_{k-1}, \bmv \rangle -
  \langle \bforce^\ipl_k, \bmv \rangle
  \langle \widehat{\bforce}{}^\ipl_k \bmv \rangle}
 {\langle \bforce^\ipl_{\mrm{in},k+1} - \bforce^\ipl_{\mrm{in},k}, \bmv \rangle}
\end{equation}
and with assumption ii) it follows that
\begin{equation}
 \langle \bforce^\ipl, \bmv \rangle \approx \langle \bforce^\ipl_{k+1}, \bmv \rangle \overset{\eqref{eq:flexbc.fp_iter}}{=} \langle \bforce^\ipl_k + \alpha_{k+1} \bforce^\ipl_{\mrm{in},k+1}, \bmv \rangle \approx
 \frac
 {\langle \widehat{\bforce}{}^\ipl_{k+1}, \bmv \rangle
  \langle \bforce^\ipl_{k-1}, \bmv \rangle -
  \langle \bforce^\ipl_k, \bmv \rangle
  \langle \widehat{\bforce}{}^\ipl_k \bmv \rangle}
 {\langle \bforce^\ipl_{\mrm{in},k+1} - \bforce^\ipl_{\mrm{in},k}, \bmv \rangle}.
\end{equation}
Equating the latter relation and solving for $\alpha_{k+1}$ yields
\begin{equation}\label{eq:alpha}
 \alpha_{k+1} :=
 - \frac
 {\langle \bforce^\ipl_k - \bforce^\ipl_{k-1}, \bmv \rangle}
 {\langle \bforce^\ipl_{\mrm{in},k+1} - \bforce^\ipl_{\mrm{in},k}, \bmv \rangle}
 =
 - \alpha_k
 \frac
 {\langle \bforce^\ipl_{\mrm{in},k}, \bmv \rangle}
 {\langle \bforce^\ipl_{\mrm{in},k+1} - \bforce^\ipl_{\mrm{in},k}, \bmv \rangle}.
\end{equation}

We now choose $\bmv = \bforce^\ipl_{\mrm{in},k+1} - \bforce^\ipl_{\mrm{in},k}$, which is commonly referred to as Aitken's or Lemar\'{e}chal's method \citep{aitken_bernoullis_1927,lemarechal_methode_1971,ramiere_iterative_2015}.
Intuitively the method can be understood as follows: if the residual between two subsequent iterations does not change too much its direction, the right hand side of \eqref{eq:alpha} will become $>$\,1 if the projected residual reduces, or $<$\,1 if the projected residual increases. Thus, the hope is to accelerate the convergence by moving faster ($\alpha_{k+1}$\,$>$\,1) in the direction of the residual, or slower ($\alpha_{k+1}$\,$<$\,1) in order to prevent oscillations.
We found this a convenient choice for defect propagation to escape the danger of the atomistic problem getting trapped around saddle points for too long as will be demonstrated in Section \ref{sec:results}.

In practice, when $\alpha_{k+1}$ becomes $<$\,1, premature convergence may occur if the pad displacements $\bdispl^\p_{k+1} = \bdispl^\p_k - \alpha_{k+1}\greenOp^\pipl[\bforce_{\mrm{in},k+1}]$ are very small.
However, we have found that this can be overcome by adding a convergence criterion for the incompatibility forces $\bforce^\ipl_\mrm{in}$.
That is, if $\alpha_{k+1}$ is small, then the norm of $\bforce^\ipl_\mrm{in}$ after updating $\bdispl^\p$ will be large and the algorithm will not trigger premature convergence.
The residual-based transformation \eqref{eq:flexbc.fp_iter} of $\bforce^\ipl_{k+1}$ can thus be understood as a weakening of the compatibility between \textbf{(AH)} and \textbf{(H)}.
Moreover, due to the nonlinearity/nonconvexity of the problem, $\bforce^\ipl_\mrm{in}$ may change its direction from one iteration to another, leading no negative $\alpha_{k+1}$'s.
To ensure that the step $\alpha_{k+1} \bforce^\ip_{\mrm{in},k+1}$ always minimizes the incompatibility force, we reset $\alpha_{k+1}$\,$=$\,$1$ whenever $\alpha_{k+1}$\,$<$\,$0$, as shown in Algorithm \ref{algo:accel} (lines 2--4).

\begin{rem}
 In the preprint \citep{hodapp_efficient_2021}, we have proposed to use a lower bound, $\alpha_{\rm min}$, for $\alpha_{k+1}$, instead of using an additional convergence criterion for $\bforce^\ipl_\mrm{in}$.
 While no significant changes in the numerical experiments have been observed, compared with the preprint, the choice of $\alpha_{\rm min}$ appears to be less intuitive than choosing the tolerance on some norm of $\bforce^\ipl_\mrm{in}$ (which can, e.g., be chosen to be the same as the tolerance on the atomistic forces).
\end{rem}

On the other extreme, the relaxation method may accelerate too strongly, leading to large changes in the pad displacements $\bdispl^\p_{k+1} - \bdispl^\p_k$ that could cause convergence issues, in particular due to the nonconvexity of the problem.
This can be avoided by putting a bound $\Delta u_\mrm{max}$, e.g., a fraction of the lattice constant, on the maximum absolute change in the solution (Algorithm \ref{algo:accel}, lines 6--8).
So far, we have not observed such cases in our numerical experiments, so this is optionally mentioned here for the sake of completeness in order to illustrate how such possible cases can be handled in future applications of the method.

\begin{algorithm}[hbt]
 \SetAlgoSkip{bigskip}
 \LinesNumbered
 \SetKwInput{Input}{Input}
 \SetKwInput{Output}{Output}
 \SetKwBlock{Repeat}{repeat}{end}
 \setstretch{1.2}
 \setlength{\commentWidth}{0.5\textwidth}
 \newcommand{\atcp}[1]{\tcp*[r]{\makebox[\commentWidth]{#1\hfill}}}
 \caption{Residual-based relaxation (\texttt{Relax})}
 \label{algo:accel}
 \Input{incompatibility forces $\bforce^\ip_{\mrm{in},k+1}$, $\bforce^\ip_{\mrm{in},k}$ and relaxation parameter $\alpha_k$ from the previous iterations}
 $\alpha_{k+1} \,\leftarrow\, - \alpha_k \dfrac{\langle \bforce^\ip_{\mrm{in},k}, \bforce^\ip_{\mrm{in},k+1} - \bforce^\ip_{\mrm{in},k} \rangle}{\| \bforce^\ip_{\mrm{in},k+1} - \bforce^\ip_{\mrm{in},k} \|^2}$; \\
 \If{$\alpha_{k+1} < 0$}{
  $\alpha_{k+1} \,\leftarrow\, 1$;
 }
 \texttt{/* Optional:\,limit the maximum change in the solution} \\
 \If{$\| \alpha_{k+1} \greenOp^\pipl[\bforce_{\mrm{in},k+1}] \|_\infty > \Delta u_\mrm{max}$}{
  $\alpha_{k+1} \,\leftarrow\, \dfrac{\Delta u_\mrm{max}}{\| \greenOp^\pipl[\bforce_{\mrm{in},k+1}] \|_\infty}$;
 }
 \texttt{*/} \\
 \Output{relaxation parameter $\alpha_{k+1}$}
\end{algorithm}

\section{Implementation of the harmonic problem}

\subsection{Periodic lattice Green functions}
\label{sec:impl.plgf}

We now turn to the computation of the periodic lattice Green function $\bmG$.
While it is possible to compute $\bmG$ directly for a given periodicity, it appears not practical since a new Green function would have to be computed each time the periodic length $l_3$ changes.
Therefore, the strategy we will pursue in this work is as follows: we first construct a computable version of the lattice Green function $\bmG^\infty$ for the infinite (non-periodic) lattice $\latInf$, and then approximate $\bmG$ by summing up the $\bmG^\infty$'s from the periodic images.

It is usually not practical to compute $\bmG^\infty(\bmr)$ at all points. Therefore, we adopt the construction of \citep{hodapp_lattice_2019} which approximates $\bmG^\infty(\bmr)$ with the continuum Green function $\bmG^\cgf$ outside some cut-off radius $\cutOff$, i.e.,
\begin{equation}
 \bmG^\infty(\bmr) \approx \widetilde{\bmG}{}^\infty(\bmr) =
 \left\{
 \begin{aligned}
  \, \bmG^\infty(\bmr) &\qquad&& \| \bmr \| \le \cutOff, \\
  \, \bmG^\cgf(\bmr)   &      && \text{else.}
 \end{aligned}
 \right.
\end{equation}
Details on how we compute $\bmG^\infty(\bmr)$ are given in Appendix \ref{sec:appdx.Glgf}.

With $\widetilde{\bmG}{}^\infty(\bmr)$, we now approximate the periodic lattice Green function as
\begin{equation}\label{eq:plgf}
 \begin{aligned}
  \bmG(\bmr) \approx \widetilde{\bmG}(\bmr)
  &= \sum_{i=-\infty}^\infty \widetilde{\bmG}{}^\infty(r_1,r_2,r_3 + \Delta r_{3,i}) \\
  &=   \sum_{i=-m_1}^{m_2} \bmG^\infty(r_1,r_2,r_3 + \Delta r_{3,i})
     + \sum_{i=-\infty}^{-m_1-1} \bmG^\cgf(r_1,r_2,r_3 + \Delta r_{3,i})
     + \sum_{i=m_2+1}^\infty \bmG^\cgf(r_1,r_2,r_3 + \Delta r_{3,i}),
 \end{aligned}
\end{equation}
with $\Delta r_{3,i} = i\cdot l_3$;
the indices $m_1$ and $m_2$ depend on the precise choice of $\cutOff$.
However, those last two sums cannot be computed due to the fact that the Green function is proportional to $\|\bmr\|^{-1}$, and so they diverge.
On the other hand, $\bmG^\infty \simeq \bmG^\infty + \bmC$, that is, the Green function is unique only up to an otherwise arbitrary constant $\bmC$.
So, we can subtract $\bmG^\infty(\bmZero)$ without changing $\grad{}{}{\bmu^\p}$. Hence, we set
\begin{equation}\label{eq:plgf_conv}
 \widetilde{\bmG}(\bmr) = \sum_{i=-\infty}^\infty \widetilde{\bmG}{}^\infty(r_1,r_2,r_3 + \Delta r_{3,i}) - \widetilde{\bmG}{}^\infty(0,0,\Delta r_{3,i}).
\end{equation}
To see why the sum \eqref{eq:plgf_conv} is convergent, we recall that the continuum Green function can be written as a function \citep{barnett_precise_1972}
\begin{equation}\label{eq:cgf}
 \bmG^\cgf(\bmr) = \|\bmr\|^{-1} \bmF(\widehat{\bmr}),
\end{equation}
where $\bmF(\widehat{\bmr})$ is a tensor-valued function depending only on the direction $\widehat{\bmr} = \bmr/\|\bmr\|$ of $\bmr$, so the first term $\|\bmr\|^{-1}$ characterizes the asymptotic behavior.
For $i \le -m_1 - 1$ and $i \ge m_2+1$, expanding \eqref{eq:cgf} around $\bmr = \bmZero$ yields
\begin{equation}
 \bmG^\cgf(r_1,r_2,r_3 + \Delta r_{3,i}) = \bmG^\cgf(0,0,\Delta r_{3,i}) + \Delta r_{3,i}^{-2} \bmF'(\widehat{\bmr}) + \clO(\Delta r_{3,i}^{-3}),
\end{equation}
where $\bmF'(\widehat{\bmr})$ is some other tensor-valued function depending only on $\widehat{\bmr}$.
Since the sums $\sum_{i=\infty}^{-m_1-1} \Delta r_{3,i}^{-2}$ and $\sum_{i=m_2+1}^\infty \Delta r_{3,i}^{-2}$ are convergent, the representation \eqref{eq:plgf_conv} is also convergent.

To evaluate \eqref{eq:plgf_conv} in practice, we need to truncate the sum after a finite number of terms.
However, a direct summation converges rather slow and does not exploit the regularity properties of the asymptotic Green function.
Therefore, we will make use of a series acceleration method in order to reduce the number of required terms to be computed.
A method which is known to work well for series with terms $\propto$\,$\Delta r_{3,i}^{-2}$ is the Richardson extrapolation which assumes that any partial sum $\widetilde{\bmG}_n(\bmr)$ admits the series expansion
\begin{equation}\label{eq:plgf_poly}
 \widetilde{\bmG}_n(\bmr) = \sum_{i=-n}^n \widetilde{\bmG}^\infty(r_1,r_2,r_3 + \Delta r_{3,i}) - \widetilde{\bmG}{}^\infty(0,0,\Delta r_{3,i})
              = Q_0 + Q_1 n^{-1} + Q_2 n^{-2} + Q_3 n^{-3} + ...
\end{equation}
such that $\widetilde{\bmG} = \underset{n \to \infty}{\lim} \widetilde{\bmG}_n(\bmr) = Q_0$.
Then, the Richardson extrapolation attempts to approximate $\widetilde{\bmG}(\bmr)$ by truncating the right hand side of \eqref{eq:plgf_poly} after the term $Q_N n^{-N}$, and fitting the terms $\widetilde{\bmG}_n(\bmr), ..., \widetilde{\bmG}_{n+N}(\bmr)$ to the series expansion of order $N$. This yields a linear system of $N$\,+\,1 equations from which $Q_0$ can be deduced as (cf. \citep{bender_advanced_1999})
\begin{equation}\label{eq:plgf_rich}
 Q_0 = \sum_{i=0}^N \frac{\widetilde{\bmG}_{n+i}(\bmr) (n+i)^N (-1)^{i+N}}{i!(N-i)!} \approx \widetilde{\bmG}(\bmr).
\end{equation}
The total number of periodic images that need to be computed using the formulation \eqref{eq:plgf_rich} is thus $n_\mrm{img} = n+N$.

\begin{figure}[t]
 \begin{minipage}{0.5\textwidth}
  \centering
  \includegraphics[width=0.9\textwidth]{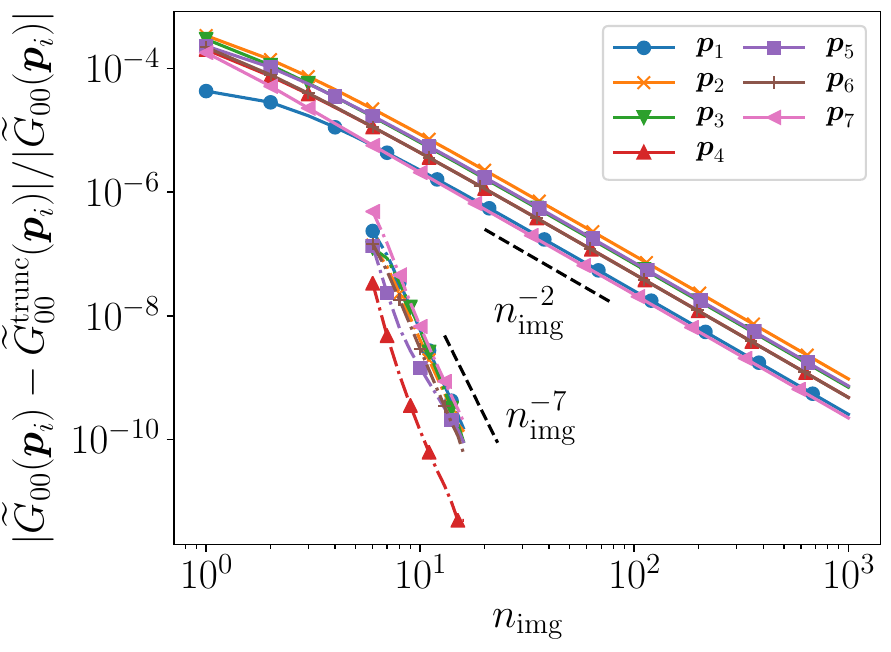}
 \end{minipage}\hfill
 \begin{minipage}{0.5\textwidth}
  \centering
  \includegraphics[width=0.9\textwidth]{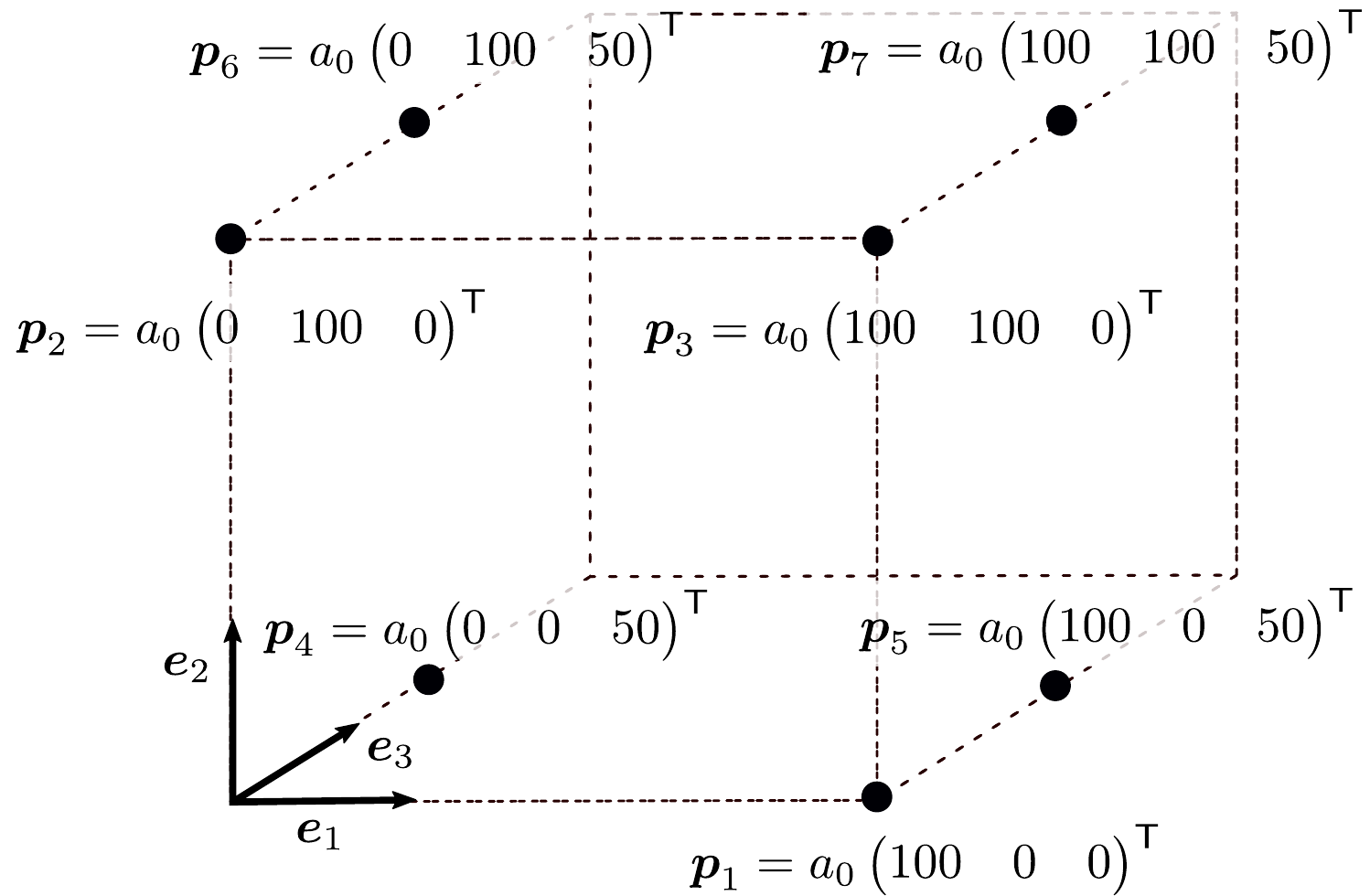}
 \end{minipage}
 \caption{Relative error $\vert \widetilde{G}_{00}(\bmp_i) - \widetilde{G}{}_{00}^\mrm{trunc}(\bmp_i) \vert / \vert \widetilde{G}_{00}(\bmp_i) \vert$ in the truncated sum $\widetilde{G}{}_{00}^\mrm{trunc}(\bmp_i)$ as a function of the number of periodic images $n_\mrm{img}$ when using direct summation (continuous lines) and Richardson extrapolation (dashed-dotted lines) for the selected points $\bmp_1$--$\bmp_7$ shown on the right ($a_0$ is the lattice constant). For the Richardson extrapolation it holds $n_\mrm{img} = n+N$ with $N$\,$=$\,$5$}
 \label{fig:err_plgf}
\end{figure}

The relative errors in the first component of the Green function, corresponding to the embedded-atom potential from Section \ref{sec:results.bowout}, are shown in Figure \ref{fig:err_plgf} for some collection of points. The points are chosen to be representative for the problem studied in Section \ref{sec:results}.
It can be seen that the error decays much faster when using the Richardson extrapolation in comparison with the direct summation. In all cases, a total number of six images, with $N$\,$=$\,$5$, is already sufficient for the relative error to be less than $10^{-6}$.
Although the direct summation also appears adequate with only a small number of images, we point out that there are situations where we require a highly accurate $\widetilde{\bmG}$.
One example is a dislocation subject to a very small applied stress so that small errors in the pad displacements can strongly impact the motion of the dislocation (see the example in Section 5.3 in \citep{hodapp_lattice_2019}).
In such cases series acceleration is valuable given the fact that $\widetilde{\bmG}$ may have to be evaluated millions of times while building the Green matrix (cf. following section).

\subsection{Hierarchical matrix approximation of the Green matrix}

An efficient implementation of Algorithm \ref{algo:fbc} requires an efficient computation of the harmonic problem {\bf{(H)}}.
To see this, let us denote the matrix-vector multiplication associated with \eqref{eq:flexbc.up=Gpipl*fipl} as
\begin{equation}\label{eq:impl.up=Gpipl*fipl_mvm}
 \uu^\p = \widetilde{\uuG}{}^\pipl \uf^\ipl, \qquad \text{with} \quad
 \uu^\p \in \real^{3\vert\lat^\p\vert}, \; \widetilde{\uuG}{}^\pipl \in \real^{3\vert\lat^\p\vert \times 3\vert\lat^\ipl\vert}, \; \uf^\ipl \in \real^{3\vert\lat^\ipl\vert},
\end{equation}
using the approximate periodic lattice Green functions.
The Green matrix $\widetilde{\uuG}{}^\pipl$ is then assembled as follows
\begin{equation}
 \widetilde{\uuG}{}^\pipl =
 \begin{pmatrix}
  \widetilde{\bmG}(\bmr_{1,1})                 & \cdots & \widetilde{\bmG}(\bmr_{1,\vert\lat^\ipl\vert}) \\
  \vdots                                       & \ddots & \vdots \\
  \widetilde{\bmG}(\bmr_{\vert\lat^\p\vert,1}) & \cdots & \widetilde{\bmG}(\bmr_{\vert\lat^\p\vert,\vert\lat^\ipl\vert})
 \end{pmatrix},
 \qquad \text{where} \quad \bmr_{i,j} = \batoB_j - \bato_i, \quad \bato_i \in \lat^\p, \; \batoB_j \in \lat^\ipl.
\end{equation}
Thus, $\widetilde{\uuG}{}^\pipl$ is dense and the complexity of computing \eqref{eq:impl.up=Gpipl*fipl_mvm} is of $\clO(\vert\lat^\ipl\vert \vert\lat^\p\vert)$.
To relate this complexity to the atomistic problem, we first note that for basic geometries, such as the rectangular domain in Figure \ref{fig:dom_decomp}, we have that $\vert\lat^\a\vert \lesssim \vert\lat^\ipl\vert^{3/2}$, and so $\vert\lat^\ipl\vert \gtrsim \vert\lat^\a\vert^{2/3}$.
Therefore, solving the harmonic problem is at \emph{least of} $\clO(\vert\lat^\a\vert^{4/3})$.
Thus, assuming that solving the atomistic problem scales linear with $\vert\lat^\a\vert$, the harmonic problem will become more expensive than the atomistic problem.
Another issue is the cost for storing the $\widetilde{\uuG}{}^\pipl$ matrix.
Due to the quadratic complexity the memory requirement quickly grows to orders of several hundred gigabytes and beyond.
This is inefficient, even on larger computing clusters, and also eliminates the possibility of using the method on laptops and smaller workstations.
In other works on the FBC method (e.g., \citep{rao_greens_1998}), eq. \eqref{eq:impl.up=Gpipl*fipl_mvm} has been computed by looking up $\widetilde{\bmG}$ ``on-the-fly'' without storing the matrix $\widetilde{\uuG}{}^\pipl$ explicitly---but in this case the computing complexity is even higher and so this is only adequate for small problems.

Fortunately, all these aforementioned issues can be remedied by using a hierarchical matrix \citep[$\scH$-matrix,][]{tyrtyshnikov_mosaic-skeleton_1996,hackbusch_sparse_1999} approximation of $\widetilde{\uuG}{}^\pipl$, in the following denoted by $\widetilde{\uuG}{}_\scH^\pipl$.
$\scH$-matrices are data-sparse approximations of the original matrix with controllable accuracy, exploiting the asymptotic smoothness of the off-diagonal entries $\widetilde{\green}_{kl}(\bmr_{i,j})$ \citep{borm_introduction_2003}, i.e.,
\begin{equation}\label{eq:impl.smoothness}
 \forall\, a,b \in \nat_0^3 \qquad
 \vert \grad{\xi}{a}{\grad{\eta}{b}{\widetilde{\green}_{kl}}}(\bmr_{i,j}) \vert \le C \| \bmr_{i,j} \|^{- \vert a \vert - \vert b \vert} \vert \widetilde{\green}_{kl}(\bmr_{i,j}) \vert,
\end{equation}
where $C$ is some constant independent of $\bmr_{i,j}$, but depending on $a,b$.
Approximating $\widetilde{\uuG}{}^\pipl$ with $\widetilde{\uuG}{}_\scH^\pipl$ ensures that all relevant matrix operations, that is, \emph{building} $\widetilde{\uuG}{}_\scH^\pipl$, \emph{storing} $\widetilde{\uuG}{}_\scH^\pipl$, and \emph{multiplying} $\widetilde{\uuG}{}_\scH^\pipl$ by a vector, scale \emph{linear-logarithmic} with the number of pad atoms $\vert \lat^\p \vert$.
We will demonstrate this below using a rectangular domain exemplified in Figure \ref{fig:dom_decomp} (which will also be used in the computational results section).
Our construction of $\widetilde{\uuG}{}_\scH^\pipl$ is largely standard and we therefore refer the reader to Appendix \ref{sec:appdx.GpiplH} for those details.

\begin{figure}[t]
 \begin{minipage}{0.5\textwidth}
  \centering
  (a)
 \end{minipage}\hfill
 \begin{minipage}{0.5\textwidth}
  \centering
  (b)
 \end{minipage}\\[0.5em]
 \begin{minipage}{0.5\textwidth}
  \centering
  \includegraphics[width=0.9\textwidth]{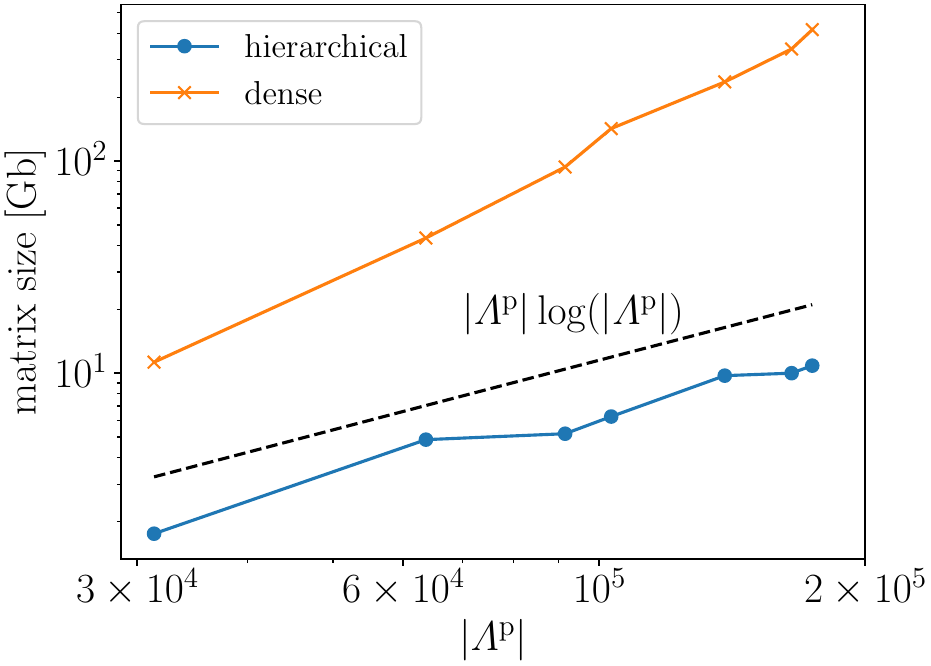}
 \end{minipage}\hfill
 \begin{minipage}{0.5\textwidth}
  \centering
  \includegraphics[width=0.9\textwidth]{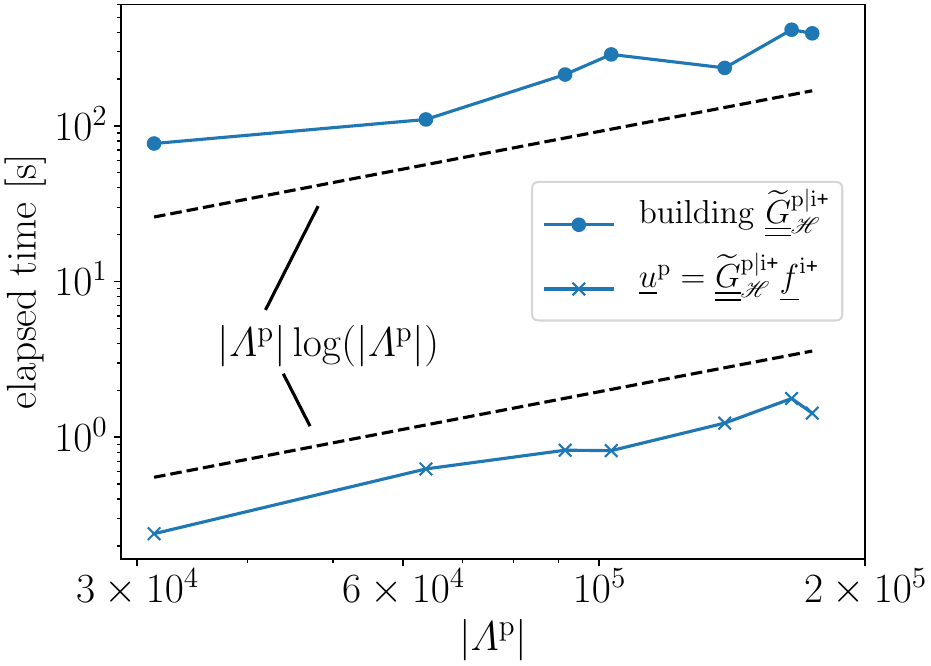}
 \end{minipage}
 \caption{(a) Matrix size vs. number of pad atoms. (b) Elapsed times for building $\widetilde{\uuG}{}_\scH^\pipl$ and computing $\widetilde{\uuG}{}_\scH^\pipl\uf^\ipl$ for a random $\uf^\ipl$}
 \label{fig:sinclair_scaling}
\end{figure}

In the following we use a $\widetilde{\uuG}{}_\scH^\pipl$ with a relative error in the Frobenius norm $\|  \widetilde{\uuG}{}^\pipl - \widetilde{\uuG}{}_\scH^\pipl \|_\mrm{fro} / \|  \widetilde{\uuG}{}^\pipl \|_\mrm{fro}$ of the order of $10^{-5}$.
The linear scaling of the matrix size as a function of the number of pad atoms $\vert \lat^\p \vert$ is shown in Figure \ref{fig:sinclair_scaling} (a).
Therein, the last point corresponds to a total number of roughly 170\,000 pad atoms and 1 mio. real atoms.
Extrapolating the matrix size from this point to 800\,000 pad atoms and 10 mio. real atoms, yields approximately 47\,GB for storing the $\scH$-matrix and 9\,TB for storing the dense matrix.
Clearly, the FBC method is thus unfeasible with dense matrix algebra for such large systems.

Further, the linear scaling of the elapsed times for building $\widetilde{\uuG}{}_\scH^\pipl$ and computing $\widetilde{\uuG}{}_\scH^\pipl\uf^\ipl$ as a function of $\vert \lat^\p \vert$ is shown in Figure \ref{fig:sinclair_scaling} (b).
Thereby, for the considered data points, the time for computing $\widetilde{\uuG}{}_\scH^\pipl\uf^\ipl$ is of the order of 0.1\,ms.
This is approximately the same amount of time as required for computing the forces on all real atoms using the embedded-atom method potential from Section \ref{sec:results.bowout}.
Therefore, the time spent for solving the harmonic problem is negligible compared to one energy minimization (which usually requires $\clO(100)$ or more force computations).
While we remark here that the time required for building the matrix is roughly two orders of magnitude higher than for computing $\widetilde{\uuG}{}_\scH^\pipl\uf^\ipl$, it should be noted that this operation is only carried out once before the beginning of the simulation.
Hence, this cost is not significant, also with respect to the fact that the same matrix can be reused for many simulations.


\section{Computational results}
\label{sec:results}

\subsection{Simulation setup}
\label{sec:results.setup}

In order to validate the proposed methodology and demonstrate its capabilities, we have selected three distinct test problems,
i) a bow-out of a nominally straight edge dislocation, ii) a precipitate strengthening mechanism, and iii) cross-slip of a nominally straight screw dislocation.
The three test problems will be discussed in the Sections \ref{sec:results.bowout}--\ref{sec:results.cross_slip}.

As an interatomic potential, we use the average embedded-atom method potential from \citep{varvenne_average-atom_2016,hodapp_lattice_2019} for isotropic fcc aluminum-magnesium%
\footnote{the potential has been tuned to produce isotropic elastic constants by varying the magnesium concentration}
\begin{equation}
 \Eato(\{\bdispl_{\atoB} - \bdispl_{\ato}\}) = \sum_{\batoB \in \intRgAto} \mphi^\mathrm{avg}(\bdispl_{\atoB} - \bdispl_{\ato}) + \sum_{X} c_X F_X(\bar{\mrho}_{\ato}(X,\{\bdispl_{\atoB} - \bdispl_{\ato}\})),
\end{equation}
where $\mphi^\mathrm{avg}$ is the average pair potential, $c_X$ is the concentration of atom type $X$, and $F_X$ is the embedding function of the average electron density $\bar{\mrho}_{\ato}(X,\{\bdispl_{\atoB} - \bdispl_{\ato}\})$ of atom $\bato$.
The potential has the lattice constant $a_0$\,=\,4.109\,\AA, the shear modulus $\mu$\,=\,21\,GPa, and the Poisson ration $\nu$\,=\,0.355.
For details concerning the computation of the periodic lattice Green function the reader is referred to Section \ref{sec:impl.plgf}.

Our version of the FBC method (Algorithm \ref{algo:fbc}) is implemented in an in-house code, with interface to LAMMPS \citep[\href{https://lammps.sandia.gov/}{lammps.sandia.gov},][]{plimpton_fast_1995} that we employ to relax the atoms in $\latA$ using its Hessian-free Newton-Raphson solver.
In each iteration $k$+1 we solve the atomistic problem in line 3 up to a tolerance of $0.025 \cdot \| \var{}{}{\Etot^\a}(\bdispl_k) \|$.
Algorithm \ref{algo:fbc} terminates when $\| \var{}{}{\Etot^\a}(\bdispl_k) \| < 10^{-2}\,\mrm{eV}/\text{\AA}$ and $\| \bforce^\ip_{\mrm{in},k} \|_\infty < 10^{-3}\,\mrm{eV}/\text{\AA}$ (if relaxation is enabled) after updating the pad atoms.

\subsection{Example 1: Bow-out}
\label{sec:results.bowout}

\subsubsection{Problem description}

\begin{figure}[hbt]
 \centering
 \includegraphics[width=0.9\textwidth]{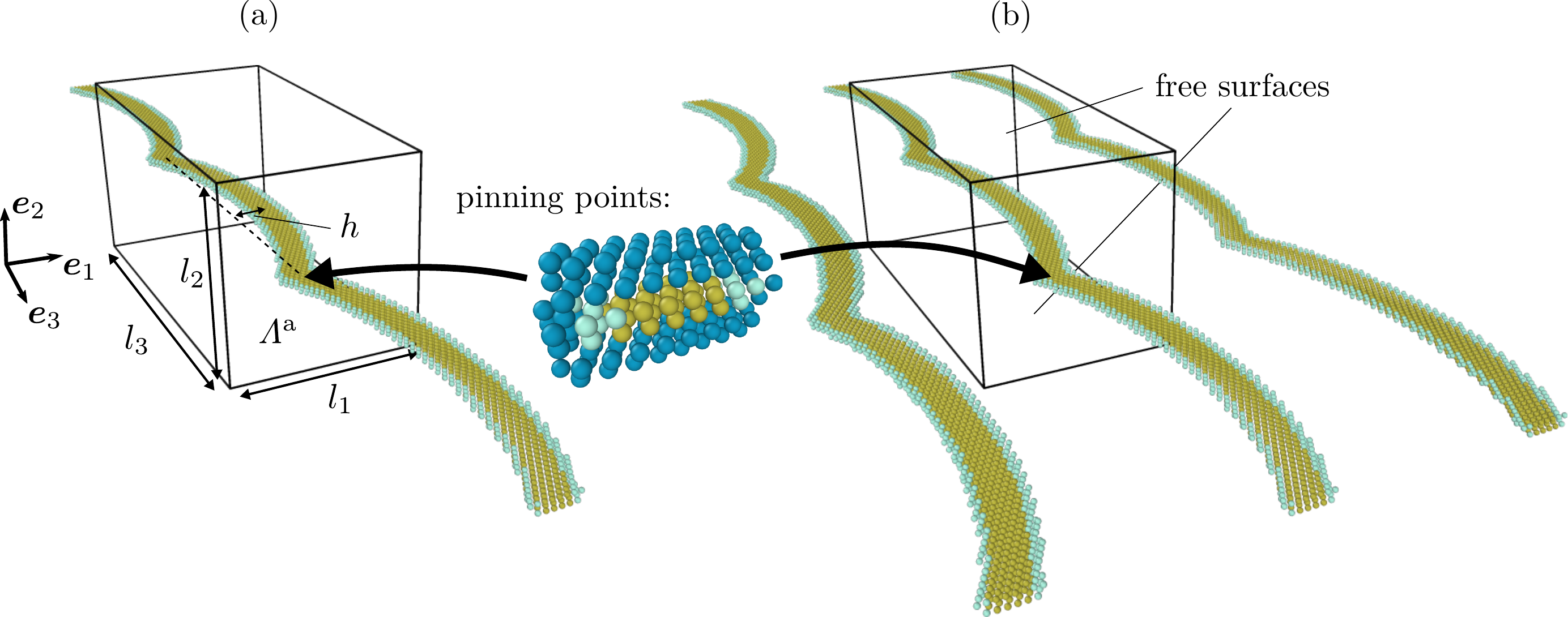}
 \caption{Schematic illustration of the bow-out problem using (a) the FBC method, and (b) the PAD method}
 \label{fig:bowout_problem}
\end{figure}

We first consider the problem of an initially straight [121] edge dislocation periodically bowing between obstacles, mimicked here by two pinning points as shown in Figure \ref{fig:bowout_problem} (a).
This is a well-defined quasi-static problem and therefore ideally suited for testing purposes, but also relevant from a practical point of view in order to, e.g., compute the line tension of a dislocation, a parameter that is required in strengthening models (see, e.g., \citep{argon_strengthening_2007}).

As an atomistic domain we use a rectangular cell of size $l_1$\,$\times$\,$l_2$\,$\times$\,$l_3$.
Our simulation setup is then as follows:
We first apply the elastic solution of a straight edge dislocation \citep{hirth_theory_1982} and relax the core structure.
Subsequently, we fix a cluster of atoms of size 26$\times$14$\times$8\,\AA$^3$ near the periodic boundaries (cf. Figure \ref{fig:bowout_problem} (a)) to pin the dislocation.
Next, we take the relaxed solution, denoted by $\bdispl_\mrm{edge}$, and superimpose it with the solution $\widehat{\bdispl}$ for a constant applied shear stress $\tau_\mrm{app}$ to obtain the initial guess $\bdispl_0 = \bdispl_\mrm{edge} + \widehat{\bdispl}$, with $\widehat{\bdispl}(\bato) = (\tau_\mrm{app}/\mu) \ato_2 \bme_1$.

We investigate the bow-out problem for two periodic lengths/applied stresses: i) $l_3$\,=\,$100\,\text{\AA}$/$\tau_\mrm{app}$\,=\,$140\,\mrm{MPa}$, and ii) $l_3$\,=\,$200\,\text{\AA}$/$\tau_\mrm{app}$\,=\,$35\,\mrm{MPa}$.
In both cases the dislocation bows over a distance of $\approx$\,20\,\AA.
The lateral dimensions are chosen to be $l_1$\,=\,70\,{\AA}, $l_2$\,=\,30\,{\AA}, and the dislocation is placed in $\latA$ such that it always remains 15--20\,{\AA} from the interface throughout the simulation.

We will compare the FBC method with the periodic array of dislocations (PAD) method \citep{daw_embedded-atom_1993,osetsky_atomic-level_2003}.
The PAD method uses periodic boundary conditions in the $\rmx_1$-direction and free surfaces on the upper and lower boundaries on the $\rmx_2$-direction (see Figure \ref{fig:bowout_problem} (b)).
Additional details of our setup can be found in Appendix \ref{sec:appdx.PAD}.

\subsubsection{Accuracy}
\label{sec:results.bowout.accuracy}

We judge the accuracy of the FBC method based on the final position of the dislocation line as this is the quantity of interest when performing such a study in practice, e.g., for computing the dislocation line tension.
A detailed analysis of the error in the energy norm can be found in several preceding publications \citep{hodapp_flexible_2018,hodapp_lattice_2019,hodapp_analysis_2021}.

First, we investigate the necessary size of the atomistic domain in terms of $l_1$ and $l_2$.
We therefore run the simulation using a large cell size with $l_1$\,=\,$120\,\text{\AA}$ and $l_2$\,=\,$115\,\text{\AA}$ and compare the results when using the domain described in the previous section, denoted by ``minimal cell'' in the following.
The origin of the large cell is thereby shifted along the $\rmx_1$-direction so that the distance of the bowing dislocation line and both interfaces is approximately equal.

\begin{figure}[hbt]
 \begin{minipage}{0.5\textwidth}
  \centering
  (a) $\tau_\mrm{app}$\,$=$\,140\,MPa
 \end{minipage}\hfill
 \begin{minipage}{0.5\textwidth}
  \centering
  (b) $\tau_\mrm{app}$\,$=$\,35\,MPa
 \end{minipage}\\[0.5em]
 \begin{minipage}{0.5\textwidth}
  \centering
  \includegraphics[width=0.9\textwidth]{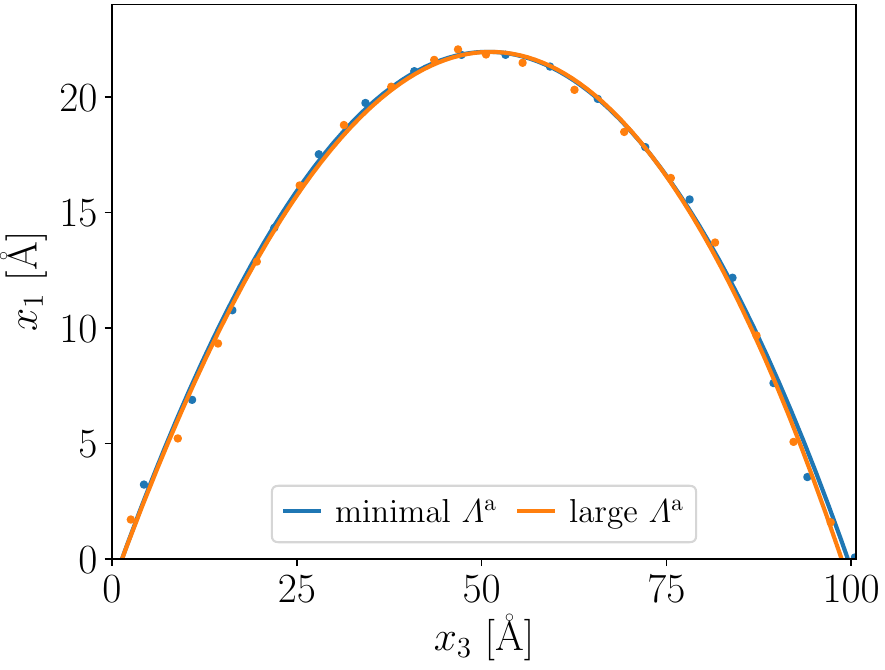}
 \end{minipage}\hfill
 \begin{minipage}{0.5\textwidth}
  \centering
  \includegraphics[width=0.9\textwidth]{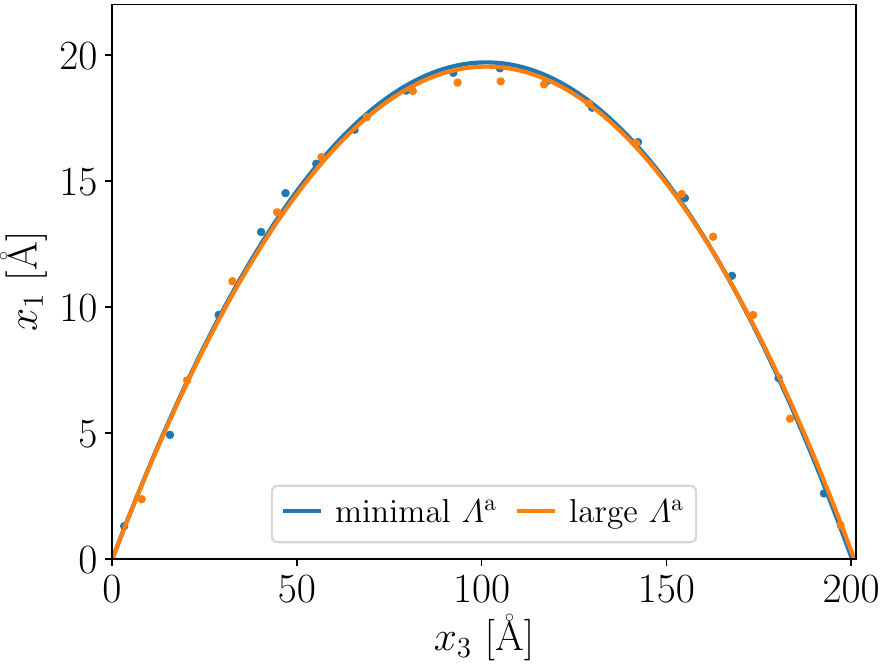}
 \end{minipage}
 \caption{Comparison of the dislocation lines when using a large atomistic cell size with $l_1$\,=\,$120\,\text{\AA}$ and $l_2$\,=\,$115\,\text{\AA}$, and a minimal atomistic cell size with $l_1$\,=\,$70\,\text{\AA}$ and $l_2$\,=\,$30\,\text{\AA}$. (a) $l_3$\,=\,$100\,\text{\AA}$. (b) $l_3$\,=\,$200\,\text{\AA}$}
 \label{fig:disloc_fbc}
\end{figure}

To compare the dislocations lines, we have taken the nodes obtained from a dislocation detection algorithm \citep{stukowski_visualization_2010,stukowski_automated_2012} and fitted a polynomial to them.
There is almost no visual difference between the final positions of the dislocations for both cell sizes as shown in Figure \ref{fig:disloc_fbc}.
The small gap in the nodes for $l_3$\,=\,$200\,\text{\AA}$ and $\tau_\mrm{app}$\,=\,35\,MPa is likely due to minor inaccuracies in the lattice Green function (due to a too small $r_\mrm{cut}$) which must be very precise for lower applied stresses (cf. \citep{hodapp_lattice_2019}). Since the gap is less than one fifth of the Burgers vector, i.e., the dislocations are essentially in the same position, we are not further concerned with it and refer for all following considerations to the results obtained with the minimal cell.

Next, we compare the FBC method to a ``pure atomistic'' calculation using the PAD boundary conditions.
Choosing the right cell size for the PAD method is, however, far more  involved than for FBCs.
This is due to the fact that in the PAD method spurious image stresses arise due to the free surface and the periodic images in the $\rmx_1$-direction.
\citet{szajewski_analysis_2015} have derived an effective image stress on a dislocation bowing into a circular arc as a function of $h$
\begin{equation}\label{eq:imgstress}
 \tau_{\rm img, eff}(h) = C_1 \bar{V}^{-1} l_3^{-2} h,
\end{equation}
where $C_1$ is a constant depending on the elastic constants, the Burgers vector and the relative length $\bar{L} = l_1/l_2$, and $\bar{V} = l_1l_2l_3/l_3^3 = l_1l_2/l_3^2$ is a normalized volume.
In their work, Szajewski and Curtin have shown that $\tau_{\rm img, eff}(h)$ is minimal for edge dislocations when $\bar{L} \approx 0.8$.
Thus, we henceforth assume tacitly that $\bar{L} = 0.8 \equiv \text{constant}$.

In the following we investigate the impact of $\tau_{\rm img, eff}(h)$ and, in particular, $\bar{V}$ on the difference between the dislocation line obtained with the PAD method and the exact (image-stress-free) FBC solution.
Therefore, we recall that the relation between the applied stress and $h$ is, according to linear elasticity (cf. \citep{shenoy_finite-sized_1997}),
\begin{equation}\label{eq:appstress}
 \tau_{\rm app}(h) = C_2 l_3^{-2} h,
\end{equation}
where $C_2$ is another constant which depends on the elastic constants and the Burgers vector. Hence, the ratio $\tau_{\rm img, eff}(h) / \tau_{\rm app}(h)$ scales as $\bar{V}^{-1}$.
To keep this ratio constant when increasing the cell length $l_3$, we need to keep $\bar{V}$ constant---simply extending the cell in the $\rmx_3$-direction, as we have done for the FBCs, can lead to large errors, even though the dislocation moves in essence over the same distance regardless of the value of $l_3$.

\begin{figure}[t]
 \begin{minipage}{0.5\textwidth}
  \centering
  (a) $\tau_\mrm{app}$\,$=$\,140\,MPa
 \end{minipage}\hfill
 \begin{minipage}{0.5\textwidth}
  \centering
  (a) $\tau_\mrm{app}$\,$=$\,35\,MPa
 \end{minipage}\\[0.5em]
 \begin{minipage}{0.5\textwidth}
  \centering
  \includegraphics[width=0.9\textwidth]{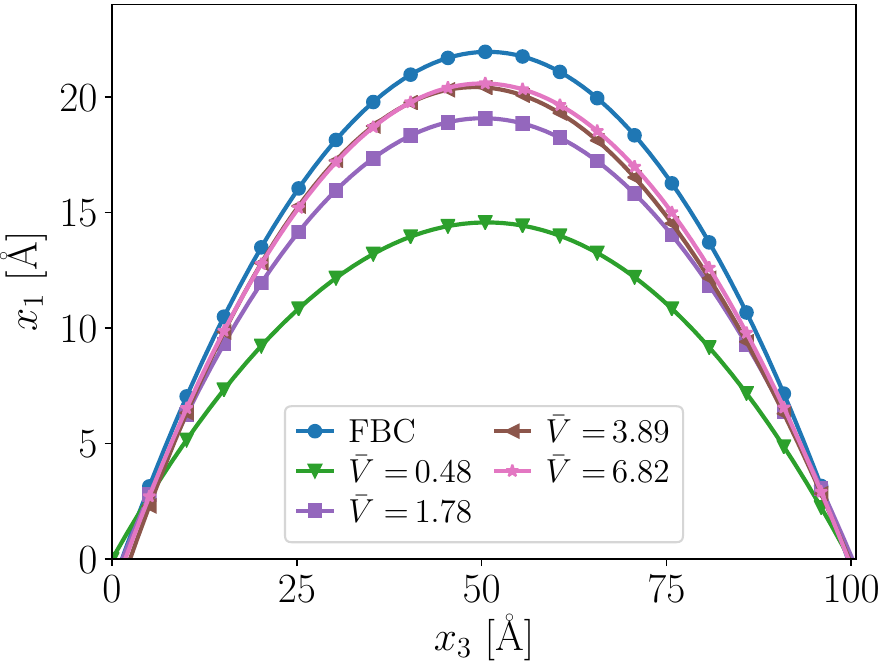}
 \end{minipage}\hfill
 \begin{minipage}{0.5\textwidth}
  \centering
  \includegraphics[width=0.9\textwidth]{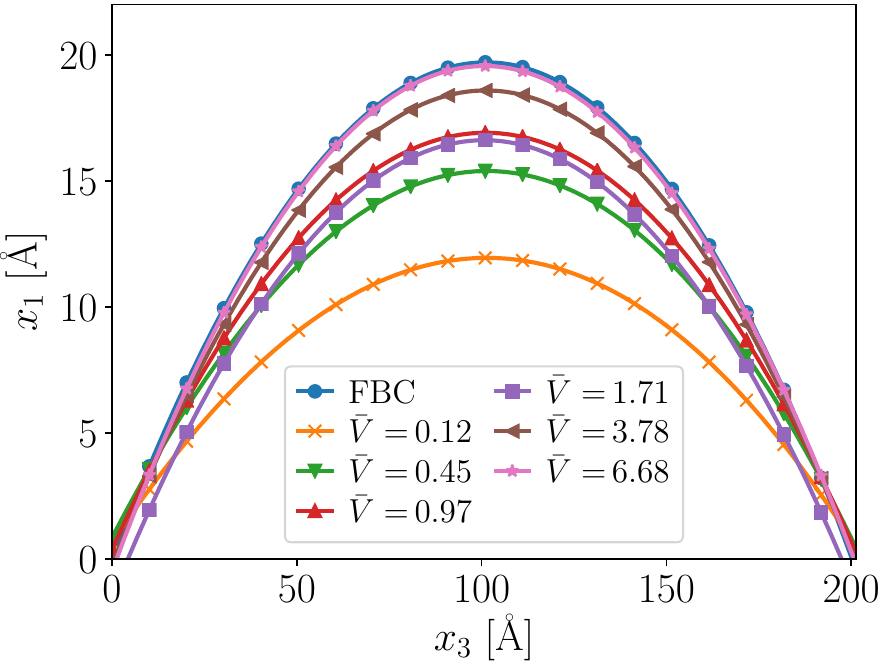}
 \end{minipage}
 \caption{{Comparison of the dislocation lines when using i) the FBC method, or ii) the PAD method for different normalized volumes $\bar{V} = l_1l_2/l_3^2$. (a) $l_3$\,=\,$100\,\text{\AA}$. (b) $l_3$\,=\,$200\,\text{\AA}$}}
 \label{fig:disloc_fbc_pad}
\end{figure}

The results are shown in Figure \ref{fig:disloc_fbc_pad}.
Comparing the dislocation lines confirms the previously stated estimate (compare, e.g., the dislocation lines for $\bar{V} = 3.89$ in (a) and $\bar{V} = 3.78$ in (b)).

\begin{figure}[hbt]
 \centering
 \includegraphics[width=0.5\textwidth]{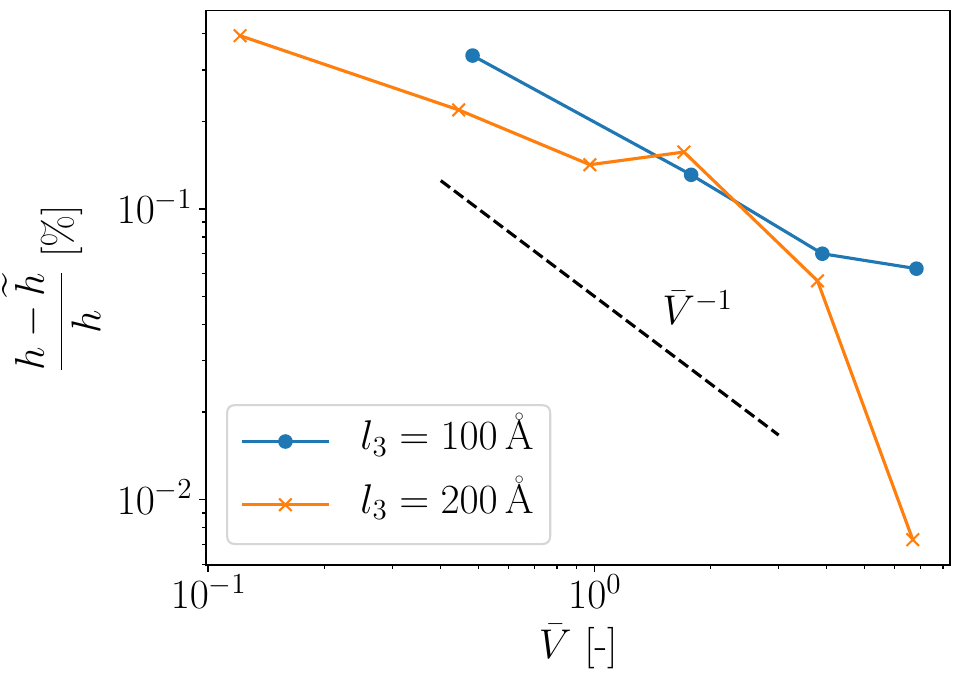}
 \caption{Relative error of the PAD method in the maximum dislocation bow-out as a function of the normalized volume $\bar{V} = l_1l_2/l_3^2$}
 \label{fig:disloc_error}
\end{figure}

To make this dependence on $\bar{V}$ more precise, we estimate the relative error in the maximum bow-out using \eqref{eq:imgstress} and \eqref{eq:appstress} by
\begin{equation}
 \frac{h - \widetilde{h}}{h} \approx \frac{h - (h - h_{\rm img})}{h} = \frac{h_{\rm img}}{h} \simeq \bar{V}^{-1},
\end{equation}
where $h_{\rm img} = \tau_{\rm img,eff} l_3^2 / C_2$ is the portion of the bow-out due to the effective image stress.
The dependence of the error on $\bar{V}$ is manifested in Figure \ref{fig:disloc_error} where the error in the maximum bow-out $h$ is shown as a function of the normalized volume $\bar{V}$: The two curves for $l_3$\,=\,100\,{\AA} and $l_3$\,=\,200\,{\AA} more or less coincide.
The preasymptotic scaling of the error is even slightly worse than $\bar{V}^{-1}$, but note that the estimates are based on continuum linear elasticity and do not take intrinsically atomistic effects into account (e.g., Peierls stress, partial dislocations, etc.).

So, to fully ensure that the error of the PAD method remains constant when increasing $l_3$, we need to increase the product $l_1l_2$ such that $\bar{V}$ remains constant.
We will see in the following section that keeping $\bar{V}$ constant can drastically degrade the efficiency of the PAD method.

\subsubsection{Efficiency}
\label{sec:results.bowout.efficiency}

We first compare the efficiency of the FBC method with and without residual-based relaxation.
For clarity we will denote the FBC method with relaxation by rFBC method.

\begin{figure}[t]
 \begin{minipage}{0.5\textwidth}
  \centering
  (a)
 \end{minipage}\hfill
 \begin{minipage}{0.5\textwidth}
  \centering
  (b)
 \end{minipage}\\[0.5em]
 \begin{minipage}{0.5\textwidth}
  \centering
  \includegraphics[width=0.9\textwidth]{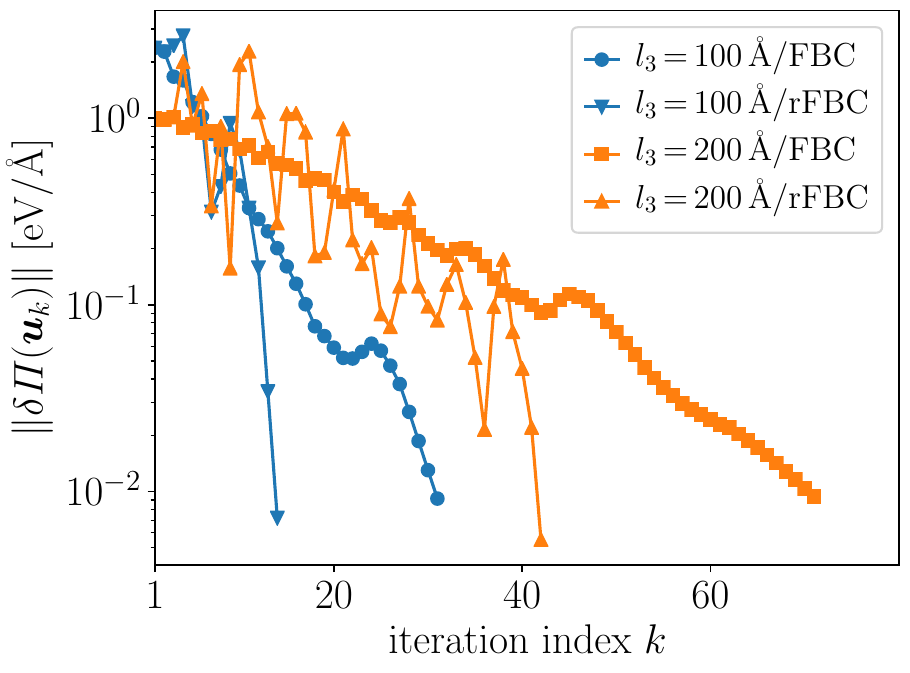}
 \end{minipage}\hfill
 \begin{minipage}{0.5\textwidth}
  \centering
  \includegraphics[width=0.83\textwidth]{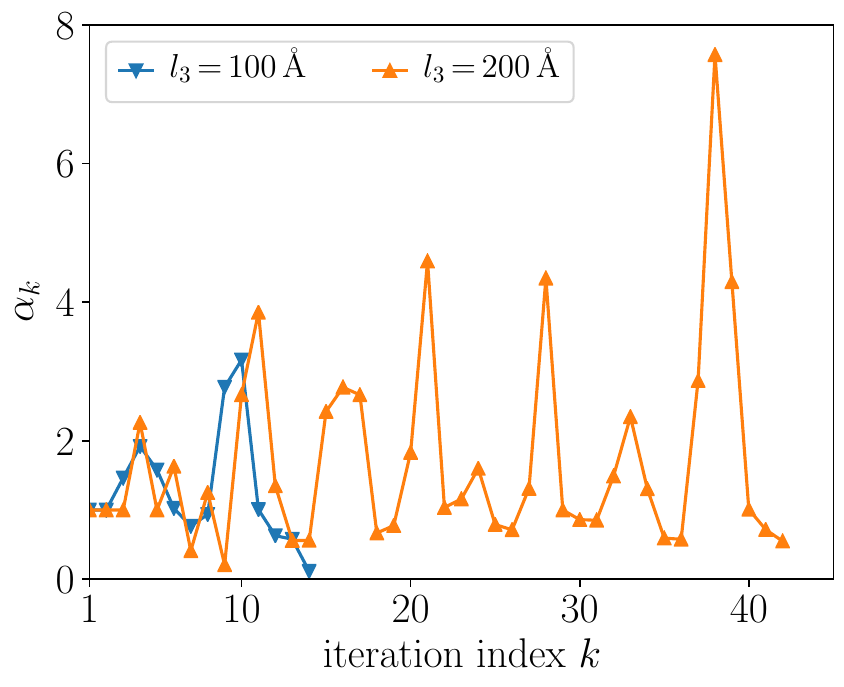}
 \end{minipage}
 \caption{(a) Force norm vs. iteration index for $l_3$\,=\,100\,{\AA} and $l_3$\,=\,200\,{\AA} using FBCs and rFBCs. (b) Relaxation parameter $\alpha$ vs. iteration index for the two simulations from (a) using rFBCs}
 \label{fig:fnorm2&alpha_vs_it}
\end{figure}

In Figure \ref{fig:fnorm2&alpha_vs_it} (a) the force norm is shown as a function of the iteration index for both problems ($l_3$\,=\,100\,\AA\;and $l_3$\,=\,200\,\AA) using FBCs and rFBCs.
Overall, rFBCs require much fewer iterations compared to FBCs, 14 vs. 31 for $l_3$\,=\,100\,\AA, and 42 vs. 71 for $l_3$\,=\,200\,\AA.
This can be rationalized by inspection of the oscillatory parts corresponding to situations when the dislocations are about to cross a Peierls barrier, i.e., saddle points of the atomistic Hessian.
The rFBC-curves show more pronounced and more frequent oscillations and, so, the rFBC method overcomes these saddle points faster.
The oscillations are intimately linked to the relaxation parameter $\alpha$ which steadily increases when the dislocation glides smoothly through a Peierls valley (compare the peaks in the curves in Figure \ref{fig:fnorm2&alpha_vs_it} (b) with the ones from (a)).
This elevates the incompatibility force, and thus the pad nodes advance more rapidly.
In addition, rFBCs prevent oscillations around the equilibrium state (we have checked that both methods converge to the same result).

Next, we assess the efficiency of the rFBC method with respect to the PAD method.
Since the most time consuming part is usually the computation of the interatomic forces, we assess the efficiency of the rFBC method using the efficiency index $\chi(l_3)$ as the ratio of the required number of \emph{per-atom} force computations (i.e., the total number of atoms $N_\a$ times the total number of \emph{global} force computations $N_{\var{}{}{\Etot}}$) per required number of per-atom force computations $N_\rma^\mrm{rFBC} \cdot N_{\var{}{}{\Etot}}^\mrm{rFBC}$ when using the rFBC method.
We thus define
\begin{equation}
 \chi(l_3) = \frac{N_\rma \cdot N_{\var{}{}{\Etot}}}{N_\rma^\mrm{rFBC} \cdot N_{\var{}{}{\Etot}}^\mrm{rFBC}}
\end{equation}
in accordance with the values reported in Table \ref{tab:efficiency_l3=100} and \ref{tab:efficiency_l3=200}.

From Table \ref{tab:efficiency_l3=100} and \ref{tab:efficiency_l3=200} it can be seen that the efficiency of the rFBC method is remarkable.
To reach an error of $\approx$\;6--7\,\% (half a Burgers vector) in the maximum amount of bow-out, the PAD method requires a normalized volume of $\bar{V}$\,$\approx$\,3.8, and this leads to $\approx$\,32 ($l_3$\,=\,100\,\AA) and $\approx$\,70 ($l_3$\,=\,200\,\AA) more per-atom force computations for the PAD method.
To reach essentially the same accuracy than the rFBC method for $l_3$\,=\,200\,\AA, a normalized volume of $\bar{V}$\,=\,6.68 is required corresponding to more than 3 mio. atoms.
In the latter case the PAD method requires more than two orders of magnitude more force computations!

The exact computing times $t$, obtained using a Lenovo ThinkPad T470s with an Intel i5-7200U processor, are reported in Tables \ref{tab:efficiency_l3=100} and \ref{tab:efficiency_l3=200}.
We remark that the exact speed-ups are about one third of $\chi$ (which is still very efficient!).
Since the absolute time $t_\rmh$, spent for solving the harmonic problem \textbf{(H)}, amounts to less than 1\,\% of the entire simulation time (cf. $t_\rmh/t$ in Tables \ref{tab:efficiency_l3=100} and \ref{tab:efficiency_l3=200}), we attribute this to additional setup times when restarting the energy minimization via LAMMPS in every iteration.
We are planning to optimize this, together with exploring possibly more efficient variants of Algorithm \ref{algo:fbc} (cf. Section \ref{sec:outlook}), and report the results in a future publication.
Nevertheless we point out that, when using more expensive interatomic potentials with cluster functionals, the reported $\chi$'s can indeed be considered as the speed-ups.

\begin{table}[htb]
 \centering
 \begin{tabular}{|c|c|c|c|c|c|c|c|c|c|c|}
  \hline
  Method & $l_1$\,[\AA] & $l_2$\,[\AA] & $\bar{V}$\,[\AA] & $N_\rma$ & $(h-\widetilde{h})/h\,[\%]$  & $N_{\var{}{}{\Etot}}$ & $\chi$ & $t$\,[h] & $t/t_\mrm{rFBC}$ & $t_\rmh/t\,[\%]$ \\ \hline\hline
  FBC  & 70  & 30  & 0.21 & 11\,520  & $\diagup$ & 5\,720 & 2.28  & 0.13 & 2.34  & 0.91      \\ \hline 
  rFBC & 70  & 30  & 0.21 & 11\,520  & $\diagup$ & 2\,509 & 1     & 0.06 & 1     & 0.89      \\ \hline 
  PAD  & 62  & 78  & 0.48 & 29\,240  & 33.64     & 1\,749 & 1.77  & 0.04 & 0.63  & $\diagup$ \\ \hline 
  PAD  & 121 & 150 & 1.78 & 106\,240 & 13.09     & 2\,576 & 9.47  & 0.19 & 3.32  & $\diagup$ \\ \hline 
  PAD  & 179 & 221 & 3.89 & 231\,240 & 7.01      & 4\,004 & 32.03 & 0.62 & 11.13 & $\diagup$ \\ \hline 
  PAD  & 237 & 292 & 6.82 & 404\,240 & 6.22      & 4\,325 & 60.49 & 1.21 & 21.47 & $\diagup$ \\ \hline 
 \end{tabular}
 \caption{Numerical results for $l_3$\,=\,100\,\AA}
 \label{tab:efficiency_l3=100}
\end{table}

\begin{table}[H]
 \centering
 \begin{tabular}{|c|c|c|c|c|c|c|c|c|c|c|}
  \hline
  Method & $l_1$\,[\AA] & $l_2$\,[\AA] & $\bar{V}$\,[\AA] & $N_\rma$ & $(h-\widetilde{h})/h\,[\%]$  & $N_{\var{}{}{\Etot}}$ & $\chi$ & $t$\,[h] & $t/t_\mrm{rFBC}$ & $t_\rmh/t\,[\%]$ \\ \hline\hline
  FBC  & 70  & 30  & 0.05 & 23040       & $\diagup$ & 14\,408 & 1.84   & 0.66  & 1.92  & 0.79      \\ \hline 
  rFBC & 70  & 30  & 0.05 & 23040       & $\diagup$ & 7\,846  & 1      & 0.27  & 1     & 0.77      \\ \hline 
  PAD  & 62  & 78  & 0.12 & 58\,480     & 39.37     & 4\,154  & 1.34   & 0.19  & 0.55  & $\diagup$ \\ \hline 
  PAD  & 121 & 150 & 0.45 & 212\,480    & 21.88     & 3\,946  & 4.64   & 0.6   & 1.74  & $\diagup$ \\ \hline 
  PAD  & 179 & 221 & 0.97 & 462\,480    & 14.18     & 5\,048  & 12.91  & 1.54  & 4.49  & $\diagup$ \\ \hline 
  PAD  & 237 & 292 & 1.71 & 808\,480    & 15.67     & 4\,374  & 19.56  & 2.38  & 6.92  & $\diagup$ \\ \hline 
  PAD  & 353 & 434 & 3.78 & 1\,788\,480 & 5.65      & 7\,070  & 69.95  & 7.82  & 22.74 & $\diagup$ \\ \hline 
  PAD  & 469 & 577 & 6.68 & 3\,152\,480 & 0.73      & 6\,081  & 106.05 & 12.11 & 35.23 & $\diagup$ \\ \hline 
 \end{tabular}
 \caption{Numerical results for $l_3$\,=\,200\,\AA}
 \label{tab:efficiency_l3=200}
\end{table}

\subsection{Example 2: Precipitate strengthening}
\label{sec:results.precipitate_strenghtening}

\subsubsection{Problem description}

\begin{figure}[hbt]
 \centering
 \includegraphics[width=0.9\textwidth]{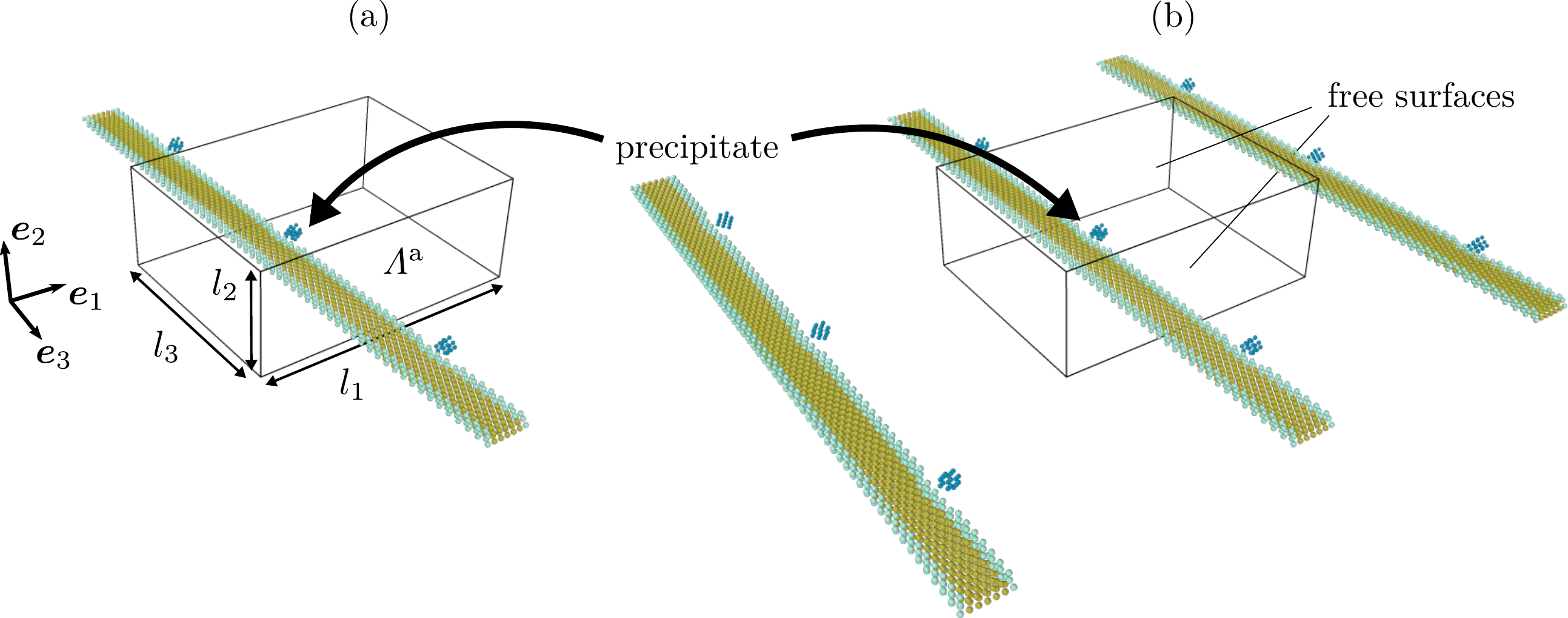}
 \caption{%
 Schematic illustration of the precipitate strengthening problem using (a) the FBC method, and (b) the PAD method}
 \label{fig:precipitate_strenghtening_problem}
\end{figure}

The second example we consider is a precipitate strengthening mechanism, more precisely, we are interested in the process of a gliding dislocation overcoming a precipitate that lies ahead of the dislocation as shown in Figure \ref{fig:precipitate_strenghtening_problem} (a).
Such kind of mechanisms are relevant for studying aging/hardening processes (see, e.g., \citep{hull_introduction_2011}).
The precipitate is mimicked here by fixing a spherical cluster of atoms ahead of the dislocation---in practice it could, however, also correspond to compounds of elements other than the main element.

\begin{figure}[hbt]
 \centering
 \includegraphics[width=0.8\textwidth]{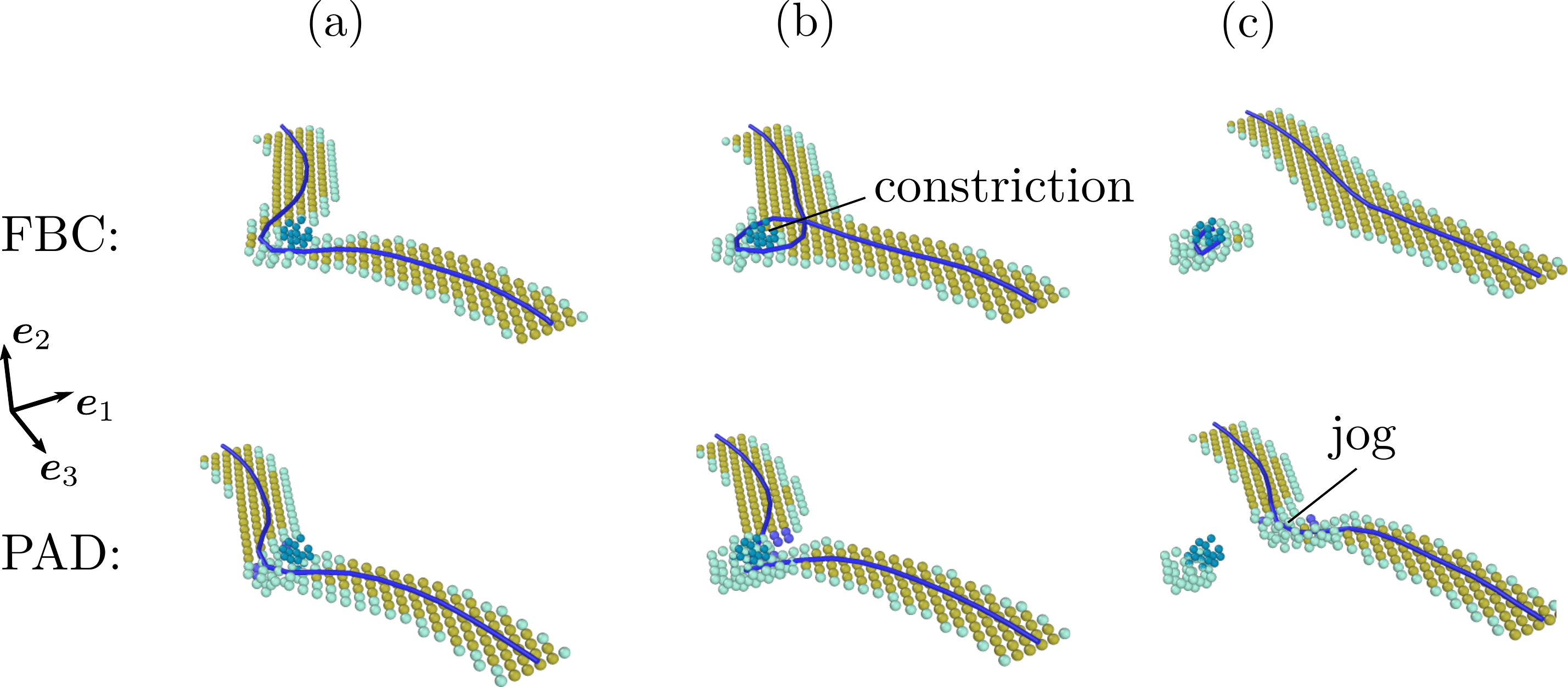}
 \caption{%
 Snapshots from the simulation of the precipitate strengthening problem using FBCs and PAD boundary conditions for the smallest domain reported in Table \ref{tab:results_precipitate_strengthening}.
 (a) Equilibrium dislocation position for some applied stress $<$\,$\tau_\mrm{ps}$.
 (b) The dislocation is on the verge of crossing the precipitate.
 The FBC method correctly predicts the Orowan looping mechanism, while the PAD method erroneously predicts a climb mechanism leading to a jogged dislocation.
 (c) The FBC dislocation is nearly straight again, having left a small loop around the precipitate, while the PAD dislocation remains jogged}
 \label{fig:precipitate_strenghtening_problem_fbc_vs_pad}
\end{figure}

Our simulation setup is as follows.
After relaxing the dislocation core structure, we fix a spherical cluster of atoms with radius $a_0$, representing the precipitate, $\approx$\,20\,{\AA} ahead of the dislocation.
We then apply a shear stress analogously to the previous section to move the dislocation.
In order to predict the precipitate strength $\tau_\mrm{ps}$, i.e., the applied stress necessary for the dislocation to overcome the precipitate, we proceed as follows.
We first apply a shear stress of 140\,MPa.
Upon convergence, we increase it by 10\,MPa and repeat this procedure until the dislocation has overcome the precipitate (cf. Figure \ref{fig:precipitate_strenghtening_problem_fbc_vs_pad} (c)).
The size of the atomistic domain and the position of the dislocation are chosen the same as for the bow-out problem with $l_3$\,=\,100\,{\AA}, with the exception that $l_2$ is increased from 70\,{\AA} to 110\,{\AA} to ensure that the dislocation still remains $\approx$\,20\,{\AA} from the interface after breaking through the precipitate (since the dislocation now moves over a longer distance as opposed to the bow-out problem).

We again compare the performance of the FBC method with the PAD method (Figure \ref{fig:precipitate_strenghtening_problem} (b)) whose setup is described in Appendix \ref{sec:appdx.PAD}.

\subsubsection{Accuracy}

For this example, we are primarily interested in the mechanism for the dislocation overcoming the precipitate.
As shown in Figure \ref{fig:precipitate_strenghtening_problem_fbc_vs_pad} (a)--(c), using the FBC method, the dislocation breaks through the precipitate as both bowing parts constrict ahead of the precipitate and leave a dislocation loop behind.
This mechanism is usually referred to as Orowan looping and is the expected mechanism for the considered problem \citep{hull_introduction_2011}.
However, using a PAD simulation with 29\,240 atoms and box lengths $l_1$\,=\,62\,{\AA} and $l_2$\,=\,78\,{\AA}, the mechanism is fundamentally different: since the bow-out is much less pronounced (see Figure \ref{fig:precipitate_strenghtening_problem_fbc_vs_pad} (a)) due to the artificial boundary effects, demonstrated in the Section \ref{sec:results.bowout}, the dislocation does not constrict, but instead overcomes the obstacle by creating a jog, i.e., the dislocation climbs \emph{under} the precipitate.
This spurious mechanism likely occurs due to the free surfaces generating stresses on the dislocation in the $\rmx_2$-direction and thus forcing the dislocation to climb.
This appears to be counterintuitive since the dislocation is 34\,{\AA} away from both free surfaces.
The FBC method, on the other hand, predicts the right mechanism using an intuitive choice of the atomistic domain by simply ensuring that the dislocation remains 15--20\,{\AA} from the boundaries.
Increasing the domain size yet reduces the influence of the free surfaces and the PAD method then predicts the same mechanism as the FBC method (see Table \ref{tab:results_precipitate_strengthening}).

Another quantity of interest is the precipitate strength $\tau_\mrm{ps}$.
To compute a very accurate value for $\tau_\mrm{ps}$, we first compute $\tau_\mrm{ps}$ as described in the previous section, restart the simulation from an applied stress $\tau_\mrm{ps}-10\,\text{MPa}$, and increment it by only 1\,MPa.
As shown in Table \ref{tab:results_precipitate_strengthening}, the FBC and PAD results are close to each other, except for the smallest PAD domain, where the difference to FBCs is $\sim$\,7\,\%.
This is due to the relatively small precipitate which has primarily been chosen to illustrate the ``mechanism pitfall'' of PAD boundary conditions---for larger precipitates the influence of the bow-out on the precipitate strength increases and the error will presumably be higher.

\subsubsection{Efficiency}

\begin{figure}[hbt]
 \begin{minipage}{0.5\textwidth}
  \centering
  (a)
 \end{minipage}\hfill
 \begin{minipage}{0.5\textwidth}
  \centering
  (b)
 \end{minipage}\\[0.5em]
 \begin{minipage}{0.5\textwidth}
  \centering
  \includegraphics[width=0.9\textwidth]{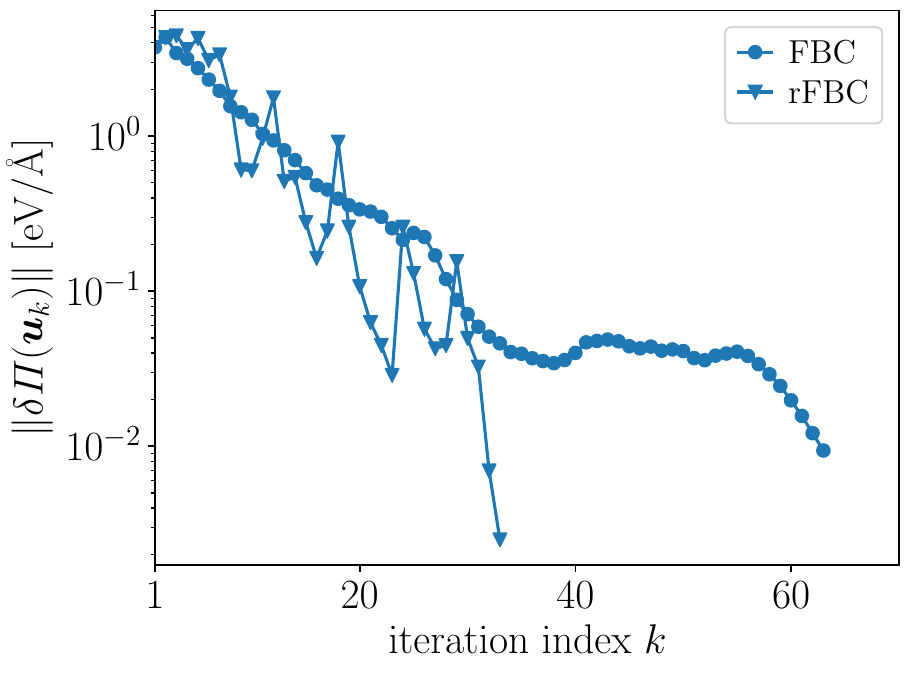}
 \end{minipage}\hfill
 \begin{minipage}{0.5\textwidth}
  \centering
  \includegraphics[width=0.83\textwidth]{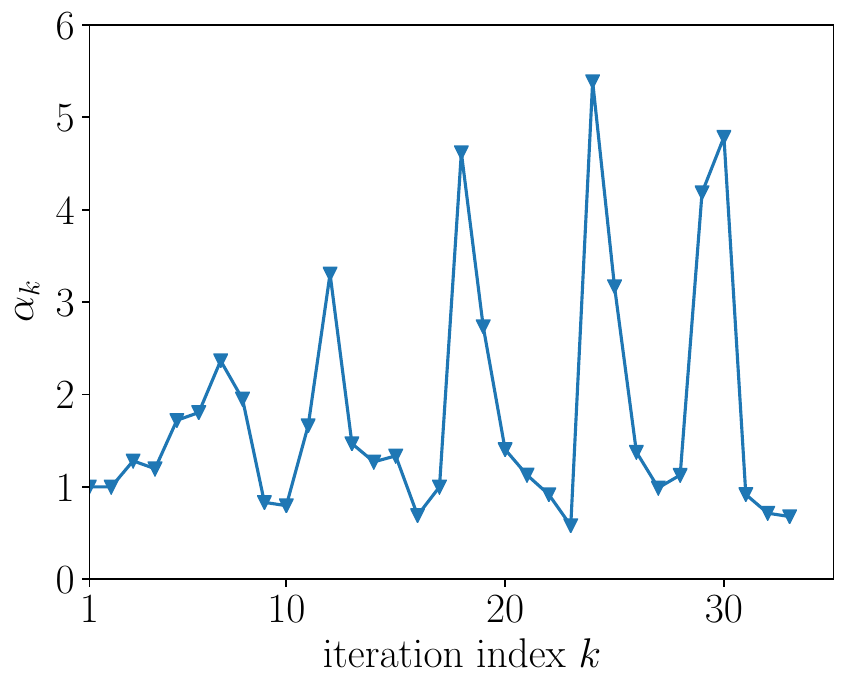}
 \end{minipage}
 \caption{%
 (a) Force norm vs. iteration index for the precipitate strengthening problem using FBCs and rFBCs.
 (b) Relaxation parameter $\alpha$ vs. iteration index for the simulation from (a) using rFBCs}
 \label{fig:precipitate_strengthening_problem_fnorm2&alpha_vs_it}
\end{figure}

For comparing the efficiency of FBCs, rFBCs, and PAD boundary conditions, we consider the problem up to an applied stress of 170\,MPa, which is slightly below $\tau_\mrm{ps}$ to ensure a well-defined quasi-static problem.
Comparing FBCs and rFBCs, the improvement is similar to the bow-out problem:
the number of global iterations is reduced by a factor of $\approx$\,2 (33 rFBC vs. 63 FBC iterations, see Figure \ref{fig:precipitate_strengthening_problem_fnorm2&alpha_vs_it}), the reduction in the required number of per-atom force evaluations is $\chi$\,$\approx$\,1.8, and the effective speedup of rFBCs over FBCs is $\approx$\,2.4 (see Table \ref{tab:results_precipitate_strengthening}).
From Table \ref{tab:results_precipitate_strengthening} it can be concluded that the rFBC method is much more efficient than the PAD method.
In comparison with the largest PAD domain, the rFBC method reduces the number of force evaluations by factor of $\sim$\,33 and is effectively $\sim$\,12 times faster than the PAD method.
The largest PAD domain is also the one that would likely be selected in practice for further studies with a similar problem setup (e.g., using other precipitate geometries) to rule out any spurious effects.
Thus, the practical speedup for a problem with similar dimensions to the ones considered here is therefore $\sim$\,12---and likely increases to $>$\,30 when using more expensive interatomic potentials and/or removing redundant setup times in the current implementation, as outlined in Section \ref{sec:results.bowout.efficiency}, since the time spent for solving the harmonic problem is again less than 1\,\% of the entire simulation time.
For problems with larger periodic lengths we hence anticipate that the speedup increases to several orders of magnitude due to the unfavorable cubic scaling of PAD boundary conditions (cf. Section \ref{sec:results.bowout.accuracy}).

\begin{table}[htb]
 \centering
 \begin{tabular}{|c|c|c|c|c|c|c|c|c|c|c|c|}
  \hline
  Method & $l_1$\,[\AA] & $l_2$\,[\AA] & $\bar{V}$\,[\AA] & $N_\rma$ & Mechanism & $\tau_\mrm{ps}$\,[MPa] & $N_{\var{}{}{\Etot}}$ & $\chi$ & $t$\,[h] & $t/t_\mrm{rFBC}$ & $t_\rmh/t\,[\%]$ \\ \hline\hline
  FBC  & 110 & 30  & 0.33 & 18\,240  & constriction  & 175 & 13\,098 & 1.83  & 0.57 & 2.43  & 0.82      \\ \hline 
  rFBC & 110 & 30  & 0.33 & 18\,240  & constriction  & 176 & 7\,155  & 1     & 0.23 & 1     & 0.7       \\ \hline 
  PAD  & 62  & 78  & 0.48 & 29\,240  & jog formation & 187 & 3\,703  & 0.83  & 0.07 & 0.3   & $\diagup$ \\ \hline 
  PAD  & 121 & 150 & 1.78 & 106\,240 & constriction  & 179 & 5\,722  & 4.66  & 0.4  & 1.71  & $\diagup$ \\ \hline 
  PAD  & 179 & 221 & 3.89 & 231\,240 & constriction  & 177 & 6\,923  & 12.27 & 1.23 & 5.28  & $\diagup$ \\ \hline 
  PAD  & 237 & 292 & 6.82 & 404\,240 & constriction  & 177 & 10\,729 & 33.23 & 2.86 & 12.23 & $\diagup$ \\ \hline 
 \end{tabular}
 \caption{%
 Numerical results for the precipitate strengthening problem}
 \label{tab:results_precipitate_strengthening}
\end{table}

\subsection{Example 3: Pile-up induced cross-slip}
\label{sec:results.cross_slip}

\subsubsection{Problem description}

\begin{figure}[t]
 \centering
 \includegraphics[width=0.9\textwidth]{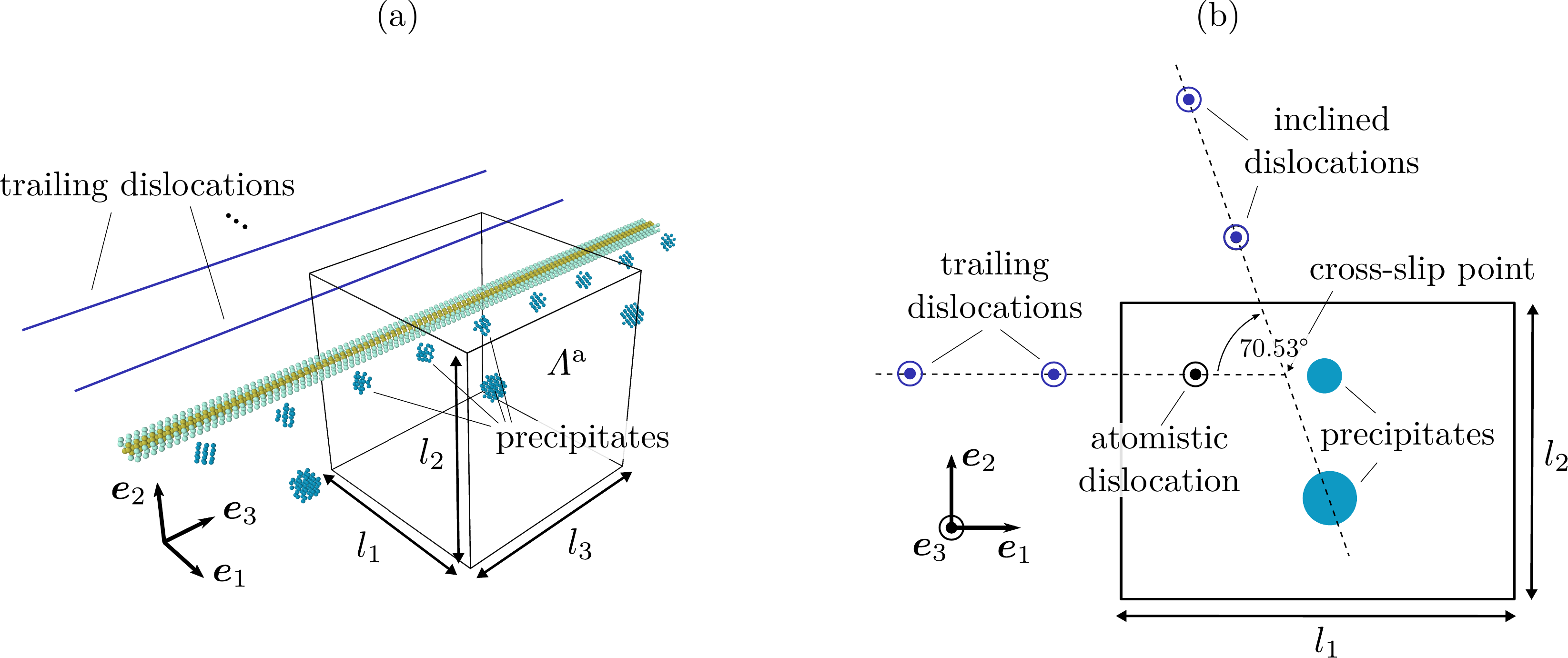}
 \caption{%
 Schematic illustration of the cross-slip problem}
 \label{fig:cross_slip_problem}
\end{figure}

\begin{figure}[t]
 \centering
 \includegraphics[width=0.8\textwidth]{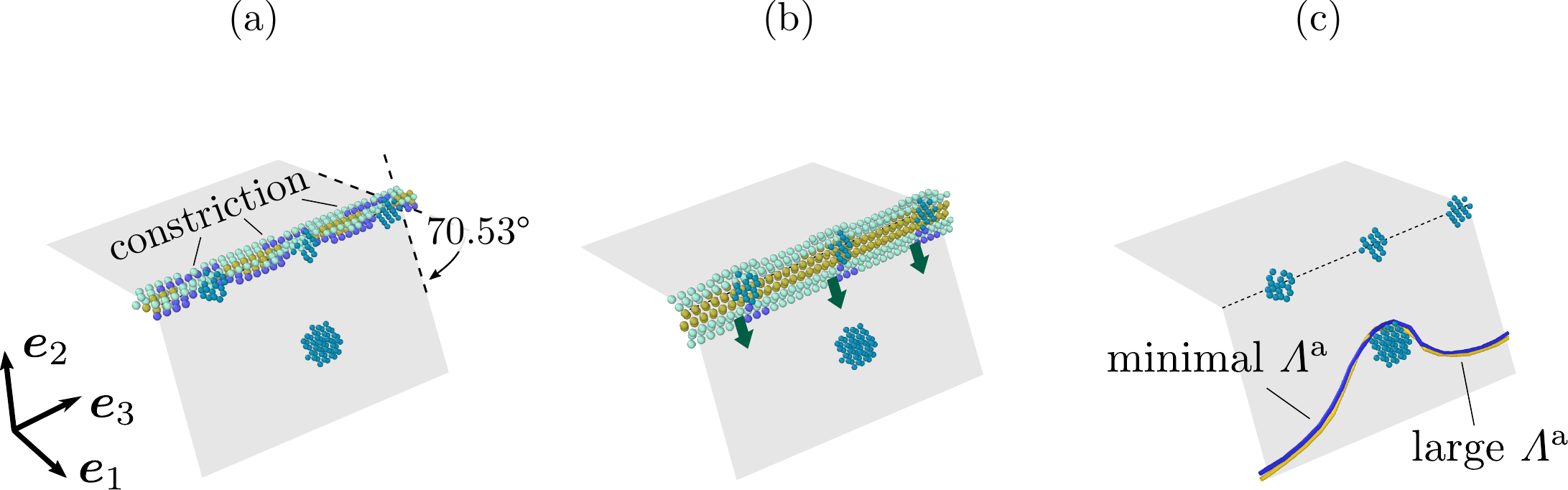}
 \caption{%
 Snapshots from the simulation of the cross-slip problem.
 (a) The dislocation constricts on the primary slip plane to move to the cross-slip plane.
 (b) The dislocation has cross-slipped and moves towards the precipitate on the inclined plane.
 (c) Final equilibrium position of the dislocation using a minimal and a large $\latA$ (cf. Section \ref{sec:results.cross_slip.accuracy}); the dislocations are visualized using the dislocation detection algorithm implemented in OVITO \citep{stukowski_visualization_2010,stukowski_automated_2012}}
 \label{fig:cross_slip_problem_evolution}
\end{figure}

The aim of the third example is to demonstrate further capabilities of the FBC method by simulating a problem that \emph{can not} (or at least by far not as easily) be studied using the PAD method.
To that end, we consider a [10$\bar{1}$] screw dislocation cross-slipping at an array of precipitates (see Figure \ref{fig:cross_slip_problem} (a))---but instead of a constant applied stress being the driving force for this mechanism, a pile-up of trailing dislocations is added in the continuum domain, influencing the motion of the atomistic dislocation in $\lat^\a$ through their long-range elastic interactions.
This is typically a more realistic scenario for processes taking place in real materials than merely considering an externally applied stress (cf. \citep{seeger_mechanism_1957}).
To make the problem even more challenging, we trigger another mechanism after the dislocation has cross-slipped by adding an additional array of dislocations on the inclined plane as shown in Figure \ref{fig:cross_slip_problem} (b), pushing the atomistic dislocation towards another precipitate lying ahead of the cross-slipped dislocation on the inclined slip plane.
The number of inclined dislocations is chosen so that the atomistic dislocation just does not break through the latter precipitate.
Snapshots of the simulation are shown in Figure \ref{fig:cross_slip_problem_evolution} (a)--(c) to illustrate this whole process.
The precipitates are again mimicked here as fixed spherical clusters of atoms, but could, in practical applications, likewise be replaced with more realistic geometries, e.g., Guinier-Preston zones (cf., e.g., \citep{singh_atomistic_2011}).

As in the previous example, we choose the periodic length of the atomistic domain to be $l_3$\,=\,100\,{\AA}.
Further, we choose $l_1$\,=\,75\,{\AA}, $l_2$\,=\,80\,{\AA}, ensuring that the dislocation always stays $\approx$\,20\,{\AA} from the interface throughout the simulation.
Upon relaxation of the dislocation core, we fix three equally-spaced spherical precipitates on the primary slip plane $\approx$\,20\,{\AA} ahead of the dislocation, and one on the inclined slip plane at a distance of $\approx$\,30\,{\AA} from the cross-slip point.
The precipitates on the primary plane have a radius of $a_0$, as in the previous example, and the precipitate on the inclined plane has a radius of $1.5\cdot a_0$.
Next, we take the relaxed solution, denoted by $\bdispl_\mrm{screw}$, and superimpose it with the elastic solution $\bdispl_\mrm{screw}^\mrm{el}$ \citep{hirth_theory_1982} of the trailing and inclined dislocations such that the initial guess is given by
$\bdispl_0 = \bdispl_\mrm{screw} + \sum_{i=1}^{n_\mrm{trail}} \bdispl_{\mrm{screw},i}^\mrm{el} + \sum_{j=1}^{n_\mrm{incl}} \bdispl_{\mrm{screw},j}^\mrm{el}$.
Thereby, the distance between the atomistic and trailing dislocations, as well as the distance between the cross-slip point and inclined dislocations, is chosen to be 50\,{\AA}.
The number of trailing dislocations is chosen to be $n_\mrm{trail}$\,=\,9 since this was found to be the minimum number for the atomistic dislocation to cross-slip.
The number of inclined dislocations is chosen to be $n_\mrm{incl}$\,=\,2 since this was found to be the maximum number for the atomistic dislocation to not break through the precipitate on the inclined plane.
The ``elastic'' dislocations are considered as fixed in the following, yet we remark that it is also possible to equilibrate their positions using discrete dislocation dynamics (cf. \citep{anciaux_coupled_2018}).

\subsubsection{Accuracy}
\label{sec:results.cross_slip.accuracy}

To judge the accuracy of the FBC method, we compare, as done in Section \ref{sec:results.bowout.accuracy} for the bow-out problem, the results obtained using the atomistic domain as described in the previous section, denoted in the remainder of this section by ``minimal $\latA$'', with the result obtained using a ``large $\latA$''.
The side lengths of the large $\latA$ are increased in the positive $\rmx_1$- and the negative $\rmx_2$-direction so that the dislocation line at equilibrium remains $\approx$\,55--65\,{\AA} from the interface.
Since spurious forces on dislocations decay quadratically with the distance from the interface (cf. \citep{hodapp_coupled_2018,hodapp_lattice_2019}), the large-$\latA$-results should be free of any artificial boundary effects.
From Figure \ref{fig:cross_slip_problem_evolution} (c), it can be seen that the final positions of the dislocations for both domains are almost indistinguishable.
The small visible difference is less than half a Burgers vector, so the dislocations are essentially in the same position.
In conclusion, this example demonstrates---once again---the excellent accuracy of the FBC method for an \emph{intuitive choice} of the atomistic domain.

\subsubsection{Efficiency}

\begin{figure}[hbt]
 \begin{minipage}{0.5\textwidth}
  \centering
  (a)
 \end{minipage}\hfill
 \begin{minipage}{0.5\textwidth}
  \centering
  (b)
 \end{minipage}\\[0.5em]
 \begin{minipage}{0.5\textwidth}
  \centering
  \includegraphics[width=0.9\textwidth]{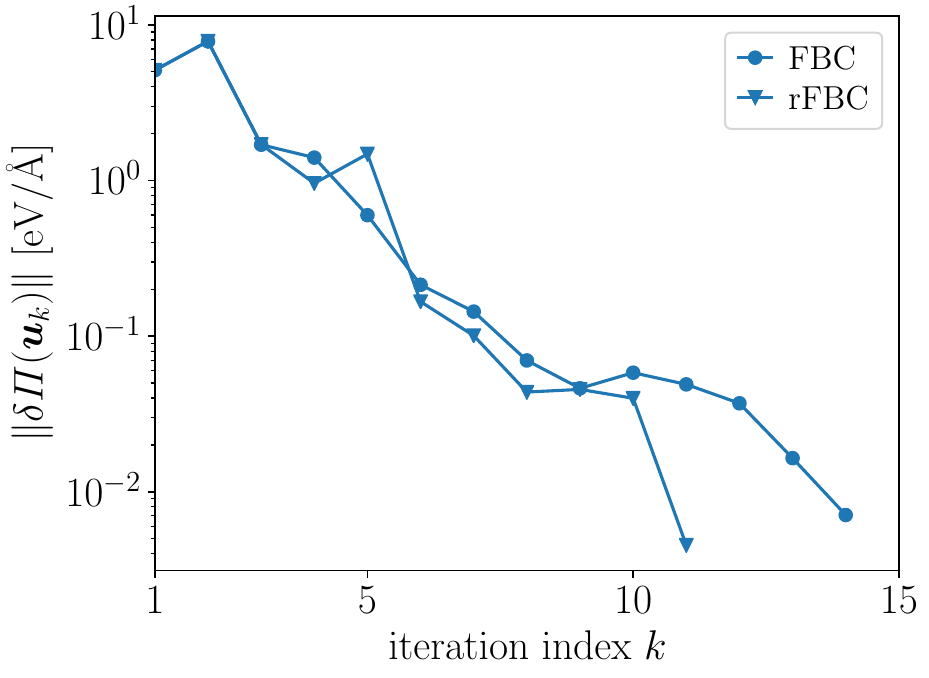}
 \end{minipage}\hfill
 \begin{minipage}{0.5\textwidth}
  \centering
  \includegraphics[width=0.83\textwidth]{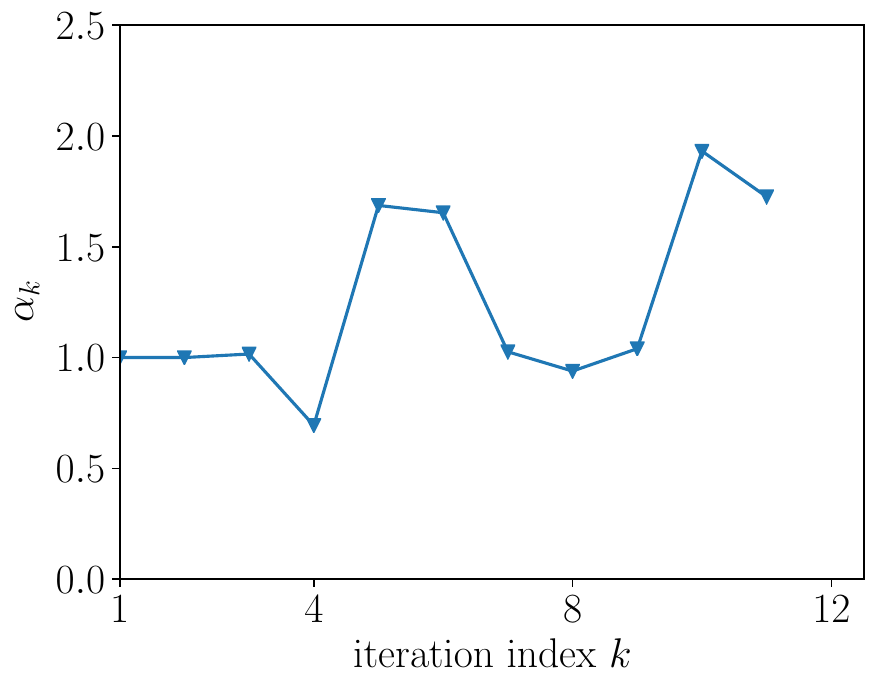}
 \end{minipage}
 \caption{%
 (a) Force norm vs. iteration index for the cross-slip problem using FBCs and rFBCs.
 (b) Relaxation parameter $\alpha$ vs. iteration index for the simulation from (a) using rFBCs}
 \label{fig:cross_slip_problem_fnorm2&alpha_vs_it}
\end{figure}

We then performed the simulation using rFBCs.
Both, FBCs and rFBCs, converge to the same result and the efficiency of rFBCs is again better than for FBCs:
rFBCs require fewer global iterations (11 vs. 14, see Figure \ref{fig:cross_slip_problem_fnorm2&alpha_vs_it}), fewer per-atom force evaluations ($\chi$\,$\approx$\,1.1), and achieve an effective speedup of $\approx$\,1.24 over FBCs (see Table \ref{tab:results_cross-slip}).
Hence, the rFBC method can also be considered as a robust and efficient boundary condition for more involved, multi-mechanism problems.
The reduced gain in efficiency, compared with the previous examples, is due to the fact that solving the problem requires less iterations and most time is spent during the 1--2 iterations where the cross-slip mechanism takes place.
This is likely because the artificial boundary effects are reduced for screw dislocations since the Burgers vector points in the direction of periodicity (the same observations has been made in the works of \citet{dewald_analysis_2006} and \citet{pavia_parallel_2015} who analyzed spurious forces on straight dislocations in the vicinity of A/C interfaces).

\begin{table}[htb]
 \centering
 \begin{tabular}{|c|c|c|c|c|c|c|c|c|c|c|c|}
  \hline
  Method & $l_1$\,[\AA] & $l_2$\,[\AA] & $\bar{V}$\,[\AA] & $N_\rma$ & $N_{\var{}{}{\Etot}}$ & $\chi$ & $t$\,[h] & $t/t_\mrm{rFBC}$ & $t_\rmh/t\,[\%]$ \\ \hline\hline
  FBC  & 75 & 80 & 0.33 & 33\,286 & 8\,639 & 1.1 & 0.45 & 1.24 & 0.16 \\ \hline 
  rFBC & 75 & 80 & 0.33 & 33\,286 & 7\,852 & 1   & 0.36 & 1    & 0.14 \\ \hline 
 \end{tabular}
 \caption{%
 Numerical results for the cross-slip problem}
 \label{tab:results_cross-slip}
\end{table}


\section{Discussion}

We have presented an efficient implementation of the flexible boundary condition (FBC) method for single-periodic problems.
Our implementation is supported by various tools from numerical analysis, in particular, a hierarchical matrix algebra rendering the computational cost for updating the ``flexible boundary'' negligible small compared to an atomistic energy minimization---an issue that previously hindered applying the method to large-scale problems.
As a result, the present work demonstrates that FBCs are now a viable and much more efficient alternative to the, according to the author's best knowledge, currently most commonly used method for computing such problems, the periodic array of dislocations (PAD) method.

To assess the performance of the FBC method we first considered a dislocation bow-out problem in an fcc lattice.
For this class of problems, using an atomistic cell of periodic length $l_3$ and lateral lengths $l_1$ and $l_2$, the PAD method requires to keep the normalized volume $\bar{V} = l_1l_2/l_3^2$ constant to ensure the same level of accuracy for a constant bowing distance.
In contrast, we have shown that, with FBCs, it suffices to vary the periodic length $l_3$, and keeping $l_1$ and $l_2$ constant.
Hence, the necessary number of real atoms for the FBC method scales linearly with $l_3$ whereas the necessary number of real atoms for the PAD method scales cubically.
This leads to an orders of magnitude higher efficiency of the FBC method over the PAD method.
The poor efficiency of the PAD method stems from the spurious boundary effects due the free surfaces and image dislocations that not only deteriorate the accuracy of dislocation motion but also of other quantities of interest, such as mechanisms.
This has been shown using the example of precipitate strengthening, where a too small PAD domain was erroneously leading to dislocation climb, rather than correctly Orowan looping, as the driving mechanism for the dislocation to overcome the precipitate.
The FBC method also naturally allows for incorporating more realistic far-field conditions, demonstrated in the third example where we considered a cross-slip mechanism of an atomistic screw dislocation, induced by a pile-up of elastic trailing dislocations.

All three examples demonstrate the excellent accuracy and robustness properties of the FBC method, as well as its superior efficiency and generality in comparison with the PAD method.
Further, our computational experiments imply in a wider sense that studies to be performed by using the FBC method can be done more intuitively than by using the PAD method:
as there are no spurious boundary effects, except for the coupling error, it is sufficient to place the atomistic domain where nonlinear material behavior is expected, that is, in vicinity of the dislocation(s) and possibly other defects (voids, precipitates, etc.).
Additionally, the PAD method requires more preliminary convergence tests before the actual study can be performed, further increasing the computational cost.

Moreover, it is anticipated that the developed FBC method can be applied to problems beyond the examples considered here. A few suggestions are
\begin{itemize}
 \item
 dislocation-solute interactions,
 \item
 nudged elastic band simulations,
 \item
 or forest hardening (cf., e.g., \citep{rao_spontaneous_2013}).
\end{itemize}
For the latter, additional complications may arise due to the forest dislocations piercing the artificial interface. Those could, however, be handled using dislocation core templates \citep{anciaux_coupled_2018,hodapp_coupled_2018}.

For potential future users we point out that the FBC method can be virtually used in a black box fashion as there are, in principle, no crucial parameters other than the standard tolerances which have to be touched.
Using FBCs would then not differ from using any of the conventional boundary conditions (fixed, free, or periodic).
Moreover, the decomposition into an atomistic and a continuum problem which are solved separately allows for an easier integration into existing molecular dynamics codes as with concurrent atomistic/continuum solvers.

\section{Further improvements and extensions}
\label{sec:outlook}

We conclude by mentioning some interesting further improvements and extensions of the present work.

\begin{itemize}
\item 
An obvious approach to further accelerate the FBC method is to replace the domain decomposition algorithm with a monolithic Newton-Krylov solver as proposed in \citep{hodapp_lattice_2019}.
However, a non-symmetric force-based coupling limits the realm of suitable linear solvers to generalized minimal residual methods, although switching to one of the state-of-the-art energy-based schemes (e.g., \citep{fang_blended_2020}) could bypass this requirement.
Moreover, the solver proposed in \citep{hodapp_lattice_2019} requires an efficient representation of the inverse of $\Lh^{\cc}$.
One could relax this requirement by weakening \eqref{eq:flexbc.Lcpl_split} and, in lieu thereof, minimize $\bdispl^\a - \bdispl_\indAHarm^\a - \bdispl_\indHarm^\a$ together with the atomistic energy, leading to the linear system (cf. \citep{hodapp_lattice_2019})
\begin{equation}
 \begin{pmatrix}
  \L^\aa & - \L^\ap\G^\pipl\Lh^\ipli \\
  \Id^\ii & - (\Id^\ii + \G^\iipl\Lh^\ipli)
 \end{pmatrix}
 \begin{bmatrix}
  \bdispl^\a \\ \bdispl_\indAHarm^\i
 \end{bmatrix}
 =
 \begin{pmatrix}
  \bforce_\mrm{ext}^\a \\ \bmZero^\i
 \end{pmatrix}
\end{equation}
which would have to be solved in every Newton iteration with respect to $\bdispl^\a$ and the anharmonic interface displacements $\bdispl_\indAHarm^\i$.
Integrating any of the mentioned techniques into existing molecular dynamics codes is nevertheless challenging and requires further investigation.
\item
A natural extension of the FBC method is to include adaptive refinement techniques to evolve the atomistic domain in case the equilibrium position of a dislocation lies outside the prechosen boundaries \citep{liao_posteriori_2020}.
In this context, an alternative option would be to use the coupled atomistic/discrete dislocations (CADD) method \citep{anciaux_coupled_2018,hodapp_coupled_2018,cho_coupled_2018} in which the atomistic domain is coupled to a discrete dislocation dynamics domain. The CADD method does not require adaptive refinement since atomistic dislocations would transform into discrete dislocations (and vice versa) when approaching the artificial interface.
\item
Another valuable extension would be to incorporate Green functions for defects other than dislocations, or other types of boundary conditions, e.g., free surfaces (see \citep{braun_asymptotic_2021} for some interesting new work in this direction).
Along these lines, a different possibility is to include a finite outer domain boundary \citep{li_atomistic-based_2012,hodapp_analysis_2021} to control the far-field behavior.
One appealing application for this could be long periodic cracks nucleating curved dislocations.
Such a method could then be combined with Sinclair's FBC algorithm for cracks \citep{sinclair_influence_1975,buze_numerical-continuation-enhanced_2021} to study, e.g., shielding of three-dimensional, semi-infinite cracks with unprecedented accuracy using realistic $K$-test geometries (cf. \citep{andric_atomistic_2019}).
\end{itemize}


\section*{Acknowledgements}

Financial support from the Fonds National Suisse (FNS), Switzerland, (project 191680) is highly acknowledged.


\section*{Appendix}

\begin{appendices}

\section{Continuum model}

\begin{figure}[hbt]
 \centering
 \includegraphics[width=0.2\textwidth]{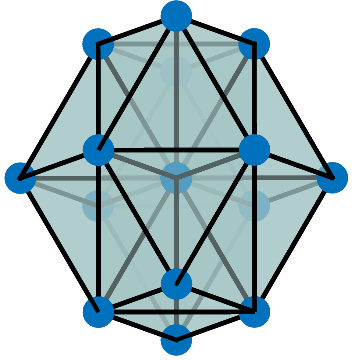}
 \caption{Partitioning of the fcc lattice into simplices}
 \label{fig:partition_fcc}
\end{figure}

To define the displacement field, we construct a periodic partition of the fcc lattice into simplices (tetrahedrons) as shown in Figure \ref{fig:partition_fcc}.
The partition of atom $\bato$ is then the set of simplices adjacent to $\bato$ and the set of nodes of those simplices is the local interaction range $\intRgAto^\mrm{h}$.
Let then $\phi_{\ato}$ be the standard $\bbP$1 nodal interpolant with compact support on the partition of $\bato$.
The displacement $\bdispl$ and its gradient are then defined $\forall\,\bmx\in\real^3$ as
\begin{align}\label{eq:appdx.def_u+ugrad}
 \bdispl(\bmx) = \sum_{\bato \in \intRgAto^\mrm{h}} \phi_{\ato}(\bmx)\bdispl(\bato),
 &&
 \grad{}{}{\bdispl}(\bmx) = \sum_{\bato \in \intRgAto^\mrm{h}} \grad{}{}{\phi_{\ato}}(\bmx) \otimes \bdispl(\bato).
\end{align}
Equation \eqref{eq:appdx.def_u+ugrad} is then used to construct the local force constant tensor $\bmK_\mrm{h}$ in \eqref{eq:flexbc.Eato_h}.

We remark that such a partitioning is not unique, but results (solution, convergence rates, etc.)  do not alter qualitatively.

\section{Computation of the lattice Green function \texorpdfstring{$\bmG^\infty(\bmr)$}{}}
\label{sec:appdx.Glgf}

The fundamental lattice Green function is the set of solutions $\bdispl^k : \latInf \rightarrow \euclD$, $k=1, ..., 3$, to
\begin{equation}\label{eq:appdx.fund_problem}
 \Lh[\bdispl^k](\bato) = \sum_{\batoB \in \intRgAto^\rmh} \bK_\rmh(\bato - \batoB) \bdispl^k(\batoB) = \bforce^k(\bato)
 \qquad \text{in} \; \latInf, \qquad \text{with} \quad
 \force^k_i(\bato) = \delta_{ik}\delta(\bato) = \left\{\,
 \begin{aligned}
  1 & \quad \text{if} \; (i=k) \wedge (\bato = \bmZero), \\
  0 & \quad \text{else},
 \end{aligned}
 \right.
\end{equation}
or component-wise
\begin{equation}\label{eq:appdx.fund_problem2}
 \forall\, k = 1, ..., 3 \qquad \delta_{ik}\delta(\bato) = \sum_{\bato \in \intRgAto^\rmh} K_{\rmh,ij}(\bato - \batoB) \displ_j^k(\batoB) = K_{\rmh,ij} \ast \displ_j^k,
\end{equation}
where $\ast$ is the convolution operator.

To solve this problem, we make use of the semi-discrete Fourier transform $\clF$ and its inverse $\inv{\clF}$ defined for lattice functions $f(\bmr)$ such that
\begin{alignat}{4}
 &\forall\,\bmk \in \clB \qquad &
 \FT{f}(\bmk) &= \sum_{\bmr \in \latInf} f(\bmr) e^{-\rmi(\bmk^\sT \cdot \bmr)}, \\
 &\forall\,\bmr \in \latInf \qquad &
 \FTinv{\FT{f}}(\bmr) &= \frac{1}{\vert \clB \vert}\int \FT{f}(\bmk) e^{\rmi(\bmk^\sT \cdot \bmr)} \,\rmd \clB = f(\bmr),
\end{alignat}
where $\clB$ is the Brillouin zone of $\latInf$ and $\vert \clB \vert$ its volume.

We now apply $\clF$ on both sides of \eqref{eq:appdx.fund_problem2} such that
\begin{equation}
 \FT{\delta_{ik}\delta(\bmr)} = \delta_{ik} = \FT{K_{\rmh,ij} \ast \displ_j^k} = \FT{K_{\rmh,ij}} \cdot \FT{\displ_j^k},
\end{equation}
where we have used the fact that the convolution operator is a multiplication in Fourier space.
Inverting $\FT{K_{\rmh,ij}}$ and applying $\inv{\clF}$ then yields the components of the lattice Green (tensor) function
\begin{equation}
 G^\infty_{ik}(\bmr) = \FTinv{\inv{\FT{K_{\rmh,ij}}}}.
\end{equation}

For the numerical examples presented here we compute $\bmG^\infty(\bmr)$ for all $\bmr$ up to a cut-off radius $\cutOff$ of five lattice constants using numerical integration over the Brillouin zone as described in \citep{hodapp_lattice_2019}.
Outside of $\cutOff$ we replace $\bmG^\infty(\bmr)$ with the continuum Green function $\bmG^\cgf(\bmr)$. Assuming that lattice and continuum Green functions are asymptotically equivalent, we ensure a sufficiently smooth transition from $\bmG^\infty(\bmr)$ to $\bmG^\cgf(\bmr)$ by computing their difference
\begin{equation}
 \bmC(\bmr') \approx \bmC = \bmG^\infty(\bmr') - \bmG^\lgf(\bmr')
\end{equation}
for some large enough $\nrm{\bmr'} > \cutOff$ such that $\bmC$ is (approximately) independent of $\bmr'$, and subtracting $\bmC$ from all the computed $\bmG^\infty$'s.

We further note that $\bmG^\infty$ is typically not rotational invariant due to anisotropy or a nonsymmetric partitioning (cf. Figure \ref{fig:partition_fcc}).
Therefore, we need to rotate $\bmG^\infty$ to the frame of the computational domain.
Let $\bmA$ and $\bmA'$ be the normalized basis tensors of $\latInf$ and the rotated lattice $\latInf'$, respectively.
The rotation tensor $\bmQ$ is then defined through the requirement $\bmQ\bmA\integ = \bmA'\integ$. The lattice Green function (and likewise the continuum Green function) with respect to the rotated lattice thus reads ${\bmG^{\infty}}' = \bmQ^\sT \bmG^\infty \bmQ$.

\section{Construction of \texorpdfstring{$\widetilde{\uuG}{}_\scH^\pipl$}{}}
\label{sec:appdx.GpiplH}

$\scH$-matrices approximate off-diagonal matrix blocks via low-rank representations using sums of outer (vector) products.
That is, for any off-diagonal block $t \times s$ of size $N_t \times N_s$ with numerical rank $k$ we write
\begin{equation}\label{eq:impl.low-rank_format}
 (\widetilde{\uuG}{}^\pipl)^{t \times s} \approx (\widetilde{\uuG}{}_\scH^\pipl)^{t \times s} = \sum_{i=1}^k \uv_i \otimes \uw_i, \qquad \text{where} \quad
 \uv_i \in \real^{N_t}, \; \uw_i \in \real^{N_s},
\end{equation}
the best rank-$k$ approximation being the well-known singular value decomposition (SVD).
Then, if $k$ can be made much smaller compared to the dimensions of $(\widetilde{\uuG}{}^\pipl)^{t \times s}$ without sacrificing accuracy, the complexity for storing this block shrinks from $\clO(N_t N_s)$ to $\clO(k(N_t + N_s))$, ultimately leading to linear scaling algorithms for approximating \eqref{eq:impl.up=Gpipl*fipl_mvm}.

\begin{figure}[t]
 \centering
 \includegraphics[width=0.9\textwidth]{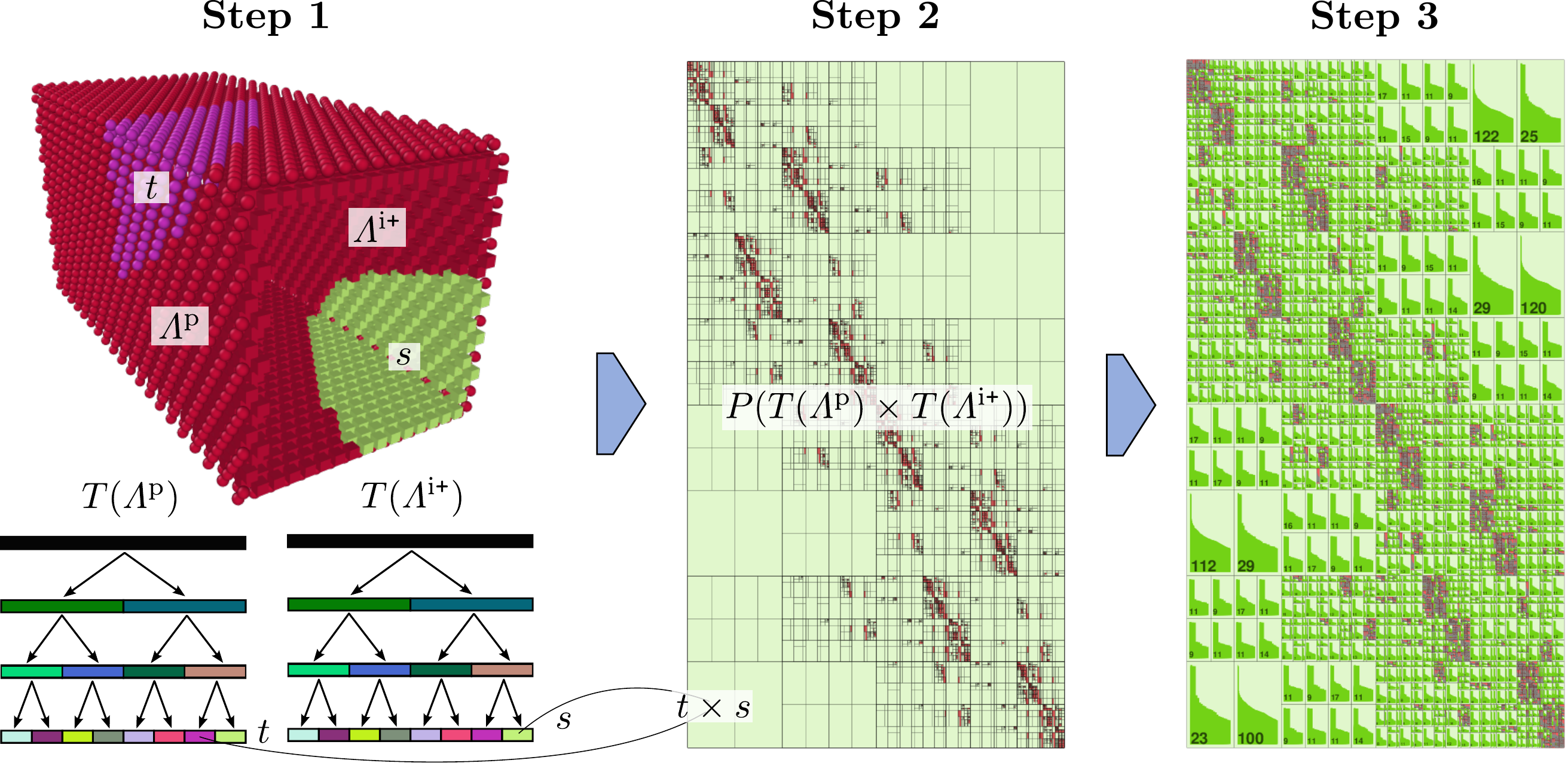}
 \caption{Required steps for constructing an $\scH$-matrix representation $\widetilde{\uuG}{}_\scH^\pipl$ of $\widetilde{\uuG}{}^\pipl$. \textbf{Step\;1.} Construction of the cluster trees for the pad and $\ipl$-atoms. \textbf{Step\;2.} Partitioning of $\widetilde{\uuG}{}_\scH^\pipl$. \textbf{Step\;3.} Computation of the low-rank representations for all matrix blocks in the partition; the numbers are the ranks of the matrix blocks and the dark blocks represent dense blocks}
 \label{fig:hmat}
\end{figure}

\begin{figure}[t!]
 \centering
 \includegraphics[width=0.95\textwidth]{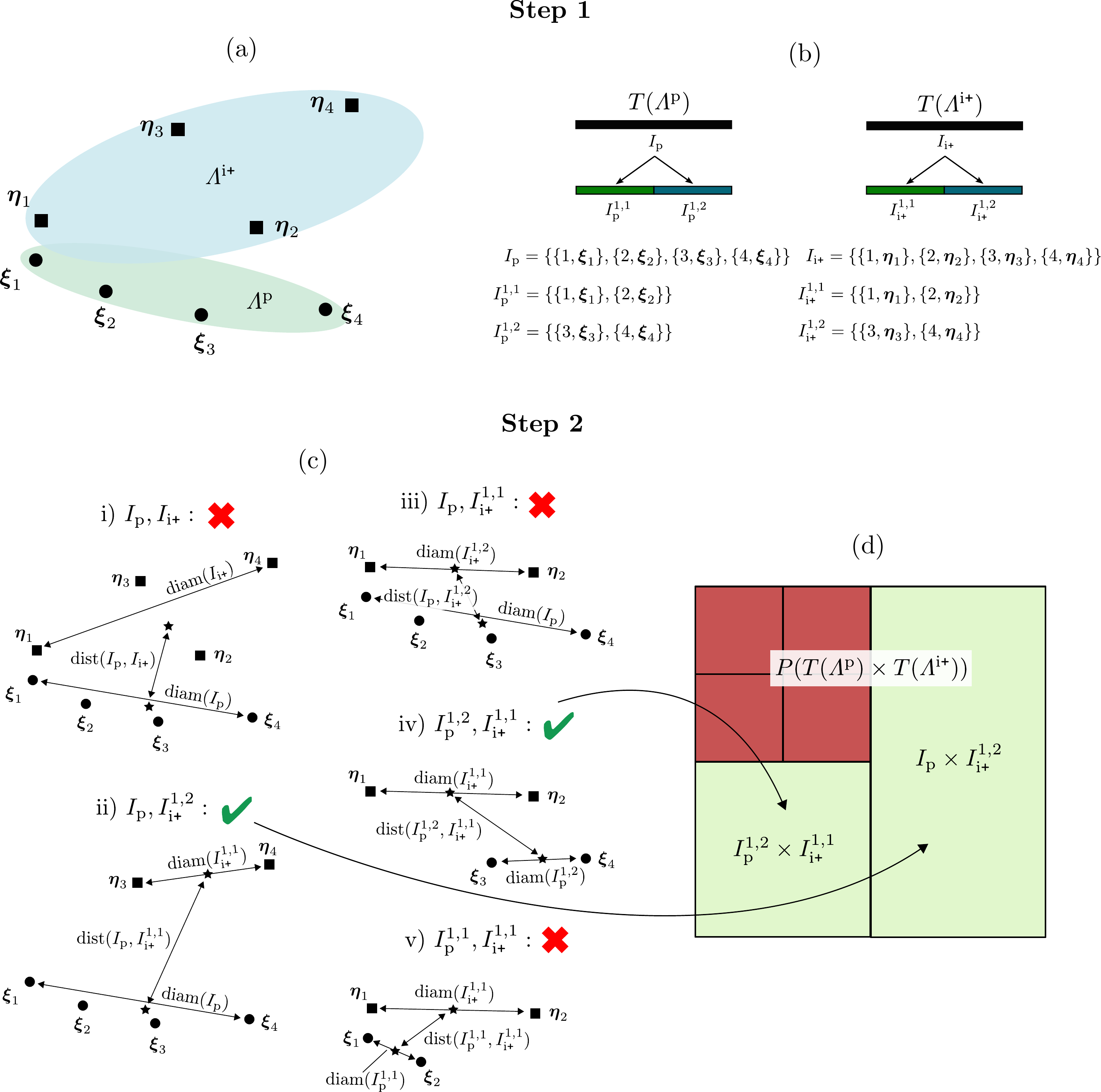}
 \caption{%
 Required steps for constructing an $\scH$-matrix block structure for the $4\times 4$ matrix \eqref{eq:impl.4x4_matrix}.
 \textbf{Step\;1.}\;Construction of the cluster trees $T(\lat^\rmp)$ and $T(\lat^\ipl)$, shown in (b), for the domains from (a), using geometric clustering.
 \textbf{Step\;2.}\;Evaluation of the admissibility criterion \eqref{eq:impl.geom_adm} for all possible combinations of elements of $T(\lat^\rmp),T(\lat^\ipl)$.
 An algorithm for such an evaluation is outlined in (c) for a $\gamma$\,$\approx$\,1.
 The final partition obtained from incrementally performing the steps i)--v) from (c) is shown in (d). Therein, the dark blocks were considered as inadmissible and therefore represent dense blocks}
 \label{fig:hmat_example}
\end{figure}

There are three steps necessary to construct an $\scH$-matrix.
In the first two steps, \textbf{Step\;1} and \textbf{Step\;2}, we construct the block structure of $\widetilde{\uuG}{}_\scH^\pipl$, i.e., we identify the blocks $t \times s$ that are admissible for low-rank compression.
In the third step, \textbf{Step\;3}, we then build the corresponding matrix blocks $(\widetilde{\uuG}{}_\scH^\pipl)^{t \times s}$ in the low-rank format \eqref{eq:impl.low-rank_format}.
The three steps are schematically depicted in Figure \ref{fig:hmat} and outlined in the following.
The first two steps will further be exemplified using a simplified problem where the domains $\lat^\p$ and $\lat^\ipl$ each consist of four atoms, $\bato_1$--$\bato_4$ and $\batoB_1$--$\batoB_4$, respectively, as shown in Figure \ref{fig:hmat_example} (a), and the matrix to be approximated contains the first diagonal component of the Green tensor, i.e., we consider the $4\times 4$ matrix
\begin{equation}\label{eq:impl.4x4_matrix}
 \begin{pmatrix}
  \widetilde{G}_{11}(\bmr_{1,1}) & \widetilde{G}_{11}(\bmr_{1,2}) & \widetilde{G}_{11}(\bmr_{1,3}) & \widetilde{G}_{11}(\bmr_{1,4}) \\
  \widetilde{G}_{11}(\bmr_{2,1}) & \widetilde{G}_{11}(\bmr_{2,2}) & \widetilde{G}_{11}(\bmr_{2,3}) & \widetilde{G}_{11}(\bmr_{2,4}) \\
  \widetilde{G}_{11}(\bmr_{3,1}) & \widetilde{G}_{11}(\bmr_{3,2}) & \widetilde{G}_{11}(\bmr_{3,3}) & \widetilde{G}_{11}(\bmr_{3,4}) \\
  \widetilde{G}_{11}(\bmr_{4,1}) & \widetilde{G}_{11}(\bmr_{4,2}) & \widetilde{G}_{11}(\bmr_{4,3}) & \widetilde{G}_{11}(\bmr_{4,4})
 \end{pmatrix},
 \qquad \text{where} \quad \bmr_{i,j} = \batoB_j - \bato_i.
\end{equation}

In \textbf{Step\;1}, we construct a hierarchical system for the row and column index sets, denoted by $I_\rmp$ and $I_\ipl$, the so-called cluster trees $T(\lat^\rmp)$ and $T(\lat^\ipl)$, using \emph{geometric clustering}.
To that end, we associate each index in $I_\p,I_\ipl$ with atoms from $\lat^\p$,$\lat^\ipl$ at which the corresponding Green function is evaluated.
For example, given an entry $\widetilde{G}^\pipl_{ij} = \widetilde{G}_{11}(\batoB_j - \bato_i)$ of $\widetilde{\uuG}{}^\pipl$, the $i$-th index in $I_\p$ will be associated with the atom $\bato_i \in \lat^\p$, and the $j$-th index in $I_\ipl$ will be associated with the atom $\batoB_j \in \lat^\ipl$, etc.
For our $4\times 4$ matrix \eqref{eq:impl.4x4_matrix}, $I_\p$ and $I_\ipl$ are thus given as follows
\begin{align}
 I_\p &= \{ \{ 1,\bato_1 \}, \{ 2,\bato_2 \}, \{ 3,\bato_3 \}, \{ 4,\bato_4 \} \}, \\
 I_\ipl &= \{ \{ 1,\batoB_1 \}, \{ 2,\batoB_2 \}, \{ 3,\batoB_3 \}, \{ 4,\batoB_4 \} \}.
\end{align}
We then subdivide $I_\p$ into smaller sets $I_\p^{1,1},I_\p^{1,2}$, ..., and $I_\ipl$ analogously.
We call the collection $\{ I_\p^{1,1},I_\p^{1,2}, ... \}$ the first level in the hierarchy of $T(\lat^\rmp)$.
For geometric clustering, the subsets are chosen based on how close the indices are spatially using the associate atomic positions (cf. Figure \ref{fig:hmat_example} (a) and (b)).
The sets $I_\p^{1,1}$, $I_\p^{1,2}$, ..., are then further subdivided, leading to a second level, a third level, etc.
For our example matrix, we have chosen the cluster trees shown in Figure \ref{fig:hmat_example} (b), with $I_\p$ being subdivided into $I_\p^{1,1}=\{ \{ 1,\bato_1 \}, \{ 2,\bato_2 \} \}$, $I_\p^{1,2} = \{ \{ 3,\bato_3 \}, \{ 4,\bato_4 \} \}$, and $I_\ipl$ being subdivided into $I_\ipl^{1,1} = \{ \{ 1,\batoB_1 \}, \{ 2,\batoB_2 \} \}$, $I_\ipl^{1,2} = \{ \{ 3,\batoB_3 \}, \{ 4,\batoB_4 \} \}$.
For a general matrix $\widetilde{\uuG}{}^\pipl$, as used in this work, we have used binary space partitioning (cf. \citep{kriemann_parallel_2005}).
All these index sets eventually represent row and column indices of potentially admissible sub-blocks of $\widetilde{\uuG}{}_\scH^\pipl$, identified in the next step.
So, in general, this subdivision is repeated a number of times until the desired minimum matrix block dimension $b_\mrm{min}$ is reached.
For example, for the cluster tree in Figure \ref{fig:hmat}, the lowest depicted level is level 3 and the element $s \in T(\lat^\ipl)$ on this level corresponds to the atoms in the lower right corner in the domain above.

In \textbf{Step\;2}, we construct the partition $P$ of the $\scH$-matrix from the two cluster trees $T(\lat^\p)$ and $T(\lat^\ipl)$.
To that end, we iteratively retrieve elements $t \in T(\lat^\p)$ and $s \in T(\lat^\ipl)$ (i.e., subsets of the index sets) to form a potential matrix block $t \times s$ and assess its admissibility for being low-rank, starting from level 0.
From \citep{borm_introduction_2003}, it follows that a good admissibility criterion that accounts for the smoothness of the lattice Green function \eqref{eq:impl.smoothness} is
\begin{equation}\label{eq:impl.geom_adm}
 \forall\,t\in T(\lat^\p) \wedge \forall\,s\in T(\lat^\ipl),\, \forall\,s'\in\{s,s^+,s^-\}
 \qquad \min(\operatorname{diam}(t), \operatorname{diam}(s')) \le \gamma \operatorname{dist}(t, s'),
\end{equation}
with $\gamma$\,$>$\,$0$, since the singular values of matrix blocks $t\times s$ fulfilling \eqref{eq:impl.geom_adm} decay exponentially with their distance (see, e.g., \citep{bebendorf_hierarchical_2008}).
In \eqref{eq:impl.geom_adm}, $s^+$ and $s^-$ correspond to the same index set as $s$, but associated with atoms from the neighboring periodic images, $\operatorname{diam}(t),\operatorname{diam}(s')$ denote the diameters of the index sets $t$ and $s'$, and $\operatorname{dist}(t, s')$ is the distance between the centroids of $t$ and $s'$.
The criterion \eqref{eq:impl.geom_adm} must in principle be checked for all possible combinations $t\times s$.
Such an algorithm is schematically illustrated in Figure \ref{fig:hmat_example} (c) for constructing an admissible partition of the matrix \eqref{eq:impl.4x4_matrix} (Figure \ref{fig:hmat_example} (d)).
Checking \eqref{eq:impl.geom_adm} in this way has, however, quadratic complexity and thus becomes expensive if the cluster trees are large.
Figure \ref{fig:hmat_example} (c) therefore only serves for illustration purposes---in practice more efficient algorithms exist that, e.g., replace $t$ and $s$ with supersets for which $\operatorname{diam}(t), \operatorname{diam}(s)$ and $\operatorname{dist}(t, s)$ are cheaper to evaluate, such as bounding boxes (see, e.g., example 2.2 in \citep{borm_introduction_2003}).

\begin{rem}[A physical interpretation of \eqref{eq:impl.geom_adm}]
 Criterion \eqref{eq:impl.geom_adm} ensures that i) $t$ and $s$ are sufficiently far from each other and ii) the $i$ row vectors formed by all $\bmr_{i,1} = \batoB_1 - \bato_i$, $\bmr_{i,2} = \batoB_2 - \bato_i$ etc. are sufficiently close.
 Thus, when \eqref{eq:impl.geom_adm} is met, we expect that many of the row vectors of $\widetilde{\uuG}{}^\pipl$ are approximately linear dependent.
 Hence, \eqref{eq:impl.geom_adm} can be considered as a formal statement of St. Venant's principle \citep{saint-venant_memoire_1855}.
\end{rem}

In the final step, \textbf{Step\;3}, we need to build the low-rank representations \eqref{eq:impl.low-rank_format} for all admissible matrix blocks $t\times s$ from the partition $P$.
In principle, this could be achieved by computing the SVD of every block $t \times s$.
However, this requires to diagonalize a matrix and, additionally, all the values of the matrix block must be computed.
A more efficient way to compute low-rank approximations of the block matrices is the CUR (also: CGR) decomposition \citep{goreinov_theory_1997} given by
\begin{equation}
 (\widetilde{\uuG}{}_\scH^{\pipl})^{t \times s} = (\widetilde{\uuG}{}^\pipl)^{t \times s}_{:,J} \inv{(\widetilde{\uuG}{}^\pipl)^{t \times s}_{I,J}} (\widetilde{\uuG}{}^\pipl)^{t \times s}_{I,:},
 \qquad
 (\widetilde{\uuG}{}^\pipl)^{t \times s}_{:,J} \in \real^{N_t\times k}, \,
 (\widetilde{\uuG}{}^\pipl)^{t \times s}_{I,J} \in \real^{k\times k}, \,
 (\widetilde{\uuG}{}^\pipl)^{t \times s}_{I,:} \in \real^{k\times N_s},
\end{equation}
where $I,J$ are well-chosen subsets of the index sets $t,s$ of length $k$.
The best approximation, which is not too far from the SVD approximation, is attained when the absolute value of the determinant of $(\widetilde{\uuG}{}^\pipl)^{t \times s}_{I,J}$ is maximal over all possible $I$'s and $J$'s \citep{goreinov_theory_1997}.
What makes CUR attractive is its composition of matrices which are formed from rows and columns of the original matrix $(\widetilde{\uuG}{}^\pipl)^{t \times s}$, and so there exist fast algorithms to compute it.
Here, we use the adaptive cross approximation \citep[ACA,][]{bebendorf_approximation_2000} which, starting from $\uuR^0 = (\uuG^\pipl)^{t \times s}$, iteratively subtracts a rank-1 matrix from the remainder $\uuR^k$ as follows%
\footnote{we iteratively cross out the $i_k$-th row and the $j_k$-th column of $(\uuG^\pipl)^{t \times s}$}
\begin{equation}\label{eq:impl.aca}
 \uuR^k = \uuR^{k-1} - \frac{(\uuR^{k-1})_{:,j_k} (\uuR^{k-1})_{i_k,:}}{(\uuR^{k-1})_{i_k,j_k}},
 \qquad
 (\uuR^{k-1})_{:,j_k} \in \real^{N_t\times 1}, \,
 (\uuR^{k-1})_{i_k,:} \in \real^{1\times N_s},
\end{equation}
where $i_k,j_k$ are row and column indices, respectively.
Typically, we stop iterating on \eqref{eq:impl.aca} when the norm of $\uuR^k - \uuR^{k-1}$ is smaller than some prescribed tolerance $\mepsilon$.
The approximate matrix in the low-rank format is then given by
$(\widetilde{\uuG}{}_\scH^\pipl)^{t \times s} = \sum_k (1 / (\uuR^{k-1})_{i_k,j_k}) (\uuR^{k-1})_{:,j_k} (\uuR^{k-1})_{i_k,:}$.
The crucial point in the ACA to achieve fast convergence (or convergence at all) is the computation of $i_k$ and $j_k$.
Ideally they should be chosen so that $(\uuR^{k-1})_{i_k,j_k}$ is the maximum element of $\uuR^{k-1}$.
But this, again, requires the computation of all matrix elements, which scales quadratically.
For this purpose, we use the linear-scaling algorithm for approximating $\underset{i_k,j_k}{\max} \abs{(\uuR^{k-1})_{i_k,j_k}}$ from \citep{bebendorf_approximation_2000} which heuristically chooses $i_k$, and then only searches for the maximum element of the row vector $(\uuR^{k-1})_{i_k,:}$ to determine $j_k$.
We found this to work well for our problems.

Our implementation uses HLIBpro \citep[\href{https://www.hlibpro.com}{www.hlibpro.com},][]{borm_introduction_2003,kriemann_parallel_2005,grasedyck_parallel_2008} which has all of the previously described functionalities.
Further, all numerical experiments conducted in this work use the minimal block size $b_\mrm{min} $\,=\,20, the admissibility constant $\gamma$\,=\,2, and the accuracy bound $\mepsilon$\,=\,$10^{-5}$.

\section{PAD method}
\label{sec:appdx.PAD}

We begin by defining the computational domain of the PAD method as a subset of $\lat$ of size $l_1+b/2$\,$\times$\,$l_2$\,$\times$\,$l_3$, where $b$ is the magnitude of the Burgers vector.
From the lower half-crystal we remove a rectangular cluster of atoms of size $b$\,$\times$\,$l_2$\,$\times$\,$l_3$ next to one of the $\rmx_1$-boundaries.
Now we linearly displace the atoms along the $\rmx_1$-direction in the upper half-crystal from 0 to $-b/2$ and, vice versa, in the lower half-crystal from 0 to $b/2$.
This procedure creates an edge dislocation in the center of a computational domain of size $l_1$\,$\times$\,$l_2$\,$\times$\,$l_3$.
We then apply periodic boundary conditions in the $\rmx_1$-direction and leave free surfaces in the $\rmx_2$-direction.

Upon relaxation of the dislocation core, we fix the pinning points/precipitates as for the FBC method (cf. Sections \ref{sec:results.bowout} and Section \ref{sec:results.precipitate_strenghtening}).
To invoke a bowing of the dislocation under some homogeneous applied shear stress $\tau_\mrm{app}$, we set the external force on all atoms $\bato$ on the upper and lower surface to $\bforce_\mrm{ext}(\bato) = (l_1l_3 / N) \tau_\mrm{app} \bme_1$ and $\bforce_\mrm{ext}(\bato) = -(l_1l_3 / N) \tau_\mrm{app} \bme_1$, respectively, where $N$ is the number of upper/lower surface atoms.
We terminate the simulation when $\| \var{}{}{\Etot^\a} \| < 10^{-2}\,\mrm{eV}/\text{\AA}$.

\end{appendices}


\section*{References}
\bibliographystyle{elsarticle-harv}
\bibliography{flexbc-pbc}

\begin{thebibliography}{69}
\expandafter\ifx\csname natexlab\endcsname\relax\def\natexlab#1{#1}\fi
\expandafter\ifx\csname url\endcsname\relax
  \def\url#1{\texttt{#1}}\fi
\expandafter\ifx\csname urlprefix\endcsname\relax\def\urlprefix{URL }\fi

\bibitem[{Aitken(1927)}]{aitken_bernoullis_1927}
Aitken, A.~C., 1927. On {Bernoulli}'s {Numerical} {Solution} of {Algebraic}
  {Equations}. Proceedings of the Royal Society of Edinburgh 46, 289--305.
\newline\urlprefix\url{https://www.cambridge.org/core/product/identifier/S0370164600022070/type/journal_article}

\bibitem[{Aldakheel et~al.(2021)Aldakheel, Noii, Wick, Allix, and
  Wriggers}]{aldakheel_multilevel_2021}
Aldakheel, F., Noii, N., Wick, T., Allix, O., Wriggers, P., Dec. 2021.
  Multilevel global{\textendash}local techniques for adaptive ductile
  phase-field fracture. Computer Methods in Applied Mechanics and Engineering
  387, 114175.
\newline\urlprefix\url{https://linkinghub.elsevier.com/retrieve/pii/S0045782521005065}

\bibitem[{Anciaux et~al.(2018)Anciaux, Junge, Hodapp, Cho, Molinari, and
  Curtin}]{anciaux_coupled_2018}
Anciaux, G., Junge, T., Hodapp, M., Cho, J., Molinari, J.-F., Curtin, W., Sep.
  2018. The {Coupled} {Atomistic}/{Discrete}-{Dislocation} method in 3d part
  {I}: {Concept} and algorithms. Journal of the Mechanics and Physics of Solids
  118, 152--171.
\newline\urlprefix\url{https://linkinghub.elsevier.com/retrieve/pii/S0022509617310098}

\bibitem[{Andric and Curtin(2019)}]{andric_atomistic_2019}
Andric, P., Curtin, W.~A., Jan. 2019. Atomistic modeling of fracture. Modelling
  and Simulation in Materials Science and Engineering 27~(1), 013001.
\newline\urlprefix\url{http://stacks.iop.org/0965-0393/27/i=1/a=013001?key=crossref.b9362ea1047954cf926f3222a957576b}

\bibitem[{Argon(2007)}]{argon_strengthening_2007}
Argon, A., Aug. 2007. Strengthening {Mechanisms} in {Crystal} {Plasticity}.
  Oxford University Press.
\newline\urlprefix\url{http://www.oxfordscholarship.com/view/10.1093/acprof:oso/9780198516002.001.0001/acprof-9780198516002}

\bibitem[{Bacon et~al.(2009)Bacon, Osetsky, and Rodney}]{bacon_chapter_2009}
Bacon, D., Osetsky, Y., Rodney, D., 2009. Chapter 88
  {Dislocation}{\textendash}{Obstacle} {Interactions} at the {Atomic} {Level}.
  In: Dislocations in {Solids}. Vol.~15. Elsevier, pp. 1--90.
\newline\urlprefix\url{https://linkinghub.elsevier.com/retrieve/pii/S1572485909015010}

\bibitem[{Barnett(1972)}]{barnett_precise_1972}
Barnett, D.~M., Feb. 1972. The precise evaluation of derivatives of the
  anisotropic elastic {Green}'s functions. Physica Status Solidi (b) 49~(2),
  741--748.
\newline\urlprefix\url{http://doi.wiley.com/10.1002/pssb.2220490238}

\bibitem[{Bebendorf(2000)}]{bebendorf_approximation_2000}
Bebendorf, M., 2000. Approximation of boundary element matrices. Numerische
  Mathematik 86, 565--589.

\bibitem[{Bebendorf(2008)}]{bebendorf_hierarchical_2008}
Bebendorf, M., 2008. Hierarchical matrices: a means to efficiently solve
  elliptic boundary value problems. No.~63 in Lecture notes in computational
  science and engineering. Springer, Berlin, oCLC: ocn220011087.

\bibitem[{Bender and Orszag(1999)}]{bender_advanced_1999}
Bender, C.~M., Orszag, S.~A., 1999. Advanced {Mathematical} {Methods} for
  {Scientists} and {Engineers} {I}. Springer New York, New York, NY.
\newline\urlprefix\url{http://link.springer.com/10.1007/978-1-4757-3069-2}

\bibitem[{B{\"o}rm et~al.(2003)B{\"o}rm, Grasedyck, and
  Hackbusch}]{borm_introduction_2003}
B{\"o}rm, S., Grasedyck, L., Hackbusch, W., May 2003. Introduction to
  hierarchical matrices with applications. Engineering Analysis with Boundary
  Elements 27~(5), 405--422.
\newline\urlprefix\url{https://linkinghub.elsevier.com/retrieve/pii/S0955799702001522}

\bibitem[{Braun et~al.(2021)Braun, Hudson, and Ortner}]{braun_asymptotic_2021}
Braun, J., Hudson, T., Ortner, C., Aug. 2021. Asymptotic {Expansion} of the
  {Elastic} {Far}-{Field} of a {Crystalline} {Defect}. arXiv:2108.04765
  [math]ArXiv: 2108.04765.
\newline\urlprefix\url{http://arxiv.org/abs/2108.04765}

\bibitem[{Buze and Kermode(2021)}]{buze_numerical-continuation-enhanced_2021}
Buze, M., Kermode, J.~R., Mar. 2021. A numerical-continuation-enhanced flexible
  boundary condition scheme applied to {Mode} {I} and {Mode} {III} fracture.
  Physical Review E 103~(3), 033002, arXiv: 2008.12822.
\newline\urlprefix\url{http://arxiv.org/abs/2008.12822}

\bibitem[{Campa{\~n}{\'a} and M{\"u}ser(2006)}]{campana_practical_2006}
Campa{\~n}{\'a}, C., M{\"u}ser, M.~H., Aug. 2006. Practical
  {Green}{\textquoteright}s function approach to the simulation of elastic
  semi-infinite solids. Physical Review B 74~(7), 075420.
\newline\urlprefix\url{https://link.aps.org/doi/10.1103/PhysRevB.74.075420}

\bibitem[{Cho et~al.(2018)Cho, Molinari, Curtin, and
  Anciaux}]{cho_coupled_2018}
Cho, J., Molinari, J.-F., Curtin, W.~A., Anciaux, G., Sep. 2018. The coupled
  atomistic/discrete-dislocation method in 3d. {Part} {III}: {Dynamics} of
  hybrid dislocations. Journal of the Mechanics and Physics of Solids 118,
  1--14.
\newline\urlprefix\url{https://linkinghub.elsevier.com/retrieve/pii/S0022509617310128}

\bibitem[{Clouet(2020)}]{andreoni_ab_2020}
Clouet, E., 2020. Ab {Initio} {Models} of {Dislocations}. In: Andreoni, W.,
  Yip, S. (Eds.), Handbook of {Materials} {Modeling}. Springer International
  Publishing, Cham, pp. 1503--1524.
\newline\urlprefix\url{http://link.springer.com/10.1007/978-3-319-44677-6_22}

\bibitem[{Curtin and Miller(2003)}]{curtin_atomistic/continuum_2003}
Curtin, W.~A., Miller, R.~E., 2003. Atomistic/continuum coupling in
  computational materials science. Modelling and Simulation in Materials
  Science and Engineering 11~(3), 33--68.
\newline\urlprefix\url{https://doi.org/10.1088%2F0965-0393%2F11%2F3%2F201}

\bibitem[{Daw et~al.(1993)Daw, Foiles, and Baskes}]{daw_embedded-atom_1993}
Daw, M.~S., Foiles, S.~M., Baskes, M.~I., Mar. 1993. The embedded-atom method:
  a review of theory and applications. Materials Science Reports 9~(7-8),
  251--310.
\newline\urlprefix\url{https://linkinghub.elsevier.com/retrieve/pii/092023079390001U}

\bibitem[{Dedner et~al.(2017)Dedner, Ortner, and Wu}]{dedner_coupling_2017}
Dedner, A.~S., Ortner, C., Wu, H., Sep. 2017. Coupling {Atomistic},
  {Elasticity} and {Boundary} {Element} {Models}. arXiv:1709.05977 [math]ArXiv:
  1709.05977.
\newline\urlprefix\url{http://arxiv.org/abs/1709.05977}

\bibitem[{Dewald and Curtin(2006)}]{dewald_analysis_2006}
Dewald, M., Curtin, W.~A., Apr. 2006. Analysis and minimization of dislocation
  interactions with atomistic/continuum interfaces. Modelling and Simulation in
  Materials Science and Engineering 14~(3), 497--514.
\newline\urlprefix\url{http://stacks.iop.org/0965-0393/14/i=3/a=011?key=crossref.e545a9873cccc7b248423c9c6c7016cd}

\bibitem[{Ehrlacher et~al.(2016)Ehrlacher, Ortner, and
  Shapeev}]{ehrlacher_analysis_2016}
Ehrlacher, V., Ortner, C., Shapeev, A.~V., Dec. 2016. Analysis of {Boundary}
  {Conditions} for {Crystal} {Defect} {Atomistic} {Simulations}. Archive for
  Rational Mechanics and Analysis 222~(3), 1217--1268.
\newline\urlprefix\url{http://link.springer.com/10.1007/s00205-016-1019-6}

\bibitem[{Fang and Zhang(2020)}]{fang_blended_2020}
Fang, L., Zhang, L., Jun. 2020. Blended {Ghost} {Force} {Correction} {Method}
  for {3D} {Crystalline} {Defects}. arXiv:2006.06501 [cs, math]ArXiv:
  2006.06501.
\newline\urlprefix\url{http://arxiv.org/abs/2006.06501}

\bibitem[{Gallego and Ortiz(1993)}]{gallego_harmonic/anharmonic_1993}
Gallego, R., Ortiz, M., Jul. 1993. A harmonic/anharmonic energy partition
  method for lattice statics computations. Modelling and Simulation in
  Materials Science and Engineering 1~(4), 417--436.
\newline\urlprefix\url{http://stacks.iop.org/0965-0393/1/i=4/a=006?key=crossref.9032252d69927d10d86054ea74cd740b}

\bibitem[{Goreinov et~al.(1997)Goreinov, Tyrtyshnikov, and
  Zamarashkin}]{goreinov_theory_1997}
Goreinov, S.~A., Tyrtyshnikov, E.~E., Zamarashkin, N.~L., 1997. A {Theory} of
  {Pseudoskeleton} {Approximations}. Linear Algebra and its Applications 261,
  1--21.

\bibitem[{Grasedyck et~al.(2008)Grasedyck, Kriemann, and
  Le~Borne}]{grasedyck_parallel_2008}
Grasedyck, L., Kriemann, R., Le~Borne, S., Sep. 2008. Parallel black box
  {H}-{LU} preconditioning for elliptic boundary value problems. Computing and
  Visualization in Science 11~(4-6), 273--291.
\newline\urlprefix\url{http://link.springer.com/10.1007/s00791-008-0098-9}

\bibitem[{Hackbusch(1999)}]{hackbusch_sparse_1999}
Hackbusch, W., 1999. A {Sparse} {Matrix} {Arithmetic} based on {H}-{Matrices}.
  {Part} {I}: {Introduction} to {H}-{Matrices}. Computing 62, 89--108.

\bibitem[{Hirth and Lothe(1982)}]{hirth_theory_1982}
Hirth, J.~P., Lothe, J., 1982. Theory of dislocations. John Wiley \& Sons.

\bibitem[{Hodapp(2018)}]{hodapp_flexible_2018}
Hodapp, M., 2018. On flexible {Green} function methods for atomistic/continuum
  coupling. {PhD} thesis, {\'E}cole polytechnique f{\'e}d{\'e}rale de Lausanne.

\bibitem[{Hodapp(2021{\natexlab{a}})}]{hodapp_analysis_2021}
Hodapp, M., Jan. 2021{\natexlab{a}}. Analysis of a {Sinclair}-{Type} {Domain}
  {Decomposition} {Solver} for {Atomistic}/{Continuum} {Coupling}. Multiscale
  Modeling \& Simulation 19~(4), 1499--1537.
\newline\urlprefix\url{https://epubs.siam.org/doi/10.1137/19M130861X}

\bibitem[{Hodapp(2021{\natexlab{b}})}]{hodapp_efficient_2021}
Hodapp, M., May 2021{\natexlab{b}}. Efficient flexible boundary conditions for
  long dislocations. arXiv:2105.08798v1 [physics]ArXiv: 2105.08798v1.
\newline\urlprefix\url{http://arxiv.org/abs/2105.08798v1}

\bibitem[{Hodapp et~al.(2019)Hodapp, Anciaux, and Curtin}]{hodapp_lattice_2019}
Hodapp, M., Anciaux, G., Curtin, W., May 2019. Lattice {Green} function methods
  for atomistic/continuum coupling: {Theory} and data-sparse implementation.
  Computer Methods in Applied Mechanics and Engineering 348, 1039--1075.
\newline\urlprefix\url{https://linkinghub.elsevier.com/retrieve/pii/S0045782519300775}

\bibitem[{Hodapp et~al.(2018)Hodapp, Anciaux, Molinari, and
  Curtin}]{hodapp_coupled_2018}
Hodapp, M., Anciaux, G., Molinari, J.-F., Curtin, W., Oct. 2018. Coupled
  atomistic/discrete dislocation method in {3D} {Part} {II}: {Validation} of
  the method. Journal of the Mechanics and Physics of Solids 119, 1--19.
\newline\urlprefix\url{https://linkinghub.elsevier.com/retrieve/pii/S0022509617310116}

\bibitem[{Hull and Bacon(2011)}]{hull_introduction_2011}
Hull, D., Bacon, D.~J., 2011. Introduction to dislocations, 5th Edition.
  Elsevier/Butterworth-Heinemann, Amsterdam, oCLC: 704891549.

\bibitem[{Knap and Ortiz(2001)}]{knap_analysis_2001}
Knap, J., Ortiz, M., Sep. 2001. An analysis of the quasicontinuum method.
  Journal of the Mechanics and Physics of Solids 49~(9), 1899--1923.
\newline\urlprefix\url{https://linkinghub.elsevier.com/retrieve/pii/S0022509601000345}

\bibitem[{Kochmann and Venturini(2014)}]{kochmann_meshless_2014}
Kochmann, D.~M., Venturini, G.~N., Apr. 2014. A meshless quasicontinuum method
  based on local maximum-entropy interpolation. Modelling and Simulation in
  Materials Science and Engineering 22~(3), 034007.
\newline\urlprefix\url{http://stacks.iop.org/0965-0393/22/i=3/a=034007?key=crossref.1c71666a6a31fc5b99b546047cf19d95}

\bibitem[{Kohlhoff and Schmauder(1989)}]{kohlhoff_new_1989}
Kohlhoff, S., Schmauder, S., 1989. A {New} {Method} for {Coupled}
  {Elastic}-{Atomistic} {Modelling}. In: Atomistic {Simulation} of {Materials}.
  Springer US, Boston, MA, pp. 411--418.

\bibitem[{Kriemann(2005)}]{kriemann_parallel_2005}
Kriemann, R., May 2005. Parallel {H}-{Matrix} {Arithmetics} on {Shared}
  {Memory} {Systems}. Computing 74~(3), 273--297.
\newline\urlprefix\url{http://link.springer.com/10.1007/s00607-004-0102-2}

\bibitem[{Lemar{\'e}chal(1971)}]{lemarechal_methode_1971}
Lemar{\'e}chal, C., 1971. Une m{\'e}thode de r{\'e}solution de certains
  syst{\`e}mes non lin{\'e}aires bien pos{\'e}s. C. R. Math. Acad. Sci. Paris,
  605--607.

\bibitem[{Li(2009)}]{li_efficient_2009}
Li, X., Sep. 2009. Efficient boundary conditions for molecular statics models
  of solids. Physical Review B 80~(10), 104112.
\newline\urlprefix\url{https://link.aps.org/doi/10.1103/PhysRevB.80.104112}

\bibitem[{Li(2012)}]{li_atomistic-based_2012}
Li, X., Jun. 2012. An atomistic-based boundary element method for the reduction
  of molecular statics models. Computer Methods in Applied Mechanics and
  Engineering 225-228, 1--13.
\newline\urlprefix\url{https://linkinghub.elsevier.com/retrieve/pii/S0045782512000825}

\bibitem[{Liao et~al.(2020)Liao, Lin, and Zhang}]{liao_posteriori_2020}
Liao, M., Lin, P., Zhang, L., Jun. 2020. A {Posteriori} {Error} {Estimate} and
  {Adaptive} {Mesh} {Refinement} {Algorithm} for {Atomistic}/{Continuum}
  {Coupling} with {Finite} {Range} {Interactions} in {Two} {Dimensions}.
  Communications in Computational Physics 27~(1), 198--226.
\newline\urlprefix\url{http://global-sci.org/intro/article_detail/cicp/13319.html}

\bibitem[{Luskin and Ortner(2013)}]{luskin_atomistic--continuum_2013}
Luskin, M., Ortner, C., May 2013. Atomistic-to-continuum coupling. Acta
  Numerica 22, 397--508.
\newline\urlprefix\url{https://www.cambridge.org/core/product/identifier/S0962492913000068/type/journal_article}

\bibitem[{Monti et~al.(2021)Monti, Pastewka, and Robbins}]{monti_greens_2021}
Monti, J.~M., Pastewka, L., Robbins, M.~O., May 2021. Green's function method
  for dynamic contact calculations. Physical Review E 103~(5), 053305.
\newline\urlprefix\url{https://link.aps.org/doi/10.1103/PhysRevE.103.053305}

\bibitem[{Ortner and Zhang(2014)}]{ortner_energy-based_2014}
Ortner, C., Zhang, L., Sep. 2014. Energy-based atomistic-to-continuum coupling
  without ghost forces. Computer Methods in Applied Mechanics and Engineering
  279, 29--45.
\newline\urlprefix\url{https://linkinghub.elsevier.com/retrieve/pii/S0045782514002059}

\bibitem[{Osetsky and Bacon(2003)}]{osetsky_atomic-level_2003}
Osetsky, Y.~N., Bacon, D.~J., Jul. 2003. An atomic-level model for studying the
  dynamics of edge dislocations in metals. Modelling and Simulation in
  Materials Science and Engineering 11~(4), 427--446.
\newline\urlprefix\url{http://stacks.iop.org/0965-0393/11/i=4/a=302?key=crossref.b4e9b78d52f6ce9b04d26305ab55c49a}

\bibitem[{Pavia and Curtin(2015)}]{pavia_parallel_2015}
Pavia, F., Curtin, W.~A., Jul. 2015. Parallel algorithm for multiscale
  atomistic/continuum simulations using {LAMMPS}. Modelling and Simulation in
  Materials Science and Engineering 23~(5), 055002.
\newline\urlprefix\url{http://stacks.iop.org/0965-0393/23/i=5/a=055002?key=crossref.02ff7625c3b88ed47b97e830253f79d1}

\bibitem[{Plimpton(1995)}]{plimpton_fast_1995}
Plimpton, S., Mar. 1995. Fast {Parallel} {Algorithms} for {Short}-{Range}
  {Molecular} {Dynamics}. Journal of Computational Physics 117~(1), 1--19.
\newline\urlprefix\url{https://linkinghub.elsevier.com/retrieve/pii/S002199918571039X}

\bibitem[{Rami{\`e}re and Helfer(2015)}]{ramiere_iterative_2015}
Rami{\`e}re, I., Helfer, T., Nov. 2015. Iterative residual-based vector methods
  to accelerate fixed point iterations. Computers \& Mathematics with
  Applications 70~(9), 2210--2226.
\newline\urlprefix\url{https://linkinghub.elsevier.com/retrieve/pii/S0898122115004046}

\bibitem[{Rao et~al.(2013)Rao, Dimiduk, El-Awady, Parthasarathy, Uchic, and
  Woodward}]{rao_spontaneous_2013}
Rao, S., Dimiduk, D., El-Awady, J., Parthasarathy, T., Uchic, M., Woodward, C.,
  Aug. 2013. Spontaneous athermal cross-slip nucleation at screw dislocation
  intersections in {FCC} metals and {L1} $_{\textrm{2}}$ intermetallics
  investigated via atomistic simulations. Philosophical Magazine 93~(22),
  3012--3028.
\newline\urlprefix\url{http://www.tandfonline.com/doi/abs/10.1080/14786435.2013.799788}

\bibitem[{Rao et~al.(1998)Rao, Hernandez, Simmons, Parthasarathy, and
  Woodward}]{rao_greens_1998}
Rao, S., Hernandez, C., Simmons, J.~P., Parthasarathy, T.~A., Woodward, C.,
  Jan. 1998. Green's function boundary conditions in two-dimensional and
  three-dimensional atomistic simulations of dislocations. Philosophical
  Magazine A 77~(1), 231--256.
\newline\urlprefix\url{http://www.tandfonline.com/doi/abs/10.1080/01418619808214240}

\bibitem[{Saint-Venant and {others}(1855)}]{saint-venant_memoire_1855}
Saint-Venant, B.~d., {others}, 1855. M{\'e}moire sur la torsion des prismes.
  M{\'e}moires des Savants {\'e}trangers 14, 233--560.

\bibitem[{Seeger(1957)}]{seeger_mechanism_1957}
Seeger, A., 1957. The {Mechanism} of {Glide} and {Work} {Hardening} in
  {Face}-{Centered} {Cubic} and {Hexagonal} {Close}-{Packed} {Metals}. In:
  Dislocations and {Mechanical} {Properties} of {Crystals}. Ed. by J. Fisher et
  al., John Wiley \& Sons, Inc., pp. 243--329.

\bibitem[{Shenoy and Phillips(1997)}]{shenoy_finite-sized_1997}
Shenoy, V.~B., Phillips, R., Aug. 1997. Finite-sized atomistic simulations of
  screw dislocations. Philosophical Magazine A 76~(2), 367--385.
\newline\urlprefix\url{http://www.tandfonline.com/doi/abs/10.1080/01418619708209981}

\bibitem[{Shimokawa et~al.(2004)Shimokawa, Mortensen, Schi{\o}tz, and
  Jacobsen}]{shimokawa_matching_2004}
Shimokawa, T., Mortensen, J.~J., Schi{\o}tz, J., Jacobsen, K.~W., Jun. 2004.
  Matching conditions in the quasicontinuum method: {Removal} of the error
  introduced at the interface between the coarse-grained and fully atomistic
  region. Physical Review B 69~(21), 214104.
\newline\urlprefix\url{https://link.aps.org/doi/10.1103/PhysRevB.69.214104}

\bibitem[{Sinclair(1971)}]{sinclair_improved_1971}
Sinclair, J.~E., Dec. 1971. Improved {Atomistic} {Model} of a bcc {Dislocation}
  {Core}. Journal of Applied Physics 42~(13), 5321--5329.
\newline\urlprefix\url{http://aip.scitation.org/doi/10.1063/1.1659943}

\bibitem[{Sinclair(1975)}]{sinclair_influence_1975}
Sinclair, J.~E., Mar. 1975. The influence of the interatomic force law and of
  kinks on the propagation of brittle cracks. Philosophical Magazine 31~(3),
  647--671.
\newline\urlprefix\url{http://www.tandfonline.com/doi/abs/10.1080/14786437508226544}

\bibitem[{Sinclair et~al.(1978)Sinclair, Gehlen, Hoagland, and
  Hirth}]{sinclair_flexible_1978}
Sinclair, J.~E., Gehlen, P.~C., Hoagland, R.~G., Hirth, J.~P., Jul. 1978.
  Flexible boundary conditions and nonlinear geometric effects in atomic
  dislocation modeling. Journal of Applied Physics 49~(7), 3890--3897.
\newline\urlprefix\url{http://aip.scitation.org/doi/10.1063/1.325395}

\bibitem[{Singh et~al.(2011)Singh, Mateos, and Warner}]{singh_atomistic_2011}
Singh, C., Mateos, A., Warner, D., Mar. 2011. Atomistic simulations of
  dislocation{\textendash}precipitate interactions emphasize importance of
  cross-slip. Scripta Materialia 64~(5), 398--401.
\newline\urlprefix\url{https://linkinghub.elsevier.com/retrieve/pii/S135964621000744X}

\bibitem[{Stukowski(2010)}]{stukowski_visualization_2010}
Stukowski, A., Jan. 2010. Visualization and analysis of atomistic simulation
  data with {OVITO}{\textendash}the {Open} {Visualization} {Tool}. Modelling
  and Simulation in Materials Science and Engineering 18~(1), 015012.
\newline\urlprefix\url{https://iopscience.iop.org/article/10.1088/0965-0393/18/1/015012}

\bibitem[{Stukowski et~al.(2012)Stukowski, Bulatov, and
  Arsenlis}]{stukowski_automated_2012}
Stukowski, A., Bulatov, V.~V., Arsenlis, A., Dec. 2012. Automated
  identification and indexing of dislocations in crystal interfaces. Modelling
  and Simulation in Materials Science and Engineering 20~(8), 085007.
\newline\urlprefix\url{http://stacks.iop.org/0965-0393/20/i=8/a=085007?key=crossref.f9ca36d5353ac80ccfbbfd992a3fd702}

\bibitem[{Szajewski and Curtin(2015)}]{szajewski_analysis_2015}
Szajewski, B.~A., Curtin, W.~A., Mar. 2015. Analysis of spurious image forces
  in atomistic simulations of dislocations. Modelling and Simulation in
  Materials Science and Engineering 23~(2), 025008.
\newline\urlprefix\url{http://stacks.iop.org/0965-0393/23/i=2/a=025008?key=crossref.499111fdd52d253e7f34ec0df4903766}

\bibitem[{Tadmor et~al.(1996)Tadmor, Ortiz, and
  Phillips}]{tadmor_quasicontinuum_1996}
Tadmor, E.~B., Ortiz, M., Phillips, R., Jun. 1996. Quasicontinuum analysis of
  defects in solids. Philosophical Magazine A 73~(6), 1529--1563.
\newline\urlprefix\url{http://www.tandfonline.com/doi/abs/10.1080/01418619608243000}

\bibitem[{Tan et~al.(2019)Tan, Woodward, and Trinkle}]{tan_dislocation_2019}
Tan, A. M.~Z., Woodward, C., Trinkle, D.~R., Mar. 2019. Dislocation core
  structures in {Ni}-based superalloys computed using a density functional
  theory based flexible boundary condition approach. Physical Review Materials
  3~(3), 033609.
\newline\urlprefix\url{https://link.aps.org/doi/10.1103/PhysRevMaterials.3.033609}

\bibitem[{Tyrtyshnikov(1996)}]{tyrtyshnikov_mosaic-skeleton_1996}
Tyrtyshnikov, E., Jun. 1996. Mosaic-{Skeleton} approximations. Calcolo
  33~(1-2), 47--57.
\newline\urlprefix\url{http://link.springer.com/10.1007/BF02575706}

\bibitem[{Varvenne et~al.(2016)Varvenne, Luque, N{\"o}hring, and
  Curtin}]{varvenne_average-atom_2016}
Varvenne, C., Luque, A., N{\"o}hring, W.~G., Curtin, W.~A., Mar. 2016.
  Average-atom interatomic potential for random alloys. Physical Review B
  93~(10), 104201.
\newline\urlprefix\url{https://link.aps.org/doi/10.1103/PhysRevB.93.104201}

\bibitem[{Woodward and Rao(2002)}]{woodward_flexible_2002}
Woodward, C., Rao, S.~I., May 2002. Flexible \textit{{Ab} {Initio}} {Boundary}
  {Conditions}: {Simulating} {Isolated} {Dislocations} in bcc {Mo} and {Ta}.
  Physical Review Letters 88~(21).
\newline\urlprefix\url{https://link.aps.org/doi/10.1103/PhysRevLett.88.216402}

\bibitem[{Xiao and Belytschko(2004)}]{xiao_bridging_2004}
Xiao, S., Belytschko, T., May 2004. A bridging domain method for coupling
  continua with molecular dynamics. Computer Methods in Applied Mechanics and
  Engineering 193~(17-20), 1645--1669.
\newline\urlprefix\url{https://linkinghub.elsevier.com/retrieve/pii/S004578250400026X}

\bibitem[{Xiong et~al.(2011)Xiong, Tucker, McDowell, and
  Chen}]{xiong_coarse-grained_2011}
Xiong, L., Tucker, G., McDowell, D., Chen, Y., 2011. Coarse-grained atomistic
  simulation of dislocations. J. Mech. Phys. Solids, 18.
\newline\urlprefix\url{https://linkinghub.elsevier.com/retrieve/pii/S0022509610002395}

\bibitem[{Yavari et~al.(2006)Yavari, Ortiz, and
  Bhattacharya}]{yavari_theory_2006}
Yavari, A., Ortiz, M., Bhattacharya, K., Dec. 2006. A {Theory} of {Anharmonic}
  {Lattice} {Statics} for {Analysis} of {Defective} {Crystals}. Journal of
  Elasticity 86~(1), 41--83.
\newline\urlprefix\url{http://link.springer.com/10.1007/s10659-006-9079-8}

\end{thebibliography}

\end{document}